\documentclass[prd,showkeys,floatfix,twocolumn,amsmath,amssymb,floatfix]{revtex4}
\usepackage{graphicx}
\usepackage{epstopdf}
\usepackage{enumitem}
\usepackage{multirow}
\usepackage{subfigure}
\usepackage{bm}
\usepackage[colorlinks,citecolor=blue,anchorcolor=red,menucolor=red,linkcolor=red,filecolor=red,runcolor=red,urlcolor=blue,frenchlinks=red]{hyperref}

\newcommand{\feynp}[1]{#1\kern-0.45em/}
\def\beq{\begin{equation}}
\def\enq{\end{equation}}
\def\beqa{\begin{eqnarray}}
\def\enqa{\end{eqnarray}}

\def\be{\begin{equation}}
\def\ee{\end{equation}}
\def\bea{\begin{eqnarray}}
\def\eea{\end{eqnarray}}
\newcommand{\slashed}{\not\!}

\allowdisplaybreaks[3]

\begin{document}

\title{A short review on QCD sum rule studies of $P$-wave single heavy baryons}

\author{Xuan Luo$^1$}
\author{Shu-Wei Zhang$^1$}
\author{Hua-Xing Chen$^1$}
\email{hxchen@seu.edu.cn}
\author{Atsushi Hosaka$^{2,3}$}
\email{hosaka@rcnp.osaka-u.ac.jp}
\author{Niu Su$^4$}
\author{Hui-Min Yang$^{1,5}$}
\affiliation{$^1$School of Physics, Southeast University, Nanjing 210094, China\\
$^2$Research Center for Nuclear Physics, Osaka University, Ibaraki 567-0047, Japan\\
$^3$Advanced Science Research Center, Japan Atomic Energy Agency, Tokai 319-1195, Japan\\
$^4$College of Physics and Optoelectronic Engineering, Taiyuan University of Technology, Taiyuan 030024, China\\
$^5$School of Physics and Center of High Energy Physics, Peking University, Beijing 100871, China}

\begin{abstract}
Over the past few decades, the study of singly heavy baryons has entered a golden era, with numerous excited states observed by experimental collaborations. Various theoretical approaches have been developed to investigate their properties, with the QCD sum rule method being one of the most widely applied. This paper provides a review of these QCD sum rule studies. Over the last ten years, we have systematically studied $P$-wave singly heavy baryons using QCD sum rules and light-cone sum rules within the framework of heavy quark effective theory. These $P$-wave singly heavy baryons can explain many excited heavy baryons, including the $\Lambda_c(2595)^+$, $\Lambda_c(2625)^+$, $\Xi_c(2790)^{0/+}$, $\Xi_c(2815)^{0/+}$, $\Sigma_c(2800)^0$, $\Xi_c(2882)^0$, $\Xi_c(2923)^0$, $\Xi_c(2939)^0$, $\Xi_c(2965)^0$, $\Omega_c(3000)^0$, $\Omega_c(3066)^0$, $\Omega_c(3090)^0$, $\Omega_c(3050)^0$, $\Omega_c(3119)^0$, $\Lambda_b(5912)^0$, $\Lambda_b(5920)^0$, $\Xi_b(6087)^0$, $\Xi_b(6095)^0/\Xi_b(6100)^-$, $\Sigma_b(6097)^\pm$, $\Xi_b(6227)^-$, $\Omega_b(6316)^-$, $\Omega_b(6330)^-$, $\Omega_b(6340)^-$, and $\Omega_b(6350)^-$, etc. Furthermore, we predict additional $P$-wave singly heavy baryons, including two $\Lambda_b$ states, two $\Xi_b$ states, three $\Sigma_b$ states, three $\Xi_b^\prime$ states, two $\Omega_b$ states, two $\Lambda_c$ states, two $\Xi_c$ states, three $\Sigma_c$ states, and one $\Omega_c$ state, all with relatively narrow decay widths, making them viable candidates for experimental observation. The study of singly heavy baryons is closely related to two meaningful questions: ``\emph{What is the shortest possible lifetime of an observable particle}'' and ``\emph{How can one generally describe approximate (flavor) symmetries}''.
\end{abstract}
\pagenumbering{arabic}
%
%

\maketitle


\section{Introduction}
\label{sec1}

The strong interaction binds quarks and gluons into hadrons, much like the electromagnetic force binds electrons and nucleons into atoms. A particularly effective system for investigating the nature of this interaction is the singly heavy baryon, which consists of one heavy quark (either charm or bottom) and two light quarks (up, down, or strange), as described by the quark model~\cite{Gell-Mann:1964ewy,Zweig:1964jf}. In such baryons, the light quarks and gluons orbit around a nearly static heavy quark, resembling the structure of a hydrogen atom, where an electron revolves around a stationary proton~\cite{Korner:1994nh,Manohar:2000dt,Bianco:2003vb,Klempt:2009pi}. Just as the electromagnetic interaction leads to fine structure in the hydrogen spectrum, the strong interaction gives rise to fine structure in the spectra of singly heavy baryons~\cite{Chen:2016spr,Chen:2022asf,Copley:1979wj,Karliner:2008sv}.

Over the past few decades, the study of singly heavy baryons has entered a golden era. All ground-state ($1S$) singly heavy baryons have been experimentally confirmed~\cite{PDG2024}, with the sole exception of the $\Omega_b^*$. Furthermore, a variety of excited states have been observed by experimental collaborations. Investigating their mass spectra, production mechanisms, and decay properties is crucial for uncovering the internal structure and the QCD dynamics that govern these systems. To address these questions, numerous theoretical approaches and phenomenological models have been developed. These include various quark models~\cite{Capstick:1985xss,Tawfiq:1998nk,Ivanov:1998qe,Hussain:1999sp,Albertus:2005zy,Hwang:2006df,Ebert:2007nw,Garcilazo:2007eh,Roberts:2007ni,Zhong:2007gp,Valcarce:2008dr,Ebert:2011kk,Ortega:2012cx,Yoshida:2015tia,Nagahiro:2016nsx,Lu:2017meb,Wang:2017vnc,Wang:2017kfr,Kawakami:2018olq,Wang:2020gkn,Lu:2020ivo,Chen:2021eyk,Arifi:2020yfp,Garcia-Tecocoatzi:2023btk,He:2023gqh,Zhou:2023xxs,Peng:2024pyl,Weng:2024roa}, lattice QCD simulations~\cite{UKQCD:1996ssj,Burch:2008qx,Padmanath:2013bla,Brown:2014ena,Detmold:2015aaa,Padmanath:2017lng,Bahtiyar:2020uuj}, the $^3P_0$ decay model~\cite{Chen:2007xf,Chen:2017gnu,Zhao:2017fov,Ye:2017dra,Ye:2017yvl}, chiral perturbation theory~\cite{Cho:1994vg,Huang:1995ke,Pirjol:1997nh,Chiladze:1997ev,Blechman:2003mq,Cheng:2006dk,Lu:2014ina,Cheng:2015naa,Shi:2021kmm}, chiral unitary model~\cite{Lu:2016gev,Liang:2017ejq,Dias:2018qhp,Yu:2018yxl,Liang:2020dxr}, the Bethe-Salpeter formalism~\cite{Wang:2017smo,Guo:2007qu} and the Regge trajectories~\cite{Shah:2017jkr,Chen:2018nnr,Jia:2018vwl,Jia:2019bkr}, etc.

The method of QCD sum rules has also been widely applied in the study of singly heavy baryons, {\it e.g.}, in Refs.~\cite{Shuryak:1981fza,Bagan:1991sc,Grozin:1992td,Dai:1995bc,Groote:1996em,Huang:2000tn,Lee:2000tb,Wang:2002ts,Huang:2004vf,Liu:2007fg,Duraes:2007te,Liu:2008zi,Liu:2009sn,Aliev:2009jt,Wang:2009cr,Khodjamirian:2011jp,Liu:2014uha,Wang:2017vtv,Aliev:2018yjo,Azizi:2020tgh,Azizi:2020ljx,Wu:2021tzo,Duan:2022uzm,Nishikawa:2024lnh,Duan:2024lnw,Ozdem:2024ydl,Ozdem:2024brk}. We have conducted in Refs.~\cite{Chen:2015kpa,Mao:2015gya,Chen:2016phw,Chen:2017sci,Mao:2017wbz,Yang:2019cvw,Chen:2020mpy,Mao:2020jln,Yang:2020zrh,Yang:2020zjl,Yang:2021lce,Yang:2022oog,Tan:2023opd,Yang:2023fsc,Luo:2024jov,Wang:2024rai,Luo:2025jpn,Luo:2025pzb} a comprehensive series of investigations into the properties of $P$-wave charmed and bottom baryons using QCD sum rules~\cite{Shifman:1978bx,Shifman:1978by,Reinders:1984sr,Colangelo:2000dp,Narison:2002woh,Gimenez:2005nt,Nielsen:2009uh,Gubler:2018ctz} and light-cone sum rules~\cite{Braun:1988qv,Chernyak:1990ag,Ball:1998je,Ball:1998kk,Ball:1998sk,Ball:1998ff,Ball:2002ps,Ball:2004rg,Ball:2006wn,Ball:2007rt,Ball:2007zt}, within the framework of heavy quark effective theory (HQET)~\cite{Eichten:1989zv,Grinstein:1990mj,Falk:1990yz,Neubert:1993mb}. This paper is devoted to reviewing these QCD sum rule studies, and we refer to the reviews~\cite{PDG2024,Chen:2016spr,Chen:2022asf,Copley:1979wj,Korner:1994nh,Ivanov:1998ms,Capstick:2000qj,Bali:2000gf,Manohar:2000dt,Aoki:2001ra,Bianco:2003vb,Colangelo:2004vu,Rosner:2006jz,Guo:2017jvc,Ali:2017jda,Karliner:2008sv,Klempt:2009pi,Crede:2013kia,Cheng:2015iom,Ayala:2016vnt,Guo:2019twa,Agaev:2020zad,Chen:2021ftn,Cheng:2021qpd,Meng:2022ozq,Mai:2022eur,Gross:2022hyw,Huang:2023jec,Nieves:2024dcz,Wang:2024jyk,Crede:2024hur} for more theoretical and experimental discussions. It is also worth noting that these techniques have recently been extended to explore hypothetical hadronic systems containing a single top quark, such as topped mesons~\cite{Zhang:2025fdp} and singly topped baryons~\cite{Zhang:2025xxd}.

In the remainder of this section, we begin by classifying singly heavy baryons within the framework of HQET and summarizing the corresponding interpolating currents used in QCD sum rule analyses. Section~\ref{sec2} presents recent experimental progress. Section~\ref{sec3} outlines the theoretical methodologies of QCD sum rules and light-cone sum rules. Section~\ref{sec4} presents recent theoretical progress through the QCD sum rule method. Finally, Section~\ref{sec5} provides a summary of our conclusions.

\subsection{Categorization of singly heavy baryons}
\label{sec1.1}

To establish a clear understanding of singly heavy baryons, it is important to develop a systematic classification based on their internal symmetries and quantum numbers. In the conventional quark model, these baryons consist of one heavy quark \( Q \) (charm or bottom) and two light quarks \( q_{1,2} \) (up, down, or strange). However, with the development of quantum chromodynamics (QCD), it is now understood that baryons also contain a sea of quark-antiquark pairs and gluons, rendering their internal structure significantly more complex, when described by the bare quark and gluon degrees of freedom. Despite this complexity, the traditional quark model remains remarkably successful, as many hadronic properties can be effectively described by considering only the valence quarks that are regarded as renormalized degrees of freedom (quasi particles) of the bare quarks via the interaction with gluons. This success largely stems from the fact that hadrons exhibit well-defined internal symmetries associated with color, flavor, spin, and orbital motion. The interplay of these symmetries provides a robust foundation for organizing singly heavy baryons into multiplets.

\begin{figure}[hbtp]
\begin{center}
\includegraphics[width=0.3\textwidth]{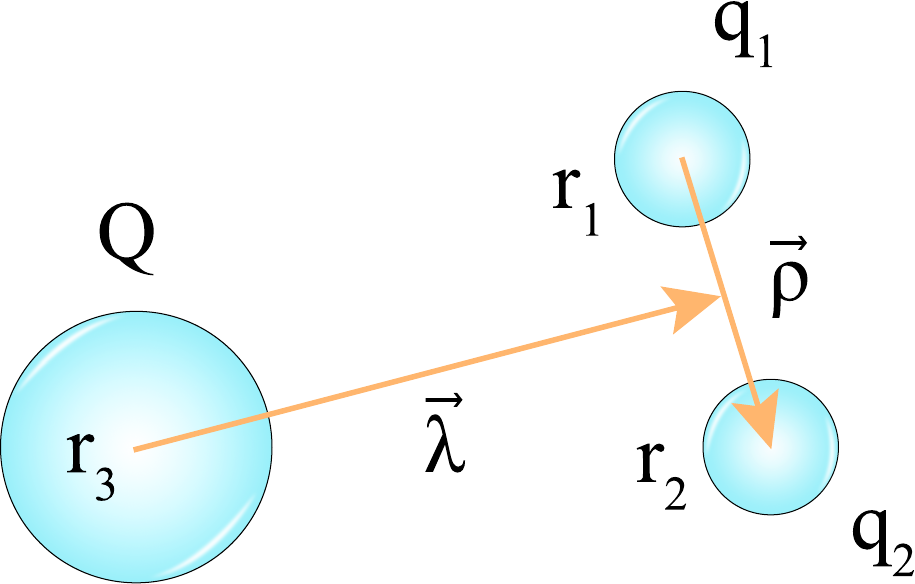}
\end{center}
\caption{Jacobi coordinates \( \vec{\lambda} \) and \( \vec{\rho} \) for a singly heavy baryon.}
\label{fig:Jacobi}
\end{figure}

The quantum numbers of a singly heavy baryon \( q_1 q_2 Q \) can be characterized using Jacobi coordinates, as illustrated in Fig.~\ref{fig:Jacobi}. The relevant angular momenta are defined as follows:
\begin{itemize}

\item The total orbital angular momentum for the internal motion of the three quarks is decomposed as
\begin{equation}
L = l_\lambda \oplus l_\rho \, ,
\end{equation}
where \( l_\lambda \) denotes the orbital angular momentum between the heavy quark and the light diquark, and \( l_\rho \) corresponds to the orbital angular momentum between the two light quarks. These are referred to as the \( \lambda \)-mode and \( \rho \)-mode excitations, respectively.

\item The total spin angular momentum is given by
\begin{equation}
S = s_Q \oplus s_{q_1} \oplus s_{q_2} = s_Q \oplus s_l \, ,
\end{equation}
where \( s_Q \) is the spin of the heavy quark, and \( s_l = s_{q_1} \oplus s_{q_2} \) denotes the total spin of the light-quark subsystem.

\item In the heavy quark limit, the total angular momentum is formed as the sum of the heavy and light components
\begin{equation}
J = s_Q \oplus \left[\left(l_\lambda \oplus l_\rho \right)_L \oplus \left(s_{q_1} \oplus s_{q_2} \right)_{s_l} \right]_{j_l} \, ,
\end{equation}
where \( j_l = L \oplus s_l = l_\lambda \oplus l_\rho \oplus s_{q_1} \oplus s_{q_2} \) represents the total angular momentum of the light degrees of freedom.

\end{itemize}

When two baryons share the same quantum numbers for light degrees of freedom \( (l_\lambda, l_\rho, L, s_l, j_l) \), they form a degenerate doublet with total angular momenta \( J = j_l \pm 1/2 \). If only a single such configuration exists, the baryon is referred to as a singlet. The classification is further constrained by the Pauli exclusion principle, which requires that the total wavefunction of the two light quarks be antisymmetric under their exchange. This condition imposes the following requirements:
\begin{itemize}

\item The color wave function of the two light quarks is always antisymmetric, corresponding to the \( \mathbf{\bar{3}}_C \) representation.

\item The flavor \( SU(3) \) structure may be either symmetric (\( \mathbf{6}_F \)) or antisymmetric (\( \mathbf{\bar{3}}_F \)) under the exchange of the two light quarks.

\item The spin configuration may also be either symmetric (\( s_l = 1 \)) or antisymmetric (\( s_l = 0 \)).

\item The spatial wave function is symmetric for even values of \( l_\rho \) (i.e., \( l_\rho = 0, 2, 4, \ldots \)) and antisymmetric for odd values of \( l_\rho \) (i.e., \( l_\rho = 1, 3, 5, \ldots \)).

\end{itemize}
Combining these symmetry considerations, singly heavy baryons can be systematically classified into various multiplets. Each multiplet is characterized by a set of quantum numbers \([ \text{Flavor}, j_l, s_l~(\lambda/\rho/\rho\rho/\rho\lambda/\lambda\lambda) ]\), and contains one or two states with total angular momentum \( J = j_l \pm 1/2 \). The labels \( \rho \) and \( \lambda \) specify the type of orbital excitation: \( \rho \) corresponds to \( l_\rho = 1 \) and \( l_\lambda = 0 \); \( \lambda \) to \( l_\rho = 0 \) and \( l_\lambda = 1 \); \( \rho\rho \) to \( l_\rho = 2 \) and \( l_\lambda = 0 \); \( \rho\lambda \) to \( l_\rho = 1 \) and \( l_\lambda = 1 \); and \( \lambda\lambda \) to \( l_\rho = 0 \) and \( l_\lambda = 2 \).

\begin{figure}[hbtp]
\begin{center}
\includegraphics[width=0.4\textwidth]{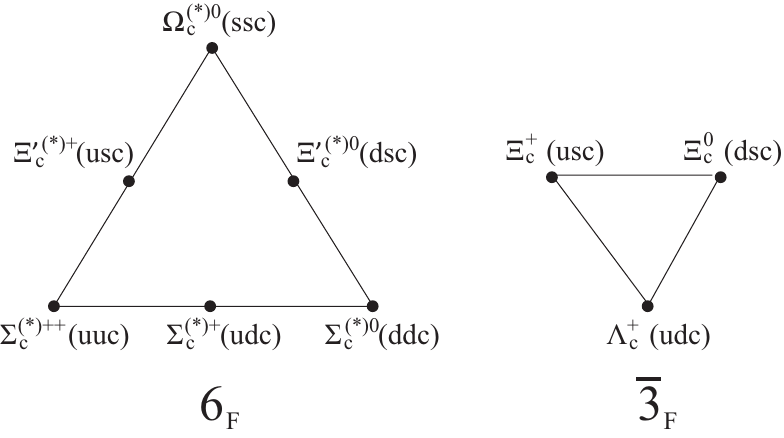}
\end{center}
\caption{Flavor \( SU(3) \) multiplets \( \mathbf{6}_F \) and \( \mathbf{\bar{3}}_F \) of the ground-state charmed baryons. The symbols \( \Xi_c \) and \( \Xi_c^{\prime} \) are used to distinguish charmed-strange baryons belonging to different flavor representations. However, the prime is typically omitted for experimentally observed states, as these cannot be directly distinguished in experiments.}
\label{fig:heavybaryon}
\end{figure}

For ground-state ($S$-wave) singly heavy baryons:
\begin{itemize}

\item The \( SU(3) \) flavor representation \( \mathbf{\bar{3}}_F \) corresponds to baryons with quantum numbers \( J^P = 1/2^+ \); due to the Pauli exclusion principle the light spin must be $s_l = 0$.

\item The \( SU(3) \) flavor representation \( \mathbf{6}_F \) gives rise to two spin states: \( J^P = 1/2^+ \) and \( J^P = 3/2^+ \); due to the Pauli exclusion principle the light spin must be $s_l = 1$.

\end{itemize}
These multiplets are illustrated in Fig.~\ref{fig:heavybaryon}. All predicted $S$-wave singly charmed and bottom baryons—except for the \( \Omega_b^* \)—have been observed experimentally~\cite{PDG2024}, thereby validating the effectiveness of this classification scheme.

\begin{figure*}[hbtp]
\begin{center}
\includegraphics[width=0.9\textwidth]{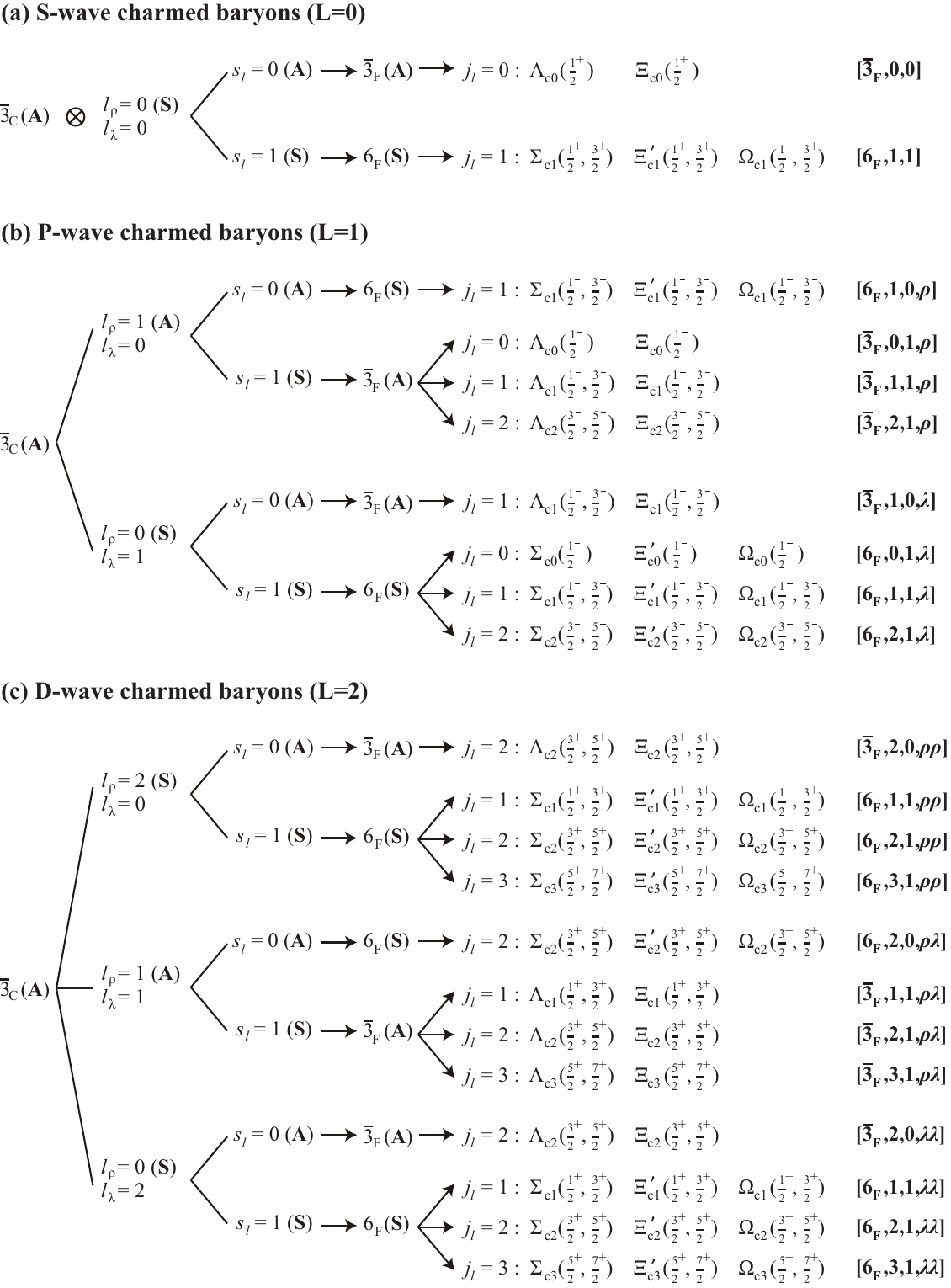}
\end{center}
\caption{Systematic categorization of $S$-, $P$-, and $D$-wave charmed baryons based on their internal quantum numbers and excitation modes.}
\label{fig:categorization}
\end{figure*}

Beyond the $S$-wave states, $P$-wave and $D$-wave excitations give rise to a richer spectrum of structures:
\begin{itemize}

\item $P$-wave baryons are organized into eight distinct multiplets.

\item $D$-wave baryons are organized into twelve distinct multiplets.

\end{itemize}
These excited-state multiplets are systematically illustrated in Fig.~\ref{fig:categorization}. This classification scheme, grounded in heavy quark effective theory (HQET, to be briefly introduced in Sec.~\ref{sec3.1}) and QCD symmetries, provides a robust framework for both theoretical predictions and experimental identification of singly heavy baryons.

\subsection{Interpolating currents}
\label{sec1.2}

A powerful tool for studying hadron spectroscopy is the use of hadronic interpolating currents (also referred to as fields or operators) constructed from quark and gluon fields. These currents are extensively employed in both lattice QCD and QCD sum rule analyses, as they can simultaneously encode the color, flavor, spin, and orbital degrees of freedom. In this subsection we systematically construct interpolating currents for the $S$- and $P$-wave singly heavy baryons.

In general, the interpolating field for a singly heavy baryon can be written as a combination of a diquark field and a heavy quark field:
\begin{eqnarray}
J(x) \sim \epsilon_{abc} \left( q^{aT}(x) \, C \, \Gamma_1 \, q^b(x) \right) \Gamma_2 \, h_v^c(x) \, .
\end{eqnarray}
Here, the operator is local depending only on a single space-time position $x$. We may include derivatives which we will discuss later. The indices \( a \), \( b \), and \( c \) denote color, and \( \epsilon_{abc} \) is the totally antisymmetric tensor in color space. The superscript \( T \) indicates transposition over Dirac indices only, and \( C \) is the charge-conjugation matrix. The field \( q(x) \) represents a light quark at position \( x \), which may be \( u(x) \), \( d(x) \), or \( s(x) \). The field \( h_v(x) \) denotes the effective heavy quark field in the heavy quark limit, with \( v \) representing the velocity of the heavy quark. The Fierz transformation has been applied to place the heavy quark field at the rightmost position without loss of generality. Additionally, we also need \( \gamma_t^\mu = \gamma^\mu - v\!\!\!\slash\, v^\mu \), \( g_t^{\alpha_1\alpha_2} = g^{\alpha_1\alpha_2} - v^{\alpha_1} v^{\alpha_2} \) as the transverse metric tensor, and \( D^\mu_t = D^\mu - (D \cdot v) v^\mu \), where \( D^\mu = \partial^\mu - i g A^\mu \) is the gauge-covariant derivative.

There are two $S$-wave diquark fields. The first is the so called ``good" diquark~\cite{Jaffe:2004ph}:
\begin{eqnarray}
\epsilon_{abc} \, q^{aT}(x) \, C \, \gamma_5 \, q^b(x) \, , \quad [^1S_0] \, ,
\end{eqnarray}
which carries quantum numbers \( J^P = 0^+ \). It has orbital angular momentum \( l_\rho = 0 \), so its spatial wavefunction is symmetric (\( \mathbf{S} \)); spin angular momentum \( s_l = 0 \), implying an antisymmetric spin configuration (\( \mathbf{A} \)); and an antisymmetric color structure \( \mathbf{\bar{3}}_C \) (\( \mathbf{A} \)). Therefore, according to the Pauli exclusion principle, the flavor configuration must be antisymmetric \( \mathbf{\bar{3}}_F \) (\( \mathbf{A} \)), although this is not shown explicitly.

The second $S$-wave diquark field is the so-called ``bad" diquark~\cite{Jaffe:2004ph}:
\begin{eqnarray}
\epsilon_{abc} \, q^{aT}(x) \, C \, \gamma_\mu^t \, q^b(x) \, , \quad [^3S_1] \, ,
\end{eqnarray}
which carries quantum numbers \( J^P = 1^+ \), with \( l_\rho = 0 \) (\( \mathbf{S} \)), \( s_l = 1 \) (\( \mathbf{S} \)), color structure \( \mathbf{\bar{3}}_C \) (\( \mathbf{A} \)), and flavor configuration \( \mathbf{6}_F \) (\( \mathbf{S} \)).

These ground-state diquark fields can be employed to construct interpolating currents for ground-state singly heavy baryons, denoted as \( J^{\alpha_1\cdots\alpha_{j-1/2}}_{j, P, F, j_l, s_l}(x) \). Their explicit forms are given below:
\begin{itemize}

\item When the diquark has spin \( s_l = 0 \) (\( \mathbf{A} \)), its flavor representation is \( \bm{\bar{3}}_F \) (\( \mathbf{A} \)). The total angular momentum of the baryon is \( J = s_l \oplus s_Q = 1/2 \), corresponding to an HQET singlet with \( J^P = 1/2^+ \):
\begin{eqnarray}
J_{1/2,+,\bm{\bar{3}}_F,0,0}(x) &=& \epsilon_{abc}\, \left[q^{aT}(x)\, C\, \gamma_5\, q^b(x)\right]\, h_v^c(x)\, .
\end{eqnarray}

\item When the diquark has spin \( s_l = 1 \) (\( \mathbf{S} \)), its flavor representation is \( \bm{6}_F \) (\( \mathbf{S} \)). The total angular momentum becomes \( J = s_l \oplus s_Q = (1/2,\, 3/2) \), resulting in an HQET doublet with \( J^P = (1/2^+,\, 3/2^+) \):
\begin{eqnarray}
&& J_{1/2,+,\bm{6}_F,1,1}(x) 
\\ \nonumber && ~~~~~ = \epsilon_{abc}\, \left[q^{aT}(x)\, C\, \gamma_\mu^t\, q^b(x)\right]\, \gamma_t^\mu \gamma_5\, h_v^c(x)\, , 
\\ && J^\mu_{3/2,+,\bm{6}_F,1,1}(x) 
\\ \nonumber && ~~~~~ = \epsilon_{abc}\, \left[q^{aT}(x)\, C\, \gamma_\nu^t\, q^b(x)\right]\, \left(g_t^{\mu\nu} - \frac{\gamma_t^\mu \gamma_t^\nu}{3} \right)\, h_v^c(x)\, .
\end{eqnarray}

\end{itemize}

\begin{widetext}
More diquark fields can be constructed by introducing a derivative to the $S$-wave diquark configurations. This derivative can act either between the two light quarks—corresponding to an internal excitation with \( l_\rho = 1 \) (\( \mathbf{A} \)) and \( l_\lambda = 0 \), or between the charm quark and the light-diquark system—corresponding to \( l_\rho = 0 \) (\( \mathbf{S} \)) and \( l_\lambda = 1 \):
\begin{eqnarray}
&& \epsilon_{abc} \Big( [D^\nu q^{aT}(x)] \, C \gamma_5 \, q^b(x) - q^{aT}(x) \, C \gamma_5 \, [D^\nu q^b(x)] \Big) \, , \quad [^1P_1] \, , \quad l_\rho = 1~(\mathbf{A}) \, , \quad l_\lambda = 0 \, ,
\label{def:diquarkP1}
\\
&& \epsilon_{abc} \Big( [D^\nu q^{aT}(x)] \, C \gamma_\mu^t \, q^b(x) - q^{aT}(x) \, C \gamma_\mu^t \, [D^\nu q^b(x)] \Big) \, , \quad [^3P_0]/[^3P_1]/[^3P_2] \, , \quad l_\rho = 1~(\mathbf{A}) \, , \quad l_\lambda = 0 \, ,
\label{def:diquarkP2}
\\
&& \epsilon_{abc} \Big( [D^\nu q^{aT}(x)] \, C \gamma_5 \, q^b(x) + q^{aT}(x) \, C \gamma_5 \, [D^\nu q^b(x)] \Big) \, , \quad [^1S_0] \, , \quad l_\rho = 0~(\mathbf{S}) \, , \quad l_\lambda = 1 \, ,
\\
&& \epsilon_{abc} \Big( [D^\nu q^{aT}(x)] \, C \gamma_\mu^t \, q^b(x) + q^{aT}(x) \, C \gamma_\mu^t \, [D^\nu q^b(x)] \Big) \, , \quad [^3S_1] \, , \quad l_\rho = 0~(\mathbf{S}) \, , \quad l_\lambda = 1 \, .
\end{eqnarray}
The latter two still contain the \( S \)-wave diquarks, while the former two correspond to explicit \( P \)-wave diquarks. Eq.~(\ref{def:diquarkP1}) carries quantum numbers \( J^P = 1^- \), with \( l_\rho = 1 \) (\( \mathbf{A} \)), \( s_l = 0 \) (\( \mathbf{A} \)), color structure \( \mathbf{\bar{3}}_C \) (\( \mathbf{A} \)), and flavor configuration \( \mathbf{6}_F \) (\( \mathbf{S} \)); Eq.~(\ref{def:diquarkP2}) carries quantum numbers \( J^P = 0^-/1^-/2^- \), with \( l_\rho = 1 \) (\( \mathbf{A} \)), \( s_l = 1 \) (\( \mathbf{S} \)), color structure \( \mathbf{\bar{3}}_C \) (\( \mathbf{A} \)), and flavor configuration \( \mathbf{\bar{3}}_F \) (\( \mathbf{A} \)).

These diquark fields can be further utilized to construct the \( P \)-wave singly heavy baryon fields, denoted as \( J^{\alpha_1\cdots\alpha_{j-1/2}}_{j, P, F, j_l, s_l, \rho/\lambda}(x) \). The explicit forms of these interpolating currents are presented below:
\begin{enumerate}

\item $\rho$-mode with $l_\rho = 1$ ($\mathbf{A}$) and $l_\lambda = 0$:

\begin{enumerate}

\item $s_l = 0$ ($\mathbf{A}$), $j_l = 1$. We denote this case as $[\mathbf{6}_F, 1, 0, \rho]$. The diquark has color $\mathbf{\bar 3}_C$ ($\mathbf{A}$) and flavor $\mathbf{6}_F$ ($\mathbf{S}$), leading to a HQET doublet with $J^P = (1/2^-,\, 3/2^-)$:
\begin{eqnarray}
J_{1/2,-,\mathbf{6}_F,1,0,\rho} &=& i \epsilon_{abc} \left( [D_t^{\mu} q^{aT}] C \gamma_5 q^b -  q^{aT} C \gamma_5 [D_t^{\mu} q^b] \right) \gamma_t^{\mu} \gamma_5 h_v^c \, ,
\\ 
J^{\alpha}_{3/2,-,\mathbf{6}_F,1,0,\rho} &=& i \epsilon_{abc} \left( [D_t^{\mu} q^{aT}] C \gamma_5 q^b -  q^{aT} C \gamma_5 [D_t^{\mu} q^b] \right) \times \left( g_t^{\alpha\mu} - \frac{1}{3} \gamma_t^{\alpha} \gamma_t^{\mu} \right) h_v^c \, .
\end{eqnarray}

\item $s_l = 1$ ($\mathbf{S}$), $j_l = 0$. We denote this case as $[\mathbf{\bar 3}_F, 0, 1, \rho]$. The diquark has color $\mathbf{\bar 3}_C$ ($\mathbf{A}$) and flavor $\mathbf{\bar 3}_F$ ($\mathbf{A}$), giving rise to an HQET singlet with $J^P = 1/2^-$:
\begin{eqnarray}
J_{1/2,-,\mathbf{\bar{3}}_F,0,1,\rho} &=& i \epsilon_{abc} \left( [D_t^{\mu} q^{aT}] C \gamma_t^\mu q^b - q^{aT} C \gamma_t^\mu [D_t^{\mu} q^b] \right) h_v^c \, .
\end{eqnarray}

\item $s_l = 1$ ($\mathbf{S}$), $j_l = 1$. We denote this case as $[\mathbf{\bar 3}_F, 1, 1, \rho]$. The diquark has color $\mathbf{\bar 3}_C$ ($\mathbf{A}$) and flavor $\mathbf{\bar 3}_F$ ($\mathbf{A}$), resulting in an HQET doublet with $J^P = (1/2^-,\, 3/2^-)$:
\begin{eqnarray}
J_{1/2,-,\mathbf{\bar{3}}_F,1,1,\rho} &=& i \epsilon_{abc} \left( [D_t^{\mu} q^{aT}] C \gamma_t^\nu q^b - q^{aT} C \gamma_t^\nu [D_t^{\mu} q^b] \right) \sigma_t^{\mu\nu} h_v^c \, ,
\\ 
J^{\alpha}_{3/2,-,\mathbf{\bar{3}}_F,1,1,\rho} &=& i \epsilon_{abc} \left( [D_t^{\mu} q^{aT}] C \gamma_t^\nu q^b - q^{aT} C \gamma_t^\nu [D_t^{\mu} q^b] \right) 
\nonumber\\ && \times \left( g_t^{\alpha\mu} \gamma_t^{\nu} \gamma_5 - g_t^{\alpha\nu} \gamma_t^{\mu} \gamma_5 - \frac{1}{3} \gamma_t^{\alpha} \gamma_t^{\mu} \gamma_t^{\nu} \gamma_5 + \frac{1}{3} \gamma_t^{\alpha} \gamma_t^{\nu} \gamma_t^{\mu} \gamma_5 \right) h_v^c \, .
\end{eqnarray}

\item $s_l = 1$ ($\mathbf{S}$), $j_l = 2$. We denote this case as $[\mathbf{\bar 3}_F, 2, 1, \rho]$. The diquark has color $\mathbf{\bar 3}_C$ ($\mathbf{A}$) and flavor $\mathbf{\bar 3}_F$ ($\mathbf{A}$), producing an HQET doublet with $J^P = (3/2^-,\, 5/2^-)$:
\begin{eqnarray}
J^{\alpha}_{3/2,-,\mathbf{\bar{3}}_F,2,1,\rho} &=& i \epsilon_{abc} \left( [D_t^{\mu} q^{aT}] C \gamma_t^\nu q^b - q^{aT} C \gamma_t^\nu [D_t^{\mu} q^b] \right)
\times \left( g_t^{\alpha\mu} \gamma_t^{\nu} \gamma_5 + g_t^{\alpha\nu} \gamma_t^{\mu} \gamma_5 - \frac{2}{3} g_t^{\mu\nu} \gamma_t^{\alpha} \gamma_5 \right) h_v^c \, ,
\\ 
J^{\alpha_1\alpha_2}_{5/2,-,\mathbf{\bar{3}}_F,2,1,\rho} &=& i \epsilon_{abc} \left( [D_t^{\mu} q^{aT}] \mathbb{C} \gamma_t^{\nu} q^b - q^{aT} \mathbb{C} \gamma_t^{\nu} [D_t^{\mu} q^b] \right) \times \Gamma^{\alpha_1\alpha_2,\mu\nu} h_v^c \, .
\end{eqnarray}

\end{enumerate}

\item $\lambda$-mode with $l_\rho = 0$ ($\mathbf{S}$) and $l_\lambda = 1$:

\begin{enumerate}

\item $s_l = 0$ ($\mathbf{A}$), $j_l = 1$. We denote this case as $[\mathbf{\bar{3}}_F, 1, 0, \lambda]$. The diquark has color $\mathbf{\bar{3}}_C$ ($\mathbf{A}$) and flavor $\mathbf{\bar{3}}_F$ ($\mathbf{A}$), giving rise to an HQET doublet with $J^P = (1/2^-,\, 3/2^-)$:
\begin{eqnarray}
J_{1/2,-,\mathbf{\bar{3}}_F,1,0,\lambda} &=& i \epsilon_{abc} \left( [D_t^{\mu} q^{aT}] C \gamma_5 q^b +  q^{aT} C \gamma_5 [D_t^{\mu} q^b] \right) \gamma_t^{\mu} \gamma_5 h_v^c \, ,
\\ 
J^{\alpha}_{3/2,-,\mathbf{\bar{3}}_F,1,0,\lambda} &=& i \epsilon_{abc} \left( [D_t^{\mu} q^{aT}] C \gamma_5 q^b +  q^{aT} C \gamma_5 [D_t^{\mu} q^b] \right) \times \left( g_t^{\alpha\mu} - \frac{1}{3} \gamma_t^{\alpha} \gamma_t^{\mu} \right) h_v^c \, .
\end{eqnarray}

\item $s_l = 1$ ($\mathbf{S}$), $j_l = 0$. We denote this case as $[\mathbf{6}_F, 0, 1, \lambda]$. The diquark has color $\mathbf{\bar{3}}_C$ ($\mathbf{A}$) and flavor $\mathbf{6}_F$ ($\mathbf{S}$), leading to an HQET singlet with $J^P = 1/2^-$:
\begin{eqnarray}
J_{1/2,-,\mathbf{6}_F,0,1,\lambda} &=& i \epsilon_{abc} \left( [D_t^{\mu} q^{aT}] C \gamma_t^\mu q^b +  q^{aT} C \gamma_t^\mu [D_t^{\mu} q^b] \right) h_v^c \, .
\end{eqnarray}

\item $s_l = 1$ ($\mathbf{S}$), $j_l = 1$. We denote this case as $[\mathbf{6}_F, 1, 1, \lambda]$. The diquark has color $\mathbf{\bar{3}}_C$ ($\mathbf{A}$) and flavor $\mathbf{6}_F$ ($\mathbf{S}$), leading to an HQET doublet with $J^P = (1/2^-,\, 3/2^-)$:
\begin{eqnarray}
J_{1/2,-,\mathbf{6}_F,1,1,\lambda} &=& i \epsilon_{abc} \left( [D_t^{\mu} q^{aT}] C \gamma_t^\nu q^b +  q^{aT} C \gamma_t^\nu [D_t^{\mu} q^b] \right) \sigma_t^{\mu\nu} h_v^c \, ,
\\ 
J^{\alpha}_{3/2,-,\mathbf{6}_F,1,1,\lambda} &=& i \epsilon_{abc} \left( [D_t^{\mu} q^{aT}] C \gamma_t^\nu q^b + q^{aT} C \gamma_t^\nu [D_t^{\mu} q^b] \right)
\nonumber\\ 
&& \times \left( g_t^{\alpha\mu} \gamma_t^{\nu} \gamma_5 - g_t^{\alpha\nu} \gamma_t^{\mu} \gamma_5 - \frac{1}{3} \gamma_t^{\alpha} \gamma_t^{\mu} \gamma_t^{\nu} \gamma_5 + \frac{1}{3} \gamma_t^{\alpha} \gamma_t^{\nu} \gamma_t^{\mu} \gamma_5 \right) h_v^c \, .
\end{eqnarray}

\item $s_l = 1$ ($\mathbf{S}$), $j_l = 2$. We denote this case as $[\mathbf{6}_F, 2, 1, \lambda]$. The diquark has color $\mathbf{\bar{3}}_C$ ($\mathbf{A}$) and flavor $\mathbf{6}_F$ ($\mathbf{S}$), producing an HQET doublet with $J^P = (3/2^-,\, 5/2^-)$:
\begin{eqnarray}
J^{\alpha}_{3/2,-,\mathbf{6}_F,2,1,\lambda} &=& i \epsilon_{abc} \left( [D_t^{\mu} q^{aT}] C \gamma_t^\nu q^b + q^{aT} C \gamma_t^\nu [D_t^{\mu} q^b] \right)
\times \left( g_t^{\alpha\mu} \gamma_t^{\nu} \gamma_5 + g_t^{\alpha\nu} \gamma_t^{\mu} \gamma_5 - \frac{2}{3} g_t^{\mu\nu} \gamma_t^{\alpha} \gamma_5 \right) h_v^c \, ,
\\ 
J^{\alpha_1\alpha_2}_{5/2,-,\mathbf{6}_F,2,1,\lambda} &=& i \epsilon_{abc} \left( [D_t^{\mu} q^{aT}] \mathbb{C} \gamma_t^{\nu} q^b + q^{aT} \mathbb{C} \gamma_t^{\nu} [D_t^{\mu} q^b] \right) \times \Gamma^{\alpha_1\alpha_2,\mu\nu} h_v^c \, .
\end{eqnarray}

\end{enumerate}

\end{enumerate}
We note that all of the above interpolating fields have been explicitly projected onto components with total angular momentum \( J = 1/2 \), \( J = 3/2 \), or \( J = 5/2 \). In particular, the projection operator \( \Gamma^{\alpha_1 \alpha_2, \mu \nu} \), which is employed in the construction of the \( J = 5/2 \) interpolating field, is defined as:
\begin{eqnarray}
\Gamma^{\alpha\beta,\mu\nu} &=& g_t^{\alpha\mu} g_t^{\beta\nu} + g_t^{\alpha\nu} g_t^{\beta\mu} - \frac{2}{15} g_t^{\alpha\beta} g_t^{\mu\nu}
- \frac{1}{3} g_t^{\alpha\mu} \gamma_t^{\beta} \gamma_t^{\nu} - \frac{1}{3} g_t^{\alpha\nu} \gamma_t^{\beta} \gamma_t^{\mu}
- \frac{1}{3} g_t^{\beta\mu} \gamma_t^{\alpha} \gamma_t^{\nu} - \frac{1}{3} g_t^{\beta\nu} \gamma_t^{\alpha} \gamma_t^{\mu}
\nonumber\\
&& + \frac{1}{15} \gamma_t^{\alpha} \gamma_t^{\mu} \gamma_t^{\beta} \gamma_t^{\nu} 
+ \frac{1}{15} \gamma_t^{\alpha} \gamma_t^{\nu} \gamma_t^{\beta} \gamma_t^{\mu}
+ \frac{1}{15} \gamma_t^{\beta} \gamma_t^{\mu} \gamma_t^{\alpha} \gamma_t^{\nu}
+ \frac{1}{15} \gamma_t^{\beta} \gamma_t^{\nu} \gamma_t^{\alpha} \gamma_t^{\mu} \, .
\end{eqnarray}
\end{widetext} 
\section{Experimental status}
\label{sec2}

In this section we provide a concise review of the experimental progress on singly heavy baryons. For simplicity, we omit the word ``singly'' when referring to singly heavy, charmed, or bottom baryons, and we also omit electric charges in most cases to streamline the notation. The symbols $\Xi_c$ and $\Xi_c^\prime$ denote the charmed-strange baryons ($qsc$ with $q = u/d$), belonging to the flavor representations $\mathbf{\bar{3}}_F$ and $\mathbf{6}_F$, respectively. However, the superscript $^\prime$ is typically omitted for experimentally observed states, as they cannot be directly distinguished in experiments.

All $1S$ charmed baryons have been well established experimentally~\cite{PDG2024}, completing the two $S$-wave multiplets $[\mathbf{\bar{3}}_F, 0, 0]$ and $[\mathbf{6}_F, 1, 1]$, as shown in Fig.~\ref{fig:categorization}. The spectrum of higher charmed baryons is considerably more complex and includes:
\begin{itemize}
    \item $\Lambda_c(2595)$~\cite{CLEO:1994oxm}, $\Lambda_c(2625)$~\cite{ARGUS:1993vtm}, $\Lambda_c(2765)$~\cite{CLEO:2000mbh}, $\Lambda_c(2860)$~\cite{LHCb:2017jym}, $\Lambda_c(2880)$~\cite{CLEO:2000mbh}, $\Lambda_c(2910)$~\cite{Belle:2022hnm}, $\Lambda_c(2940)$~\cite{BaBar:2006itc};
    \item $\Sigma_c(2800)$~\cite{Belle:2004zjl};
    \item $\Xi_c(2790)$~\cite{CLEO:2000ibb}, $\Xi_c(2815)$~\cite{CLEO:1999msf}, $\Xi_c(2882)$~\cite{LHCb:2022vns}, $\Xi_c(2923)$~\cite{BaBar:2007xtc,LHCb:2020iby}, $\Xi_c(2939)$~\cite{BaBar:2007xtc,LHCb:2020iby}, $\Xi_c(2965)$~\cite{LHCb:2020iby}, $\Xi_c(2970)$~\cite{Belle:2006edu,Belle:2020tom}, $\Xi_c(3055)$~\cite{BaBar:2007zjt}, $\Xi_c(3080)$~\cite{Belle:2006edu}, $\Xi_c(3123)$~\cite{BaBar:2007zjt};
    \item $\Omega_c(3000)$~\cite{LHCb:2017uwr}, $\Omega_c(3050)$~\cite{LHCb:2017uwr}, $\Omega_c(3066)$~\cite{LHCb:2017uwr}, $\Omega_c(3090)$~\cite{LHCb:2017uwr}, $\Omega_c(3119)$~\cite{LHCb:2017uwr}, $\Omega_c(3185)$~\cite{LHCb:2023sxp}, $\Omega_c(3327)$~\cite{LHCb:2023sxp}.
\end{itemize}
Several of these states were discussed in our earlier works~\cite{Chen:2016spr,Chen:2022asf}. In this section we focus on the newly observed states: $\Lambda_c(2910)$, $\Xi_c(2882)$, $\Omega_c(3185)$, and $\Omega_c(3327)$. To provide a comprehensive overview, Table~\ref{sec2:charm} summarizes the theoretical predictions of charmed baryons from various quark models~\cite{Ebert:2011kk,Roberts:2007ni,Yoshida:2015tia,Chen:2016iyi}, including $1S$, $2S$, $1P$, $2P$, and $1D$ states. Corresponding experimental candidates are listed for comparison.

All $1S$ bottom baryons, except $\Omega_b^*$, have been experimentally observed~\cite{PDG2024}, and they also complete the $S$-wave multiplets $[\mathbf{\bar{3}}_F, 0, 0]$ and $[\mathbf{6}_F, 1, 1]$, as illustrated in Fig.~\ref{fig:categorization}. The spectrum of higher bottom baryons includes:
\begin{itemize}
    \item $\Lambda_b(5912)$~\cite{LHCb:2012kxf}, $\Lambda_b(5920)$~\cite{LHCb:2012kxf,CDF:2013pvu}, $\Lambda_b(6072)$~\cite{CMS:2020zzv,LHCb:2020lzx}, $\Lambda_b(6146)$~\cite{LHCb:2019soc}, $\Lambda_b(6152)$~\cite{LHCb:2019soc};
    \item $\Sigma_b(6097)$~\cite{LHCb:2018haf};
    \item $\Xi_b(6087)$~\cite{LHCb:2023zpu}, $\Xi_b(6095)$~\cite{LHCb:2023zpu}, $\Xi_b(6100)$~\cite{CMS:2021rvl}, $\Xi_b(6227)$~\cite{LHCb:2018vuc,LHCb:2020xpu}, $\Xi_b(6327)$~\cite{LHCb:2021ssn}, $\Xi_b(6333)$~\cite{LHCb:2021ssn};
    \item $\Omega_b(6316)$~\cite{LHCb:2020tqd}, $\Omega_b(6330)$~\cite{LHCb:2020tqd}, $\Omega_b(6340)$~\cite{LHCb:2020tqd}, $\Omega_b(6350)$~\cite{LHCb:2020tqd}.
\end{itemize}
Some of these states were previously analyzed in Refs.~\cite{Chen:2016spr,Chen:2022asf}. In this section we focus on the recently observed $\Xi_b(6087)$ and $\Xi_b(6095)$. Table~\ref{sec2:bottom} summarizes the theoretical results for bottom baryons from various quark models~\cite{Ebert:2011kk,Roberts:2007ni,Yoshida:2015tia}, including the $1S$, $2S$, $1P$, and $1D$ states. Experimental candidates are also provided for comparison.

\begin{table*}[] 
\renewcommand{\arraystretch}{1.4} 
\begin{center} 
\caption{Theoretical masses of singly charmed baryons from representative quark models~\cite{Ebert:2011kk,Roberts:2007ni,Yoshida:2015tia,Chen:2016iyi}, including the $1S$, $2S$, and $1P$ states, as well as selected $2P$ and $1D$ levels. Only $\lambda$-mode orbital excitations are considered. Possible experimental candidates are shown for comparison, with all quoted experimental masses and widths taken from the Particle Data Group (PDG)~\cite{PDG2024}, unless otherwise specified. All masses are given in MeV.}
\begin{tabular}{ c c c c c c c c } 
\hline\hline & $J^P~(nL)$ & Exp. Mass~\cite{PDG2024} & Exp. Width~\cite{PDG2024} & ~RQM~\cite{Ebert:2011kk}~ & ~NQM~\cite{Roberts:2007ni}~ & ~NQM~\cite{Yoshida:2015tia}~ & ~NQM~\cite{Chen:2016iyi}~ 
\\ \hline\hline 
$\Lambda_c$ & $1/2^+~(1S)$ & $2286.46\pm0.14$ & $\sim10^{-13}$~s & 2286 & 2268 & 2285 & 2286 
\\ $\Xi_c$ & $1/2^+~(1S)$ & $2467.94^{+0.17}_{-0.20}$ & $\sim10^{-13}$~s & 2476 & 2466 & -- & 2470 
\\ $\Sigma_c$ & $1/2^+~(1S)$ & $2452.9\pm0.4$ & $<4.6$ & 2443 & 2455 & 2460 & 2456 
\\ $\Sigma_c^{*}$ & $3/2^+~(1S)$ & $2517.5\pm2.3$ & $<17$ & 2519 & 2519 & 2523 & 2515 
\\ $\Xi_c^{\prime}$ & $1/2^+~(1S)$ & $2578.4\pm0.5$ & ? & 2579 & 2594 & -- & 2579 
\\ $\Xi_c^{*}$ & $3/2^+~(1S)$ & $2645.56^{+0.24}_{-0.30}$ & $2.14 \pm 0.19$ & 2649 & 2649 & -- & 2649 
\\ $\Omega_c$ & $1/2^+~(1S)$ & $2695.2\pm1.7$ & $\sim10^{-13}$~s & 2698 & 2718 & 2731 & -- 
\\ $\Omega_c^{*}$ & $3/2^+~(1S)$ & $2765.9\pm2.0$ & ? & 2768 & 2776 & 2779 & -- 
\\ \hline 
$\Lambda_c$ & $1/2^-~(1P)$ & $\Lambda_c(2595):2592.25\pm0.28$~\cite{CLEO:1994oxm} & $2.59 \pm 0.56$~\cite{CLEO:1994oxm} & 2598 & 2625 & 2628 & 2614 
\\ $\Lambda_c$ & $3/2^-~(1P)$ & $\Lambda_c(2625):2628.11\pm0.19$~\cite{ARGUS:1993vtm} & $<0.97$~\cite{ARGUS:1993vtm} & 2627 & 2636 & 2630 & 2639 
\\ $\Xi_c$ & $1/2^-~(1P)$ & $\Xi_c(2790):2792.4\pm0.5$~\cite{CLEO:2000ibb} & $8.9 \pm 1.0$~\cite{CLEO:2000ibb} & 2792 & 2773 & -- & 2793 
\\ $\Xi_c$ & $3/2^-~(1P)$ & $\Xi_c(2815):2816.74^{+0.20}_{-0.23}$ ~\cite{CLEO:1999msf}& $2.43 \pm 0.26$~\cite{CLEO:1999msf} & 2819 & 2783 & -- & 2820 
\\ $\Sigma_c$ & $1/2^-~(1P)$ & -- & -- & 2713 & 2748 & 2802 & 2702 
\\ $\Sigma_c$ & $1/2^-~(1P)$ &\multirow{4}{*}{$\Sigma_c(2800):2792^{+14}_{-~5}$} &\multirow{4}{*}{$62^{+64}_{-44}$} & 2799 & 2768 & 2826 & 2765 
\\ $\Sigma_c$ & $3/2^-~(1P)$ & & & 2773 & 2763 & 2807 & 2785 
\\ $\Sigma_c$ & $3/2^-~(1P)$ & & & 2798 & 2776 & 2837 & 2798 
\\ $\Sigma_c$ & $5/2^-~(1P)$ & & & 2789 & 2790 & 2839 & 2790 
\\ $\Xi^\prime_c$ & $1/2^-~(1P)$ & $\Xi_c(2882):2881.8 \pm 9.0$~\cite{LHCb:2022vns} & $12.4 \pm 7.8$~\cite{LHCb:2022vns} & 2854 & 2855 & -- & 2839 
\\ $\Xi^\prime_c$ & $1/2^-~(1P)$ & $\Xi_c(2923):2923.04\pm0.35$~\cite{LHCb:2020iby} & $7.1 \pm 2.0$~\cite{LHCb:2020iby} & 2936 & -- & -- & 2900 
\\ $\Xi^\prime_c$ & $3/2^-~(1P)$ & $\Xi_c(2939):2938.55\pm0.30$~\cite{LHCb:2020iby} & $10.2 \pm 1.4$~\cite{LHCb:2020iby} & 2912 & 2866 & -- & 2921 
\\ $\Xi^\prime_c$ & $3/2^-~(1P)$ & $\Xi_c(2965):2964.88\pm0.33$~\cite{LHCb:2020iby} & $14.1 \pm 1.6$~\cite{LHCb:2020iby} & 2935 & -- & -- & 2932 
\\ $\Xi^\prime_c$ & $5/2^-~(1P)$ & -- & -- & 2929 & 2895 & -- & 2927 
\\ $\Omega_c$ & $1/2^-~(1P)$ & -- & -- & 2966 & 2977 & 3030 & -- 
\\ $\Omega_c$ & $1/2^-~(1P)$ & $\Omega_c(3000):3000.41\pm0.22$~\cite{LHCb:2017uwr} & $4.5 \pm 0.7$ ~\cite{LHCb:2017uwr}& 3055 & 2990 & 3048 & -- 
\\ $\Omega_c$ & $3/2^-~(1P)$ & $\Omega_c(3050):3050.20\pm0.13$~\cite{LHCb:2017uwr} & $<1.2$~\cite{LHCb:2017uwr} & 3029 & 2986 & 3033 & -- 
\\ $\Omega_c$ & $3/2^-~(1P)$ & $\Omega_c(3066):3065.46\pm0.28$~\cite{LHCb:2017uwr} & $3.5 \pm 0.4$~\cite{LHCb:2017uwr} & 3054 & 2994 & 3056 & -- 
\\ $\Omega_c$ & $5/2^-~(1P)$ & $\Omega_c(3090):3090.0\pm0.5$~\cite{LHCb:2017uwr} & $8.7 \pm 1.3$ ~\cite{LHCb:2017uwr}& 3051 & 3014 & 3057 & -- 
\\ \hline 
$\Lambda_c$ & $1/2^+~(2S)$ & $\Lambda_c(2765):2766.6\pm2.4$~\cite{CLEO:2000mbh} & $\sim50$~\cite{CLEO:2000mbh} & 2769 & 2791 & 2857 & 2772 
\\ $\Xi_c$ & $1/2^+~(2S)$ & $\Xi_c(2970):2966.34^{+0.17}_{-1.00}$~\cite{Belle:2020tom} & $20.9^{+2.4}_{-3.5}$~\cite{Belle:2020tom} & 2959 & -- & -- & 2940 
\\ $\Sigma_c$ & $1/2^+~(2S)$ & -- & -- & 2901 & 2958 & 3029 & 2850 
\\ $\Sigma_c$ & $3/2^+~(2S)$ & -- & -- & 2936 & 2995 & 3065 & 2876 
\\ $\Xi_c^{\prime}$ & $1/2^+~(2S)$ & -- & -- & 2983 & -- & -- & 2977 
\\ $\Xi_c^{\prime}$ & $3/2^+~(2S)$ & -- & -- & 3026 & 3012 & -- & 3007 
\\ $\Omega_c$ & $1/2^+~(2S)$ & \multirow{1}{*}{$\Omega_c(3119):3119.1\pm1.0$}~\cite{LHCb:2017uwr} & \multirow{1}{*}{$<2.6$}~\cite{LHCb:2017uwr} & 3088 & 3152 & 3227 & -- 
\\ $\Omega_c$ & $3/2^+~(2S)$ & \multirow{1}{*}{$\Omega_c(3185):3185.7\pm1.7$} ~\cite{LHCb:2023sxp}& \multirow{1}{*}{$50\pm7$}~\cite{LHCb:2023sxp} & 3123 & 3190 & 3257 & -- 
\\ $\Lambda_c$ & $1/2^-~(2P)$ & $\Lambda_c(2910):2913.8\pm5.6$~\cite{Belle:2022hnm} & $51.8\pm20$~\cite{Belle:2022hnm} & 2983 & 2816 & 2890 & 2980 
\\ $\Lambda_c$ & $3/2^-~(2P)$ & $\Lambda_c(2940):2939.8\pm5.6$~\cite{BaBar:2006itc} & $17.5\pm5.2$~\cite{BaBar:2006itc} & 3005 & 2830 & 2917 & 3004 
\\ \hline 
$\Lambda_c$ & $3/2^+~(1D)$ & $\Lambda_c(2860):2856.1^{+2.3}_{-5.9}$~\cite{LHCb:2017jym} & $67.6^{+11.8}_{-21.6}$~\cite{LHCb:2017jym} & 2874 & 2887 & 2920 & 2843 
\\ $\Lambda_c$ & $5/2^+~(1D)$ & $\Lambda_c(2880):2881.63\pm0.24$~\cite{CLEO:2000mbh} & $5.6^{+0.8}_{-0.6}$~\cite{CLEO:2000mbh} & 2880 & 2887 & 2922 & 2851 
\\ $\Xi_c$ & $3/2^+~(1D)$ & $\Xi_c(3055):3055.9\pm0.4$~\cite{BaBar:2007zjt} & $7.8 \pm 1.9$~\cite{BaBar:2007zjt} & 3059 & -- & -- & 3033 
\\ $\Xi_c$ & $5/2^+~(1D)$ & $\Xi_c(3080):3077.2\pm0.4$~\cite{Belle:2006edu} & $3.6 \pm 1.1$~\cite{Belle:2006edu} & 3076 & 3004 & -- & 3040 
\\ \hline\hline 
\end {tabular}
\label{sec2:charm} 
\end{center} 
\end{table*} 

\begin{table*}[] 
\renewcommand{\arraystretch}{1.4} 
\begin{center} 
\caption{Theoretical masses of singly bottom baryons from representative quark models~\cite{Ebert:2011kk,Roberts:2007ni,Yoshida:2015tia}, including the $1S$, $2S$, $1P$, and selected $1D$ states. Only $\lambda$-mode orbital excitations are considered. Possible experimental candidates are shown for comparison, with all quoted experimental masses and widths taken from the Particle Data Group (PDG)~\cite{PDG2024}, unless otherwise specified. All masses are given in MeV.}
\begin{tabular}{ c c c c c c c } 
\hline \hline & $J^P~(nL)$ & Exp. Mass~\cite{PDG2024} & Exp. Width~\cite{PDG2024} & ~~RQM~\cite{Ebert:2011kk}~~ & ~~NQM~\cite{Roberts:2007ni}~~ & ~~NQM~\cite{Yoshida:2015tia}~~ 
\\ \hline \hline 
$\Lambda_b$ & $1/2^+~(1S)$ & $5619.60\pm0.17$ & $\sim10^{-12}$~s & 5620 & 5612 & 5618 
\\ $\Xi_b$ & $1/2^+~(1S)$ & $5797.0\pm0.6$ & $\sim10^{-12}$~s & 5803 & 5806 & -- 
\\ $\Sigma_b$ & $1/2^+~(1S)$ & $5810.56\pm0.25$ & $5.0 \pm 0.5$ & 5808 & 5833 & 5823 
\\ $\Sigma_b^{*}$ & $3/2^+~(1S)$ & $5830.32\pm0.27$ & $9.4 \pm 0.5$ & 5834 & 5858 & 5845 
\\ $\Xi_b^{\prime}$ & $1/2^+~(1S)$ & $5935.02\pm0.05$ & $<0.08$ & 5936 & 5970 & -- 
\\ $\Xi_b^*$ & $3/2^+~(1S)$ & $5955.33\pm0.13$ & $1.65 \pm 0.33$ & 5963 & 5980 & -- 
\\ $\Omega_b$ & $1/2^+~(1S)$ & $6046.1\pm1.7$ & $\sim10^{-12}$~s & 6064 & 6081 & 6076 
\\ $\Omega_b^{*}$ & $3/2^+~(1S)$ & -- & -- & 6088 & 6102 & 6094 
\\ \hline 
$\Lambda_b$ & $1/2^-~(1P)$ & $\Lambda_b(5912):5912.20\pm0.21$~\cite{LHCb:2012kxf} & $<0.66$~\cite{LHCb:2012kxf} & 5930 & 5939 & 5938 
\\ $\Lambda_b$ & $3/2^-~(1P)$ & $\Lambda_b(5920):5919.92\pm0.19$~\cite{LHCb:2012kxf,CDF:2013pvu} & $<0.63$ ~\cite{LHCb:2012kxf,CDF:2013pvu}& 5942 & 5941 & 5939 
\\ $\Xi_b$ & $1/2^-~(1P)$ & 
$\Xi_b(6087):6087.24\pm0.20$~\cite{LHCb:2023zpu} & $2.43\pm0.51$~\cite{LHCb:2023zpu}  & 6120 & 6090 & -- 
\\ $\Xi_b$ & $3/2^-~(1P)$ & 
$\Xi_b(6095):6095.36\pm0.15$~\cite{LHCb:2023zpu} & $0.50\pm0.33$~\cite{LHCb:2023zpu} & 6130 & 6093 & -- 
\\ $\Sigma_b$ & $1/2^-~(1P)$ & -- & -- & 6095 & 6099 & 6127 
\\ $\Sigma_b$ & $1/2^-~(1P)$ & -- & -- & 6101 & 6106 & 6135 
\\ $\Sigma_b$ & $3/2^-~(1P)$ & -- & -- & 6087 & 6101 & 6132 
\\ $\Sigma_b$ & $3/2^-~(1P)$ & $\Sigma_b(6097):6095.8\pm1.7$~\cite{LHCb:2018haf} & $31.0 \pm 5.5$~\cite{LHCb:2018haf} & 6096 & 6105 & 6141 
\\ $\Sigma_b$ & $5/2^-~(1P)$ & -- & -- & 6084 & 6172 & 6144 
\\ $\Xi^\prime_b$ & $1/2^-~(1P)$ & -- & -- & 6227 & 6188 & -- 
\\ $\Xi^\prime_b$ & $1/2^-~(1P)$ & -- & -- & 6233 & -- & -- 
\\ $\Xi^\prime_b$ & $3/2^-~(1P)$ & -- & -- & 6224 & 6190 & -- 
\\ $\Xi^\prime_b$ & $3/2^-~(1P)$ & $\Xi_b(6227):6226.9\pm2.0$~\cite{LHCb:2018vuc} & $18.1 \pm 5.7$~\cite{LHCb:2018vuc} & 6234 & -- & -- 
\\ $\Xi^\prime_b$ & $5/2^-~(1P)$ & -- & -- & 6226 & 6201 & -- 
\\ $\Omega_b$ & $1/2^-~(1P)$ & -- & -- & 6330 & 6301 & 6333 
\\ $\Omega_b$ & $1/2^-~(1P)$ &$\Omega_b(6316):6315.64\pm0.59$~\cite{LHCb:2020tqd}& $<2.8$~\cite{LHCb:2020tqd} & 6339 & 6312 & 6340 
\\ $\Omega_b$ & $3/2^-~(1P)$ &$\Omega_b(6330):6330.30\pm0.58$~\cite{LHCb:2020tqd}& $<3.1$~\cite{LHCb:2020tqd} & 6331 & 6304 & 6336 
\\ $\Omega_b$ & $3/2^-~(1P)$ &$\Omega_b(6340):6339.71\pm0.57$~\cite{LHCb:2020tqd}& $<1.5$~\cite{LHCb:2020tqd} & 6340 & 6311 & 6344 
\\ $\Omega_b$ & $5/2^-~(1P)$ &$\Omega_b(6350):6349.88\pm0.61$~\cite{LHCb:2020tqd}& $1.4^{+1.0}_{-0.8}$~\cite{LHCb:2020tqd} & 6334 & 6311 & 6345 
\\ \hline 
$\Lambda_b$ & $1/2^+~(2S)$ & $\Lambda_b(6072):6072.3\pm3.0$~\cite{LHCb:2020lzx}& $72 \pm 11$~\cite{LHCb:2020lzx} & 6089 & 6107 & 6153 
\\ $\Xi_b$ & $1/2^+~(2S)$ & -- & -- & 6266 & -- & -- 
\\ $\Sigma_b$ & $1/2^+~(2S)$ & -- & -- & 6213 & 6294 & 6343 
\\ $\Sigma_b$ & $3/2^+~(2S)$ & -- & -- & 6226 & 6308 & 6356 
\\ $\Xi_b^{\prime}$ & $1/2^+~(2S)$ & -- & -- & 6329 & -- & -- 
\\ $\Xi_b^{\prime}$ & $3/2^+~(2S)$ & -- & -- & 6342 & 6311 & -- 
\\ $\Omega_b$ & $1/2^+~(2S)$ & -- & -- & 6450 & 6472 & 6517 
\\ $\Omega_b$ & $3/2^+~(2S)$ & -- & -- & 6461 & 6478 & 6528 
\\ \hline 
$\Lambda_b$ & $3/2^+~(1D)$ & $\Lambda_b(6152):6152.51\pm0.37$~\cite{LHCb:2019soc} & $2.1 \pm 0.9$~\cite{LHCb:2019soc} & 6190 & 6181 & 6211 
\\ $\Lambda_b$ & $5/2^+~(1D)$ & $\Lambda_b(6146):6146.17\pm0.43$~\cite{LHCb:2019soc} & $2.9 \pm 1.3$~\cite{LHCb:2019soc} & 6196 & 6183 & 6212 
\\ $\Xi_b$ & $3/2^+~(1D)$ & $\Xi_b(6327):6327.28^{+0.34}_{-0.33}$~\cite{LHCb:2021ssn} & $<2.20$~\cite{LHCb:2021ssn} & 6366 & -- & -- 
\\ $\Xi_b$ & $5/2^+~(1D)$ & $\Xi_b(6333):6332.69^{+0.28}_{-0.29}$~\cite{LHCb:2021ssn} & $<1.55$~\cite{LHCb:2021ssn} & 6373 & 6300 & -- 
\\ \hline\hline 
\end {tabular} 
\label{sec2:bottom} 
\end{center} 
\end{table*}

\subsection{$\Lambda_c(2910)$ and $\Lambda_c(2940)$}

In 2022 the Belle Collaboration reported the observation of a new structure in the $\Sigma_c(2455)^{0,++} \pi^\pm$ invariant mass spectrum, with a statistical significance of $4.2\sigma$~\cite{Belle:2022hnm}. This resonance, tentatively identified as the $\Lambda_c(2910)^+$, was found to have the following mass and width:
\begin{eqnarray}
\Lambda_c(2910)^+ &:& M = 2913.8 \pm 5.6 \pm 3.8~\mathrm{MeV}, \\
\nonumber && \Gamma = 51.8 \pm 20.0 \pm 18.8~\mathrm{MeV}.
\end{eqnarray}
Its likely heavy-quark spin partner is the $\Lambda_c(2940)^+$, first observed by the BaBar Collaboration in the $D^0 p$ final state~\cite{BaBar:2006itc}. This state was subsequently confirmed by the LHCb Collaboration~\cite{LHCb:2017jym}, and its decay into $\Sigma_c(2455)^{++} \pi^-$ was also observed by Belle~\cite{Belle:2006xni}. The mass and width of the $\Lambda_c(2940)^+$ were measured to be:
\begin{eqnarray}
\Lambda_c(2940)^+ &:& M = 2944.8^{+3.5}_{-2.5} \pm 0.4^{+0.1}_{-4.6}~\mathrm{MeV}, \\
\nonumber && \Gamma = 27.7^{+8.2}_{-6.0} \pm 0.9^{+5.2}_{-10.4}~\mathrm{MeV}.
\end{eqnarray}

In Ref.~\cite{Yang:2023fsc} we analyzed the mass spectra of the $1P$- and $2P$-wave charmed baryons using QCD sum rules and light-cone sum rules within the framework of heavy quark effective theory. Our results support the assignment of $\Lambda_c(2910)^+$ and $\Lambda_c(2940)^+$ as $2P$-wave charmed baryons with quantum numbers $J^P = 1/2^-$ and $3/2^-$, respectively, belonging to the $SU(3)$ flavor representation $\mathbf{\bar{3}_F}$. This assignment is consistent with the findings of Ref.~\cite{Azizi:2022dpn}, which also employed the QCD sum rule method. Moreover, as shown in Table~\ref{sec2:charm}, this assignment agrees with quark model calculations, and the quantum number $J^P = 3/2^-$ is also favored for the $\Lambda_c(2940)^+$ according to LHCb results.

In Ref.~\cite{Zhang:2024afw} the authors analyzed the OZI-allowed two-body strong decay properties of $1P$-, $1D$-, $2S$-, and $2P$-wave $\Lambda_c$ baryons within the $j$-$j$ coupling scheme using the quark pair creation model. Their results support the interpretation of the newly observed state $\Lambda_c(2910)$ as one of the $1P$-wave $\rho$-mode states, either with $J^P = 3/2^-$ or $5/2^-$. The $2P$-wave $\lambda$-mode state with $J^P = 3/2^-$ is suggested to be a suitable assignment for the $\Lambda_c(2940)$.

An alternative interpretation considers $\Lambda_c(2940)^+$ as a hadronic molecule in the $D^\ast N$ channel~\cite{Luo:2019qkm,Zhang:2022pxc}, possibly forming an isoscalar bound state with both $J^P = 1/2^-$ and $3/2^-$ components. However, this leads to an inverted mass ordering, underscoring the importance of further experimental and theoretical studies to clarify the nature of both $\Lambda_c(2910)$ and $\Lambda_c(2940)$. In Ref.~\cite{Yue:2024paz} the authors carried out an analysis of the decay behavior of $\Lambda_c(2910)$ and $\Lambda_c(2940)$ in the $ND^\ast$ molecular framework with possible quantum numbers $J^P = 1/2^-$ and $3/2^-$. Using an effective Lagrangian approach, they evaluated the partial decay widths for the $ND$, $\Sigma_c\pi$, and $\Sigma_c^\ast\pi$ channels. Their results indicate that the preferred quantum numbers for $\Lambda_c(2910)$ and $\Lambda_c(2940)$ are $J^P = 1/2^-$ and $3/2^-$, respectively.

In Ref.~\cite{Yan:2022nxp} the authors analyzed $S$-wave pentaquark systems $qqq\bar{q}c$ with $I = 0$ and $J^P = 1/2^-, 3/2^-, 5/2^-$ using the quark delocalization color screening model. Their results indicate that $\Lambda_c(2910)$ cannot be interpreted as a molecular state, whereas $\Lambda_c(2940)$ is likely to be a molecular state with $J^P = 3/2^-$, predominantly composed of an $ND^\ast$ component.

\subsection{$\Xi_c(2882)$}

In 2023 the LHCb Collaboration reported first evidence for the $\Xi_c(2882)^0$ state in the $\Lambda_c^+ K^-$ mass spectrum of the $B^- \to \Lambda_c^+ \bar \Lambda_c^- K^-$ decay, with a local significance of $3.8\sigma$~\cite{LHCb:2022vns}. Its mass and width were measured as:
\begin{eqnarray}
\Xi_c(2882)^0 &:& \mathrm{M} = 2881.8 \pm 3.1 \pm 8.5~\mathrm{MeV}, \\
\nonumber && \Gamma = 12.4 \pm 5.2 \pm 5.8~\mathrm{MeV}.
\end{eqnarray}

In Ref.~\cite{Pooja:2025fsl} the authors calculated the masses of five $1P$-wave states of the $\Xi_c^\prime$ baryon within a quark–diquark model. Their results suggest that $\Xi_c(2882)$ is a good candidate for a $1P$-wave state with spin-parity $J^P = 1/2^-$.

\subsection{$\Omega_c(3185)$ and $\Omega_c(3327)$}

Recent results from the LHCb Collaboration confirmed the existence of two new singly charmed baryons, $\Omega_c(3185)$ and $\Omega_c(3327)$, observed in the $\Xi_c^+ K^-$ final state~\cite{LHCb:2023sxp}. Their masses and widths were measured as:
\begin{eqnarray}
\Omega_c(3185)^0 &:& \mathrm{M} = 3185.7 \pm 1.7 \pm 0.2~\mathrm{MeV}, \\
\nonumber && \Gamma = 50 \pm 7^{+10}_{-20}~\mathrm{MeV}, \\
\Omega_c(3327)^0 &:& \mathrm{M} = 3327.1 \pm 1.2 \pm 0.2~\mathrm{MeV}, \\
\nonumber && \Gamma = 20 \pm 5^{+13}_{-1.0}~\mathrm{MeV}.
\end{eqnarray}

In Ref.~\cite{Zhong:2025oti} the authors studied the $1S$-, $1P$-, $1D$-, $2S$-, and $2P$-wave $\Omega_c$ states within a semi-relativistic constituent quark model and evaluated their strong decay widths. They found that $\Omega_c(3185)$ can be interpreted as a $2S$ state with $J^P = 3/2^+$, dominated by a $\lambda$-mode radial excitation. Meanwhile, $\Omega_c(3327)$ could be identified as a $1D$ state with $J^P = 5/2^+$ or $7/2^+$. Further exploration of the $\Xi D$ invariant mass spectrum around 3.3~GeV in future experiments may help clarify its nature.  It should be pointed out that the $\Omega_c(3327)$ is also suggested to be a candidate of the $1D$-wave state in the literature~\cite{Yu:2023bxn,Luo:2023sra,Wang:2023wii,Pan:2023hwt,Jakhad:2023mni,Li:2024zze}.

In Ref.~\cite{Kucukyilmaz:2025rsd} the authors analyzed $\Omega_c(3185)$ and $\Omega_c(3327)$ within a quark–diquark model using a non-relativistic framework and a Cornell-like potential. They systematically calculated mass spectra and magnetic moments. Their results suggest that $\Omega_c(3185)$ could be interpreted either as a $2S$ state with $J^P = 1/2^+$ or $3/2^+$, or as a $1P$ state with $J^P = 1/2^-$ or $3/2^-$, depending on the diquark configuration. Similarly, $\Omega_c(3327)$ is also consistent with a $2S$ assignment, in agreement with~\cite{Karliner:2023okv}.

\subsection{$\Xi_b(6087)$ and $\Xi_b(6095)$}

In 2023 the LHCb Collaboration reported the first observation of two new bottom baryons, $\Xi_b(6087)$ and $\Xi_b(6095)$, in the $\Xi_b^0 \pi^+ \pi^-$ final state. Their masses and widths were measured as:
\begin{eqnarray}
\nonumber \Xi_b(6087)^0 &:& \mathrm{M} = 6087.24 \pm 0.20 \pm 0.06 \pm 0.5~\mathrm{MeV}, \\
&& \Gamma = 2.43 \pm 0.51 \pm 0.10~\mathrm{MeV}, \\
\nonumber \Xi_b(6095)^0 &:& \mathrm{M} = 6095.36 \pm 0.15 \pm 0.03 \pm 0.5~\mathrm{MeV}, \\
&& \Gamma = 0.50 \pm 0.33 \pm 0.11~\mathrm{MeV}.
\end{eqnarray}

In Ref.~\cite{Yang:2022oog} we interpreted $\Lambda_b(5912)^0$, $\Lambda_b(5920)^0$, $\Xi_b(6087)^0$, and $\Xi_b(6095)^0$ as $P$-wave bottom baryons in the $\rho$-mode. This interpretation differs from predictions of several quark model studies~\cite{Yoshida:2015tia,Nagahiro:2016nsx,Wang:2017kfr,Chen:2018orb,Kawakami:2019hpp,He:2021xrh}, underscoring the need for further experimental and theoretical investigation.

\section{Theoretical frameworks}
\label{sec3}

In this section we briefly introduce the theoretical frameworks employed in our study. The following subsections respectively cover heavy quark effective theory (HQET)~\cite{Eichten:1989zv,Grinstein:1990mj,Falk:1990yz,Neubert:1993mb}, QCD sum rules (QSR)~\cite{Shifman:1978bx,Shifman:1978by,Reinders:1984sr,Colangelo:2000dp,Narison:2002woh,Gimenez:2005nt,Nielsen:2009uh,Gubler:2018ctz}, and light-cone sum rules (LCSR)~\cite{Braun:1988qv,Chernyak:1990ag,Ball:1998je,Ball:2006wn,Ball:2004rg,Ball:1998kk,Ball:1998sk,Ball:1998ff,Ball:2007rt,Ball:2007zt,Ball:2002ps}. Besides the singly heavy baryons, these methods have also been extensively applied to investigate the $S$-, $P$-, $D$-, and $F$-wave states of heavy mesons~\cite{Dai:1993kt,Bagan:1991sg,Neubert:1991sp,Broadhurst:1991fc,Ball:1993xv,Huang:1994zj,Colangelo:1991ug,Dai:1996yw,Colangelo:1998ga,Zhou:2014ytp,Zhou:2015ywa}.

\subsection{Heavy quark effective theory}
\label{sec3.1}

In the heavy quark limit \( m_Q \to \infty \), a heavy quark can be characterized by a time-independent four-velocity \( v \). Its behavior resembles that of a static color source, and the dynamics of the hadron reduce to strong interactions between the light degrees of freedom and this source. As a result, in the limit \( m_Q \to \infty \), the internal dynamics of the hadron become entirely independent of the heavy quark, giving rise to heavy quark flavor and spin symmetries.

In heavy-flavor hadrons, the momentum of the heavy quark \( P_Q \) can be decomposed as
\begin{align}
P_Q = m_Q v + k,
\end{align}
where \( m_Q \) is the heavy quark mass, \( v \) is the four-velocity of the hadron, and \( k \) is the residual momentum exchanged between the heavy quark and the light degrees of freedom, typically of order \( \Lambda_{\text{QCD}} \). In the heavy quark limit \( m_Q \gg \Lambda_{\text{QCD}} \), the velocity of the heavy quark coincides with that of the hadron. Since this velocity remains invariant under strong interactions, the heavy quark propagator can be approximated by
\begin{align}
\frac{i}{\not\! P_Q - m_Q + i\epsilon} \rightarrow \frac{i}{v \cdot k + i\epsilon} P_+,
\end{align}
where \( P_+ \) is the positive-energy projection operator, satisfying
\begin{align}
P_+^2 = P_+, \quad P_+ \gamma^\mu P_+ = \frac{1 + \not\! v}{2} \gamma^\mu \frac{1 + \not\! v}{2}.
\end{align}

To derive an effective Lagrangian for the heavy quark field, let us start with the QCD Lagrangian
\begin{align}
\mathcal{L}_{\text{QCD}} = \bar{Q}(i\not\! D - m_Q)Q,
\end{align}
where \( Q \) denotes the heavy quark field and \( D_\mu \) is the QCD covariant derivative defined as \( D_\mu = \partial_\mu - ig_s A_\mu \). To separate the large mass scale, the heavy quark field can be redefined as
\begin{align}
Q(x) = e^{- i m_Q v \cdot x} Q_v(x),
\end{align}
where \( Q_v(x) \) contains only the residual momentum components. Neglecting \( \mathcal{O}(1/m_Q) \) corrections, \( Q_v(x) \) can be replaced by the effective field \( h_v(x) \). Under this field redefinition, the QCD Lagrangian for a heavy quark becomes
\begin{eqnarray}
\nonumber \mathcal{L} &\rightarrow& \bar{h}_v (m_Q \not\! v + i\not\! D - m_Q) h_v 
\\ \nonumber &=& \bar{h}_v (i\not\! D) h_v 
\\ \nonumber &=& \bar{h}_v \frac{1 + \not\! v}{2} i\not\! D \frac{1 + \not\! v}{2} h_v 
\\ &=& \bar{h}_v\, i v \cdot D\, h_v \equiv \mathcal{L}_{\text{HQET}}.
\end{eqnarray}

Now consider a system with \( N_f \) heavy quark flavors. The HQET Lagrangian generalizes to
\begin{align}
\mathcal{L} = \sum_{j=1}^{N_f} \bar{h}_j\, i v \cdot D\, h_j,
\end{align}
where the field \( h_j \) denotes the effective heavy quark field of the \( j \)-th flavor. Since the Lagrangian is the sum over $h_j$ for the $j$-independent operator $iv \cdot D$, it is invariant under flavor rotations, thereby possessing an \( SU(N_f) \) flavor symmetry. In the heavy quark limit \( m_Q \to \infty \), heavy quark systems therefore exhibit both spin and heavy flavor symmetries. However, these symmetries are broken when \( 1/m_Q \) corrections are taken into account. The breaking effects are suppressed by powers of \( \Lambda_{\text{QCD}} / m_Q \).

When considering the \( 1/m_Q \) corrections, the heavy quark field \( Q_v(x) \) is decomposed into two components:
\begin{align}
Q_v(x) = e^{-i m_Q v \cdot x}[H_v + h_v],
\end{align}
where the large (upper) component \( h_v(x) \) and the small (lower) component \( H_v(x) \) are defined as
\begin{align}
h_v(x) &= e^{i m_Q v \cdot x} \frac{1 + \slashed{v}}{2} Q_v(x), \\ \nonumber
H_v(x) &= e^{i m_Q v \cdot x} \frac{1 - \slashed{v}}{2} Q_v(x).
\end{align}

Using the relations \( \not\!{v} h_v =  h_v \) and \( \not\!{v} H_v = -H_v \), the full QCD Lagrangian can be rewritten as
\begin{eqnarray}
\nonumber \mathcal{L} &=& \bar{Q}(i\slashed{D} - m_Q) Q 
\\ \nonumber &=& (\bar{h}_v + \bar{H}_v)\left[m_Q(\slashed{v} - 1) + i\slashed{D} \right](h_v + H_v) 
\\ \nonumber &=& \bar{h}_v\, i v \cdot D\, h_v - \bar{H}_v\, (i v \cdot D + 2m_Q)\, H_v 
\\ && + \bar{h}_v\, i \slashed{D}_\perp\, H_v + \bar{H}_v\, i \slashed{D}_\perp\, h_v.
\end{eqnarray}
In this expression, \( h_v \) is independent of the heavy quark mass, while the small component \( H_v \) retains the mass-dependence degrees of freedom. By applying the Euler--Lagrange equations of motion, the full heavy quark field can be expressed in terms of the large component \( h_v \), yielding
\begin{eqnarray}
-i v \cdot D\, h_v &=& i \slashed{D}_\perp\, H_v, \\ \nonumber
(i v \cdot D + 2m_Q)\, H_v &=& i \slashed{D}_\perp\, h_v.
\end{eqnarray}
Substituting the second equation into the first gives
\begin{align}
-i v \cdot D\, h_v = i \slashed{D}_\perp\, \frac{1}{i v \cdot D + 2m_Q}\, i \slashed{D}_\perp\, h_v.
\end{align}

From the above equation of motion, we can derive the corresponding effective Lagrangian as
\begin{align}
\nonumber \mathcal{L} &= \overline{h}_v\, i v \cdot D\, h_v + \overline{h}_v\, i \slashed{D}_{\perp} \frac{1}{i v \cdot D + 2m_Q}\, i \slashed{D}_{\perp} h_v \\
\nonumber &= \overline{h}_v\, i v \cdot D\, h_v + \frac{1}{2m_Q} \overline{h}_v\, (i \slashed{D}_{\perp})^2 h_v + \mathcal{O}\left(\frac{1}{m_Q^2}\right) \\
\nonumber &= \overline{h}_v\, i v \cdot D\, h_v + \frac{1}{2m_Q} \overline{h}_v\, (i D_{\perp})^2 h_v 
\\
&+ \mathcal{C}_{mag}(\mu)\frac{g}{4 m_Q} \overline{h}_v\, \sigma_{\mu \nu} G^{\mu \nu} h_v + \mathcal{O}\left(\frac{1}{m_Q^2}\right),
\end{align}
where the gluon field strength tensor is defined as
\begin{align}
G^{\mu \nu} &= \partial^\mu A^\nu - \partial^\nu A^\mu - i g [A^\mu, A^\nu] \, ,
\\ \nonumber -i g G^{\mu \nu} &= [D^\mu, D^\nu] \, .
\end{align}
This Lagrangian, incorporating the \( 1/m_Q \) corrections, represents an expansion of the QCD Lagrangian in inverse powers of the heavy quark mass. This expression can be compared well with the Lagrangian for the non-relativistic Schrodinger equation:
\begin{itemize}

\item In the limit of heavy quark at rest, the first term corresponds to the time derivative term with color-electric (Coulomb) interaction.

\item The second term accounts for the kinetic energy of the heavy quark. This term breaks flavor symmetry due to its dependence on the heavy quark motion.

\item The third term describes the chromomagnetic interaction between the heavy quark spin and the gluon field. It breaks both heavy quark spin and flavor symmetries.

\end{itemize}

By evaluating the \( 1/m_Q \) corrections to the HQET Lagrangian, the mass of a heavy hadron can be expressed as
\begin{align}
m_{H_Q} = m_Q + \bar{\Lambda} + {\delta m} + \mathcal{O}\left({1/m_Q^2}\right), 
\end{align}
where \( \bar{\Lambda} \) denotes the leading-order contribution from \( \mathcal{L}_{\text{HQET}} \), which is independent of the heavy quark mass and spin, and \( \delta m \) encodes the subleading \( 1/m_Q \) corrections arising from the kinetic and chromomagnetic terms.

\subsection{QCD sum rules}
\label{sec3.2}

In this subsection we briefly introduce the method of QCD sum rules (QSR) within the framework of heavy quark effective theory (HQET). As a representative example, we consider the interpolating current
\begin{eqnarray}
&& J^\alpha_{3/2,-,\Omega_b^-,2,1,\lambda} 
\label{eq:example}
\\ \nonumber &=& i \epsilon_{abc} \left( [D_t^{\mu} s^{aT}] C \gamma_t^\nu s^b + s^{aT} C \gamma_t^\nu [D_t^{\mu} s^b] \right)
\\ \nonumber &\times& \left( g_t^{\alpha\mu} \gamma_t^{\nu} \gamma_5 + g_t^{\alpha\nu} \gamma_t^{\mu} \gamma_5 - \frac{2}{3} g_t^{\mu\nu} \gamma_t^{\alpha} \gamma_5 \right) h_v^c \, ,
\end{eqnarray}
which has the quark content \( ssb \), quantum numbers \( J^P = 3/2^- \), and contains an explicit $\lambda$-mode orbital excitation. This current couples to the $P$-wave state \( [\Omega_b^-(3/2^-), j_l = 2, s_l = 1,\lambda] \) ($s_l = 1$  ($\mathbf{S}$), $l_\rho = 0$ ($\mathbf{S}$), $l_\lambda = 1$, $j_l = s_l \oplus l_\rho \oplus l_\lambda = 2$, color $\mathbf{\bar{3}}_C$ ($\mathbf{A}$), and flavor $\mathbf{6}_F$ ($\mathbf{S}$)) via
\begin{eqnarray}
\langle 0| J^\alpha_{3/2,-,\Omega_b^-,2,1,\lambda}(x) | \Omega_b^-(3/2^-),2,1,\lambda \rangle = f ~ u^\alpha(x) \, .
\label{eq:decayconstant}
\end{eqnarray}
In these expressions, \( a, b, c \) are color indices, and \( \epsilon_{abc} \) is the totally antisymmetric Levi-Civita tensor in color space. The covariant derivative is defined as \( D^\mu = \partial^\mu - i g_s A^\mu \), and the transverse gamma matrix and derivative are given by \( \gamma_t^\mu = \gamma^\mu - v\!\!\!\slash\, v^\mu \) and \( D_t^\mu = D^\mu - (D \cdot v)\, v^\mu \), respectively. The transverse metric tensor is defined as \( g_t^{\alpha\beta} = g^{\alpha\beta} - v^{\alpha}v^{\beta} \). Here, \( C \) is the charge-conjugation matrix, \( T \) denotes transposition in Dirac space, \( s(x) \) is the light strange quark field, and \( h_v(x) \) represents the heavy bottom quark field in HQET. The parameter \( f \) is the decay constant, and \( u^\alpha(x) \) is a Rarita-Schwinger spinor describing a spin-\( 3/2 \) fermion.

The corresponding two-point correlation function is defined as
\begin{eqnarray}
\Pi^{\alpha\beta}(\omega) 
\label{eq:pi}
&=& i \int d^4 x\, e^{i k \cdot x} \langle 0 \big| T\big[J^\alpha_{3/2,-,\Omega_b^-,2,1,\lambda}(x)
\\ \nonumber && ~~~~~~~~~~~~~~~~~~~~~ \times
\bar J^\beta_{3/2,-,\Omega_b^-,2,1,\lambda}(0)\big] \big| 0 \rangle
\\ \nonumber &=& \left( g^{\alpha\beta} - \frac{q^\alpha q^\beta}{q^2} \right) \times \frac{1 + v\!\!\!\slash}{2} \Pi(\omega) + \cdots \, ,
\end{eqnarray}
where \( \omega = v \cdot k \) is the residual energy in HQET. We isolate the Lorentz structure corresponding to the spin-\( 3/2 \) component, while the ellipsis denotes contributions from other spin structures.

The dispersion relation for the correlation function is
\begin{eqnarray}
\Pi(\omega) = \int \frac{\rho(\omega^\prime)}{
\omega^\prime - \omega - i \epsilon} \, d\omega^\prime \, ,
\end{eqnarray}
where $\rho(\omega)$ denotes the spectral density in the limit \( m_b \to \infty \). After parameterizing it as a sum of an isolated pole and the contributions from higher states (including the continuum), we obtain
\begin{eqnarray}
\rho_{\rm phen}(\omega) &\equiv& \frac{1}{\pi}\mathrm{Im}\,\Pi(\omega)
\\ \nonumber &=& f^2 \delta(\omega - \overline{\Lambda}) + \rho_{\rm phen}^{\rm higher}(\omega)\, \theta(\omega - \omega_h) \, ,
\end{eqnarray}
where \( \omega_h \) denotes the threshold for the onset of higher states and continuum contributions, and \( \rho_{\rm phen}^{\rm higher}(\omega) \) accounts for the higher contributions. Accordingly, the correlation function $\Pi(\omega)$ can be expressed at the hadronic level as
\begin{equation}
\Pi(\omega) = \frac{f^2}{\overline{\Lambda} - \omega} + \text{higher states} \, ,
\label{eq:pole}
\end{equation}
where the parameter \( \overline{\Lambda} \) is defined by
\begin{equation}
\overline{\Lambda} \equiv \lim_{m_b \rightarrow \infty} (m - m_b) \, ,
\label{eq:leading}
\end{equation}
with \( m \) denoting the physical mass of the heavy baryon and \( m_b \) the mass of the bottom quark. The parameter \( \overline{\Lambda} \) thus characterizes the contribution from the light degrees of freedom in the heavy quark limit.

At the quark-gluon level, we employ the operator product expansion method (OPE) to evaluate the correlation function, from which the OPE spectral density \( \rho_{\rm OPE}(s) \) is extracted. Based on the principle of quark hadron duality which asserts that, in the limit \( \omega \to \infty \), the correlation function becomes dominated by perturbative contributions while nonperturbative condensate effects vanish, we arrive at the relation:
\begin{equation}
\int_{\omega_h}^\infty \frac{\rho_{\rm phen}^{\rm higher}(\omega^\prime)}{\omega^\prime - \omega - i \epsilon} \, d\omega^\prime 
= \int_{\omega_c}^\infty \frac{\rho_{\rm OPE}(\omega^\prime)}{\omega^\prime - \omega - i \epsilon} \, d\omega^\prime \, ,
\label{eq:quarkhadron}
\end{equation}
where the parameter \( \omega_c \) serves as the continuum threshold, determined through phenomenological analysis. In practice, \( \omega_c \) is chosen to lie moderately above the lowest-lying pole in order to ensure pole dominance while maintaining good convergence of the OPE series. We note that \( \omega_c \) and \( \omega_h \) are not necessarily equal.

To enhance the contribution of the lowest-lying state and suppress that from higher excitations, we perform a Borel transformation on both sides of the dispersion relation and obtain:
\begin{eqnarray}
&& f^2 e^{- \overline{\Lambda} / T} + \int_{\omega_h}^\infty \rho_{\rm phen}^{\rm higher}(\omega^\prime)\, e^{- \omega^\prime / T} \, d\omega^\prime 
\\ \nonumber && ~~~~~~~~~~~~~~~~~~~~~~~~~~~~~ = \int_{s_<}^\infty \rho_{\rm OPE}(\omega^\prime)\, e^{- \omega^\prime / T} \, d\omega^\prime \, .
\end{eqnarray}
Comparing this with Eq.~(\ref{eq:quarkhadron}), we arrive at the QCD sum rule relation:
\begin{eqnarray} 
&& \Pi(\omega_c, T) = f(\omega_c, T)^2 \times e^{- \overline{\Lambda}(\omega_c, T) / T} 
\label{eq:ope} 
\\ \nonumber &=&
\int_{s_<}^{\omega_c} [ \frac{8}{63\,\pi^4}\omega^7-\frac{4\,m_s^2}{3\,\pi^4}\omega^5+\frac{10\,m_s^4}{3\,\pi^4}\omega^3]e^{-\omega/T}d\omega
\\ \nonumber &&
- \frac{5\,\langle g_s^2 GG \rangle}{12\, \pi^4} T^4 -\frac{10 \,m_s^3\, \langle \bar s s\rangle }{3\,\pi^2}T^2
+\frac{5 \,\langle g_s^2 GG \rangle\, m_s^2}{48 \,\pi^4}T^2
\\ \nonumber &&
-\frac{5\, \langle g_s \bar s \sigma G s \rangle \, \langle \bar s s\rangle}{9}
- \frac{5\, \langle g_s^2 GG \rangle\,m_s\,\langle \bar s s\rangle}{288\,\pi^2}  \,
\\ \nonumber &&
- \frac{5\, \langle g_s \bar s \sigma G s\rangle^2}{144} {1 \over T^2} \, . 
\end{eqnarray}
Here, \( s_< = 2m_s \) is the physical threshold, and \( T \) is the Borel mass parameter.

The physical parameters can be extracted by differentiating Eq.~(\ref{eq:ope}) with respect to \( -1/T \), leading to the following expressions:
\begin{eqnarray}
\overline{\Lambda}(\omega_c, T)
&=& \frac{1}{\Pi(\omega_c, T)} \cdot \frac{\partial \Pi(\omega_c, T)}{\partial(-1/T)} \, ,
\label{eq:mass}
\\ \nonumber
f^2(\omega_c, T)
&=& \Pi(\omega_c, T) \cdot e^{\overline{\Lambda}(\omega_c, T) / T} \, .
\label{eq:coupling}
\end{eqnarray}
These results depend on two essential parameters: the continuum threshold \( \omega_c \) and the Borel mass \( T \). To constrain their allowed ranges, we impose three commonly adopted criteria. In the numerical analysis, we employ the following QCD input parameters~\cite{PDG2024,Yang:1993bp,Narison:2011xe,Narison:2018dcr,Gimenez:2005nt,Jamin:2002ev,Ioffe:2002be,Ovchinnikov:1988gk,Colangelo:2000dp}:
\begin{eqnarray}
\nonumber \langle \bar{q} q \rangle &=& -(0.24\pm0.01~\mathrm{GeV})^3 \, , 
\\ \nonumber \langle \bar{s} s \rangle &=& (0.8 \pm 0.1) \times \langle \bar{q} q \rangle \, , 
\\ \nonumber \langle g_s \bar{q} \sigma G q \rangle &=& - M_0^2 \times \langle \bar{q} q \rangle \, , 
\\ \langle g_s \bar{s} \sigma G s \rangle &=& - M_0^2 \times \langle \bar{s} s \rangle \, ,
\label{eq:condensate}
\\ \nonumber M_0^2 &=& 0.8~\mathrm{GeV}^2 \, , 
\\ \nonumber \langle g_s^2 G^2 \rangle &=& (0.48 \pm 0.14)~\mathrm{GeV}^4 \, , 
\\ \nonumber m_s &=& 0.15~\mathrm{GeV} \, , 
\\ \nonumber m_q &\approx& m_u \approx m_d \approx 0~\mathrm{MeV} \, .
\end{eqnarray}

\begin{figure*}[htbp]
\centering
\subfigure[]{\scalebox{0.42}{\includegraphics{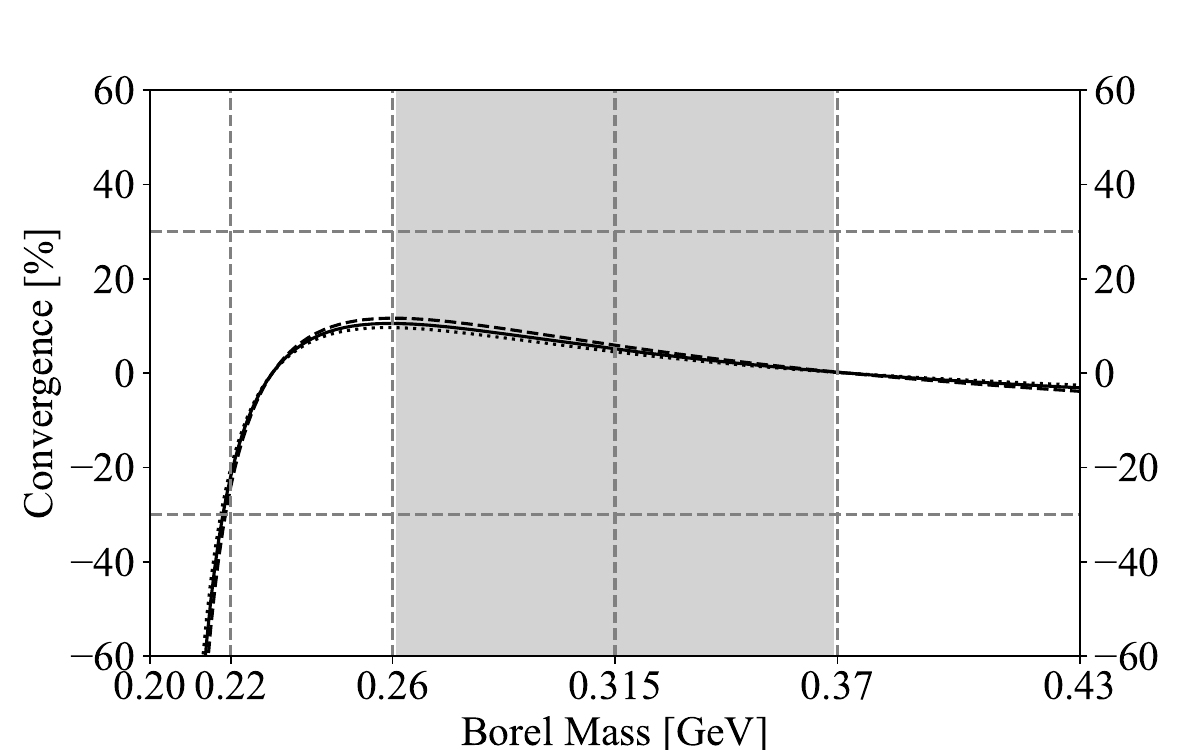}}}
\hspace{0.5cm}
\subfigure[]{\scalebox{0.42}{\includegraphics{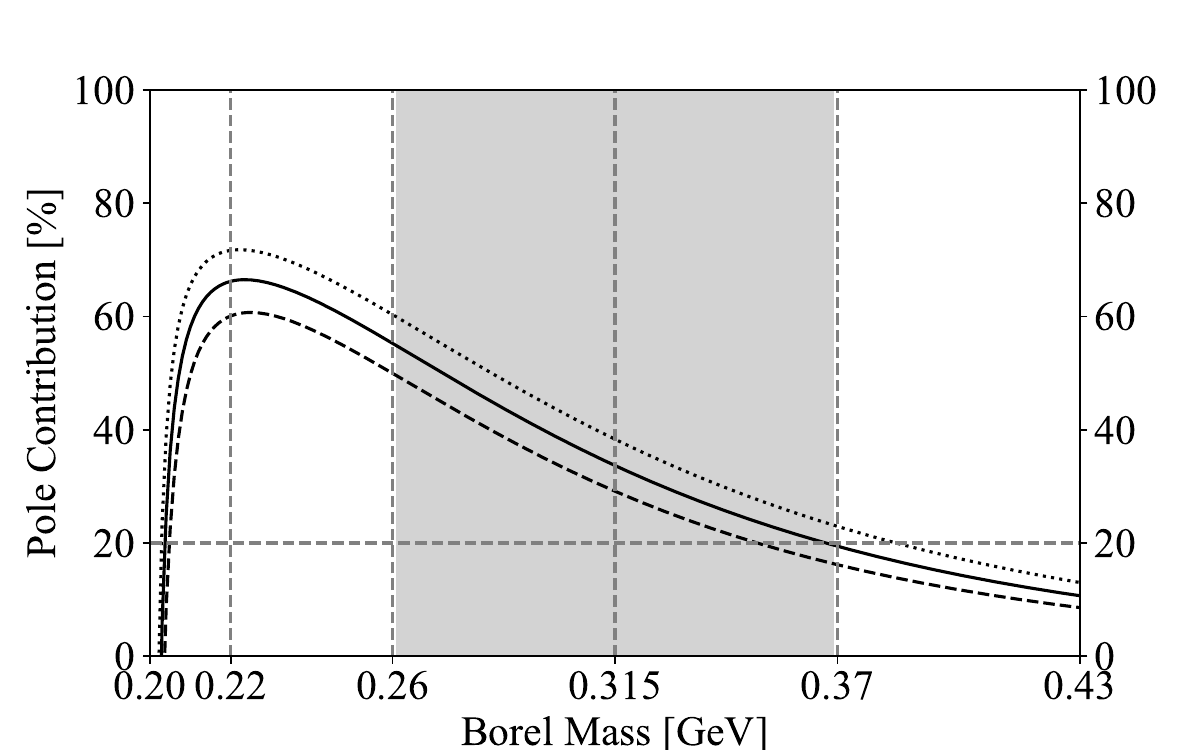}}}
\caption{Dependence of (a) the convergence ratio (CVG) and (b) the pole contribution (PC), defined in Eqs.~(\ref{eq_convergence}) and~(\ref{eq_pole}), on the Borel mass \( T \), evaluated using the interpolating current \( J^\alpha_{3/2,-,\Omega_b^-,2,1,\lambda}  \). The dashed, solid, and dotted curves correspond to \( \omega_c = 1.98 \), \( 2.08 \), and \( 2.18~\mathrm{GeV} \), respectively.}
\label{fig:pole}
\end{figure*}

\begin{figure*}[htbp]
\centering
\subfigure[]{\scalebox{0.42}{\includegraphics{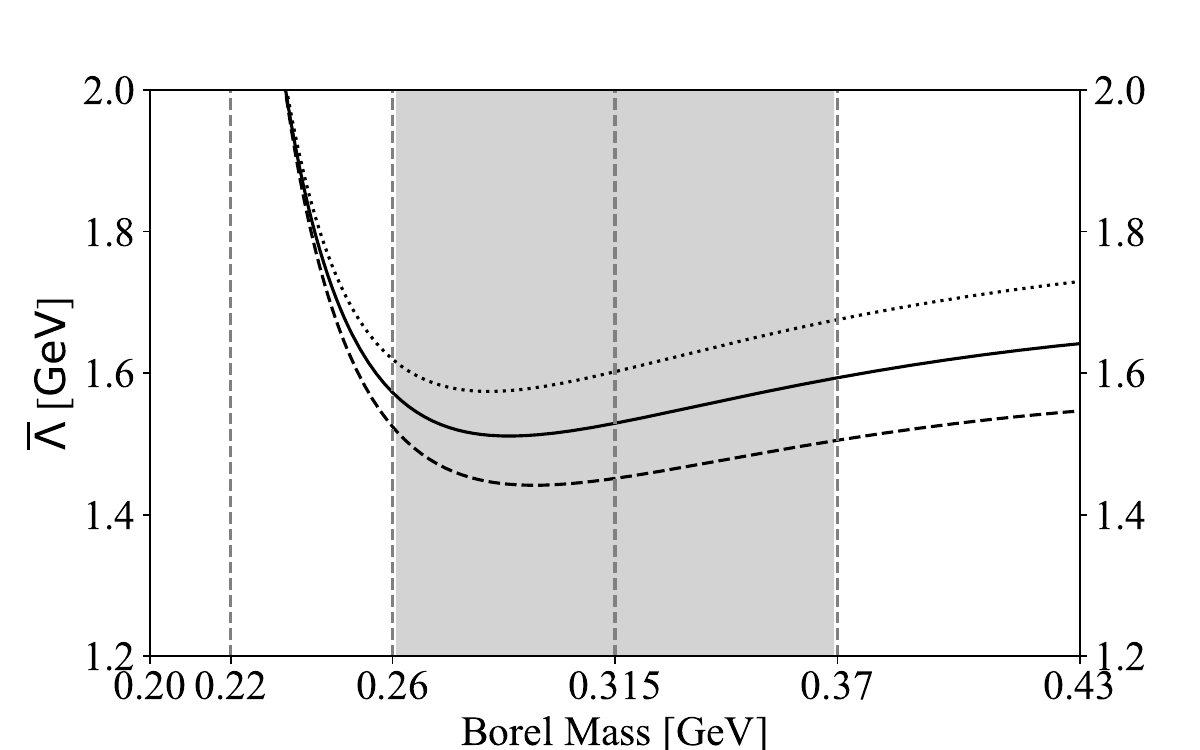}}}
\hspace{0.5cm}
\subfigure[]{\scalebox{0.42}{\includegraphics{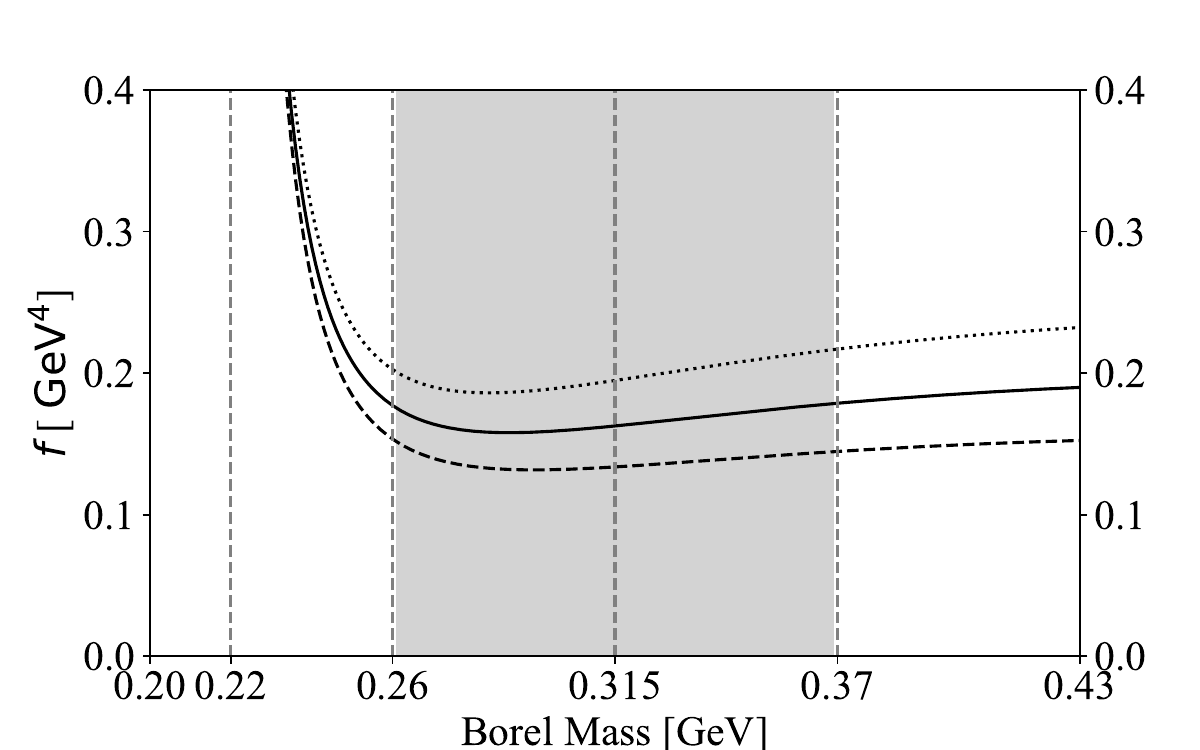}}}
\caption{Dependence of (a) the residual mass \( \overline{\Lambda} \) and (b) the decay constant \( f \) on the Borel mass \( T \), extracted using the interpolating current \( J^\alpha_{3/2,-,\Omega_b^-,2,1,\lambda}  \). The dashed, solid, and dotted curves correspond to \( \omega_c = 1.98 \), \( 2.08 \), and \( 2.18~\mathrm{GeV} \), respectively.}
\label{fig:leading}
\end{figure*}

The first criterion requires that the contribution from high-dimensional condensates remain below 30\% of the total correlation function:
\begin{equation}
\label{eq_convergence}
\mathrm{CVG} \equiv \left| \frac{ \Pi^{\text{high-order}}(\omega_c, T) }{ \Pi(\omega_c, T) } \right| \leq 30\% \, .
\end{equation}
As shown in Fig.~\ref{fig:pole}(a), this condition sets the lower bound of the Borel window to \( T_{\min} = 0.22~\mathrm{GeV} \).

The second criterion requires that the pole contribution account for at least 20\% of the total:
\begin{equation}
\label{eq_pole}
\mathrm{PC} \equiv \frac{ \Pi(\omega_c, T) }{ \Pi(\infty, T) } \geq 20\% \, .
\end{equation}
As illustrated in Fig.~\ref{fig:pole}(b), this condition imposes an upper limit of \( T_{\max} = 0.37~\mathrm{GeV} \) when \( \omega_c = 2.08~\mathrm{GeV} \). Combining both constraints, we obtain the preliminary Borel window:
\begin{equation}
0.22~\mathrm{GeV} < T < 0.37~\mathrm{GeV} \, .
\end{equation}
Note that the two values, ``$30\%$'' given in Eq.~(\ref{eq_convergence}) and ``$20\%$'' given in Eq.~(\ref{eq_pole}), are somewhat adjustable. These values were chosen in Ref.~\cite{Chen:2015kpa} to better describe the corresponding experimental results. While the obtained results do depend on these values, they remain consistent across different sets of values when taking the associated uncertainties into account.

To further constrain the Borel window, we examine the behavior of the extracted parameters \( \overline{\Lambda} \) and \( f \) as functions of the Borel mass \( T \), as shown in Fig.~\ref{fig:leading}. The convergence ratio CVG reaches its minimum near \( T_{\rm peak} = 0.26~\mathrm{GeV} \), above which both \( \overline{\Lambda} \) and \( f \) exhibit improved stability. In contrast, below this threshold, the strong \( T \)-dependence violates the Borel stability criterion and undermines the reliability of the extracted quantities. Based on this observation, we adopt the refined Borel window:
\begin{equation}
0.26~\mathrm{GeV} < T < 0.37~\mathrm{GeV} \, .
\end{equation}
Within this optimized range, the extracted physical parameters are:
\begin{eqnarray}
\overline{\Lambda} &=& 1.53\pm0.10~\mathrm{GeV} \, , \\ \nonumber
f &=& 0.162\pm0.035~\mathrm{GeV}^4\, ,
\end{eqnarray}
where the central values correspond to \( T = 0.315~\mathrm{GeV} \) and \( \omega_c = 2.08~\mathrm{GeV} \).

To incorporate \( \mathcal{O}(1/m_Q) \) corrections, we employ the effective Lagrangian of heavy quark effective theory (HQET), which is given by
\begin{eqnarray}
\mathcal{L}_{\rm eff} = \bar{h}_{v}^a\, i v \cdot D\, h_{v}^a + \frac{1}{2m_b} \mathcal{K} + \frac{1}{2m_b} \mathcal{S} \, ,
\label{eq:next}
\end{eqnarray}
where \( \mathcal{K} \) denotes the kinetic energy operator,
\begin{eqnarray}
\mathcal{K} = \bar{h}_{v}^a (i D_{t})^{2} h_{v}^a \, ,
\end{eqnarray}
and \( \mathcal{S} \) represents the chromomagnetic interaction term,
\begin{eqnarray}
\mathcal{S} = \frac{g_s}{2} C_{\rm mag} \left( \frac{m_b}{\mu} \right) \bar{h}_{v}^a \sigma_{\mu\nu} G^{\mu\nu} h_{v}^a \, .
\end{eqnarray}
The corresponding Wilson coefficient is given by
\begin{eqnarray}
C_{\rm mag} \left( \frac{m_b}{\mu} \right) = \left[ \frac{\overline{\alpha}_s(m_b)}{\overline{\alpha}_s(\mu)} \right]^{3/\beta_0} \, ,
\end{eqnarray}
where \( \beta_0 = 11 - \frac{2}{3} n_f \). At the renormalization scale \( \mu = 1~\mathrm{GeV} \), this coefficient evaluates to \( C_{\rm mag}(\mu) \approx 0.8 \) for the bottom quark.

At the hadronic level, the pole term can be expanded up to \( \mathcal{O}(1/m_Q) \) as
\begin{eqnarray}
\Pi(\omega) &=& \frac{(f + \delta f)^2}{\overline{\Lambda} + \delta m - \omega}
\label{eq:correction}
\\ \nonumber &=& \frac{f^2}{\overline{\Lambda} - \omega} - \frac{\delta m f^2}{(\overline{\Lambda} - \omega)^2} + \frac{2f \delta f}{\overline{\Lambda} - \omega} \, ,
\end{eqnarray}
where \( \delta m \) and \( \delta f \) denote the \( \mathcal{O}(1/m_Q) \) corrections to the mass and decay constant, respectively.

To evaluate \( \delta m \), we consider the following three-point correlation functions:
\begin{eqnarray}
&& \delta_O \Pi^{\alpha\beta}(\omega, \omega^\prime)
\label{eq:nextpi}
\\ \nonumber &=& i^2 \int d^4x\, d^4y\, e^{i k \cdot x - i k^\prime \cdot y}
\\ \nonumber &\times&
\langle 0 | T[J^\alpha_{3/2,-,\Omega_b^-,2,1,\lambda}(x)\, O(0)\, 
\bar{J}^\beta_{3/2,-,\Omega_b^-,2,1,\lambda}(y)] | 0 \rangle 
\\ \nonumber &=& \left(g^{\alpha\beta} - \frac{q^\alpha q^\beta}{q^2} \right) \times \frac{1 + v\!\!\!\slash}{2}  \delta_O \Pi(\omega, \omega^\prime) + \cdots \, ,
\end{eqnarray}
where \( O = \mathcal{K} \) or \( \mathcal{S} \). At the hadronic level, these correlators can be expressed as
\begin{eqnarray}
\delta_{\mathcal{K}} \Pi(\omega, \omega^\prime) &=& \frac{f^2 K}{(\overline{\Lambda} - \omega)(\overline{\Lambda} - \omega^\prime)} + \cdots \, ,
\label{eq:K}
\\
\delta_{\mathcal{S}} \Pi(\omega, \omega^\prime) &=& \frac{d_M f^2 \Sigma}{(\overline{\Lambda} - \omega)(\overline{\Lambda} - \omega^\prime)} + \cdots \, ,
\label{eq:S}
\end{eqnarray}
with the matrix elements defined as
\begin{eqnarray}
\nonumber K &\equiv& \langle \Omega_b^-(3/2^-) | \bar{h}_v^a (i D_{t})^2 h_v^a | \Omega_b^-(3/2^-) \rangle \, ,
\\ \nonumber 
d_M \Sigma &\equiv& \left\langle \Omega_b^-(3/2^-) \left| \frac{g_s}{2} \bar{h}_v^a \sigma_{\mu\nu} G^{\mu\nu} h_v^a \right| \Omega_b^-(3/2^-) \right\rangle \, ,
\\
d_M &\equiv& d_{j, j_l} \, ,
\\ \nonumber 
d_{j_l - 1/2, j_l} &=& 2 j_l + 2 \, ,
\\ \nonumber 
d_{j_l + 1/2, j_l} &=& -2 j_l \, .
\end{eqnarray}
By setting \( \omega = \omega^\prime \) and combining Eqs.~(\ref{eq:correction}), (\ref{eq:K}), and (\ref{eq:S}), we obtain the \( \mathcal{O}(1/m_Q) \) correction to the mass:
\begin{eqnarray}
\delta m = -\frac{1}{2 m_b} \left( K + d_M C_{\rm mag} \Sigma \right) \, .
\label{eq:more}
\end{eqnarray}

\begin{figure*}[hbt]
\centering
\subfigure[]{\scalebox{0.42}{\includegraphics{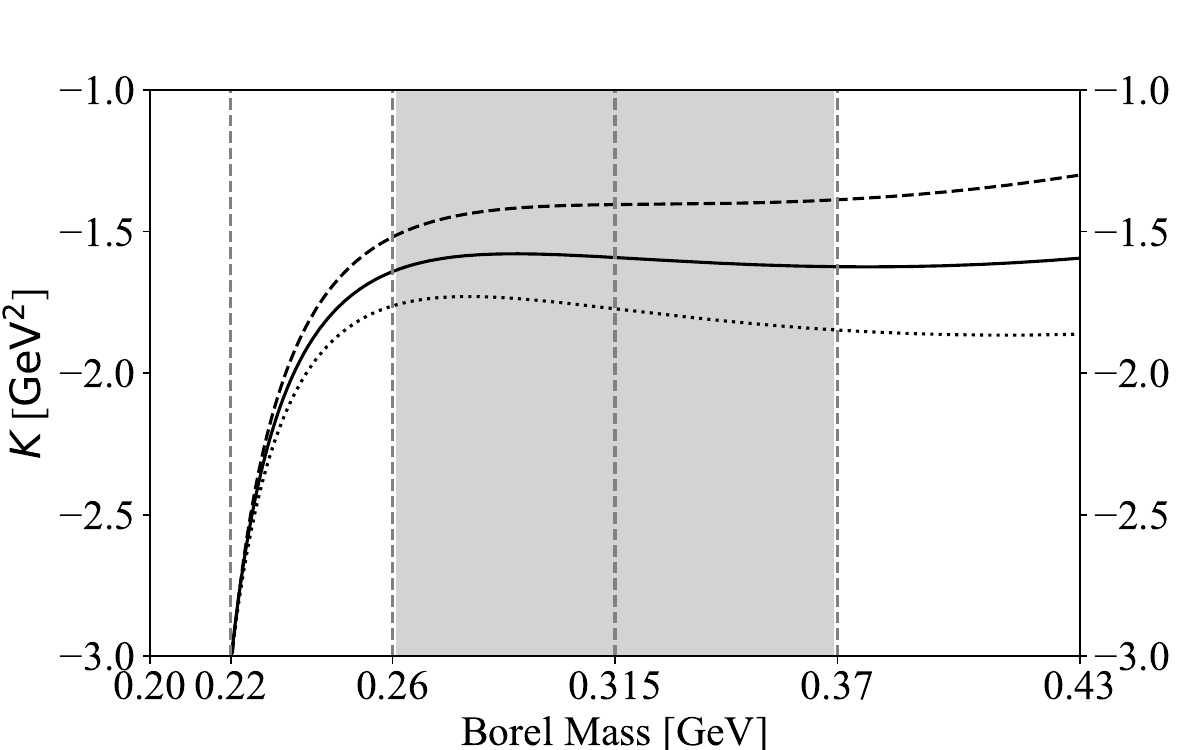}}}
\hspace{0.5cm}
\subfigure[]{\scalebox{0.42}{\includegraphics{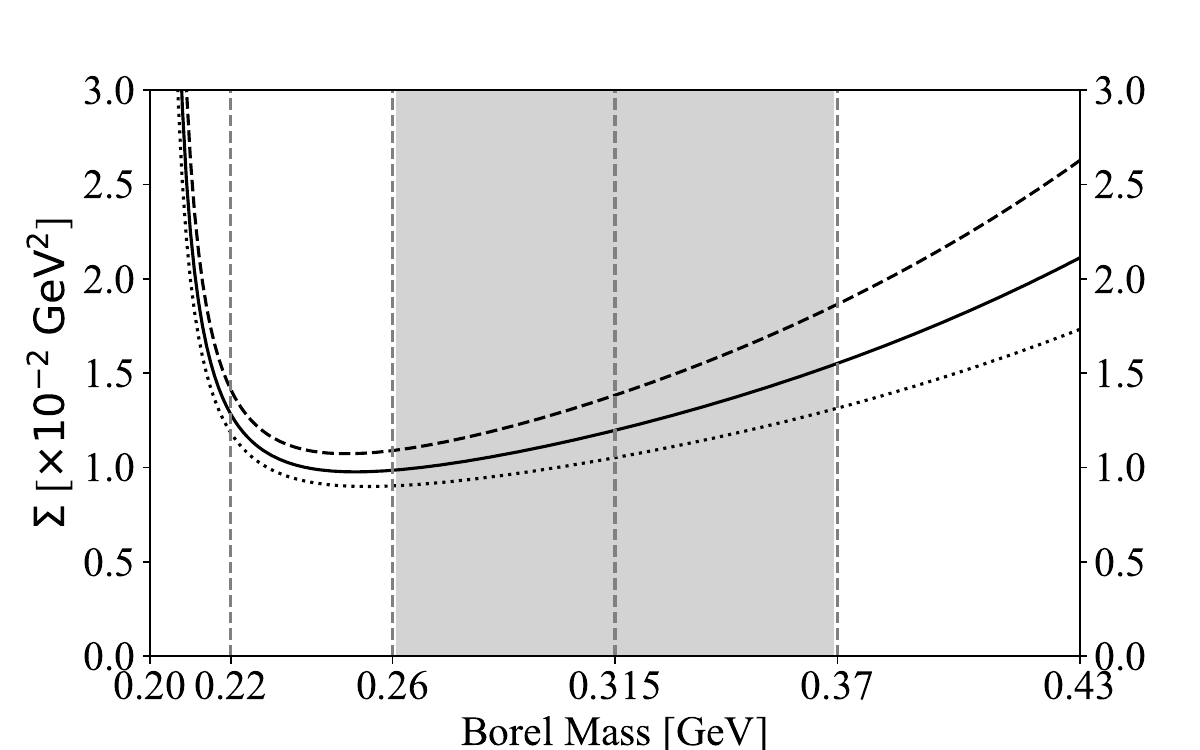}}}
\caption{Dependence of (a) \( K \) and (b) \(  \Sigma \) on the Borel mass \( T \), obtained using the interpolating current \( J^\alpha_{3/2,-,\Omega_b^-,2,1,\lambda}  \). The working region is \( 0.26~\mathrm{GeV} < T < 0.37~\mathrm{GeV} \). The dashed, solid, and dotted curves correspond to \( \omega_c = 1.98 \), \( 2.08 \), and \( 2.18~\mathrm{GeV} \), respectively.}
\label{fig:KandS}
\end{figure*}

The three-point correlation functions in Eq.~(\ref{eq:nextpi}) can also be evaluated at the quark-gluon level using the operator product expansion. By substituting Eq.~(\ref{eq:example}), performing a double Borel transformation in the variables \( \omega \) and \( \omega^\prime \), and setting \( T_1 = T_2 = T \), we obtain:
\begin{eqnarray}
 &&f^2 K\, e^{-\overline{\Lambda} / T} \label{eq:Kc}
\\ \nonumber &=& \int_{s_<}^{\omega_c}
\left[-\frac{2}{45\, \pi^4} \omega^9+\frac{74\, m_s^2 }{105\, \pi^2}\omega^7
\right] e^{-\omega/T}\, d\omega
\\ \nonumber &&
-\frac{352\, m_s\, \langle \bar{s} s \rangle}{3 \,\pi^2}T^6+\frac{205\,\langle g^2 GG \rangle}{72\, \pi^4}T^6
\\ \nonumber &&
-\frac{37\, \langle g^2 GG \rangle\,m_s^2}{72\,\pi^4}T^4-\frac{13\langle g_s \bar{s} \sigma G s \rangle^2}{72} \, 
\\ \nonumber &&
+\frac{7\,m_s\,\langle \bar{s} s \rangle\,\langle g^2 GG \rangle}{432\,\pi^2}T^2
-\frac{5\langle g^2 GG \rangle\,\langle \bar{s} s \rangle\, \langle g_s \bar{s} \sigma G s \rangle}{6912\,T^2}
\\ \nonumber &&
-\frac{\langle g^2 GG \rangle\,m_s^2\,\langle \bar{s} s \rangle^2}{10368\,T^2}
-\frac{5\,\langle g^2 GG \rangle\,\langle g_s \bar{s} \sigma G s \rangle^2}{110592\,T^4},
\\
&&f^2 \Sigma\, e^{-\overline{\Lambda} / T}
\\ \nonumber &=& \frac{5\,\langle g^2 GG \rangle\,T^6}{18\,\pi^4}-\frac{5\,\langle g^2 GG \rangle\,m_s^2\,}{96\,\pi^2}T^4
\\\nonumber &&
+\frac{5\,m_s\,\langle \bar{s} s \rangle\,\langle g^2 GG \rangle}{864\,\pi^2}T^2\, .
\label{eq:Sc}
\end{eqnarray}
The dependence of these sum rules on the Borel mass is illustrated in Fig.~\ref{fig:KandS}. Both \( K \) and \( \Sigma \) exhibit mild variation within the working region \( 0.26~\mathrm{GeV} < T < 0.37~\mathrm{GeV} \). Within this range, we extract the numerical results:
\begin{eqnarray}
K &=& -1.59\pm0.20~\mathrm{GeV}^2 \, , 
\\ \nonumber \Sigma &=& 0.012\pm0.005~\mathrm{GeV}^2 \, .
\end{eqnarray}

Based on the above results, we determine the mass of the ground-state topped baryon \( \Omega_b^-(3/2^-) \). A crucial aspect of this calculation is the choice of renormalization scheme for the heavy quark mass. In our QCD sum rule analyses of bottom baryons~\cite{Mao:2015gya,Yang:2020zrh,Wang:2024rai}, the \( 1S \) scheme has been adopted:
\begin{eqnarray}
m_b^{1S} &=& 4.66\pm0.03~\mathrm{GeV} \, .
\end{eqnarray}
Applying Eqs.~(\ref{eq:leading}) and~(\ref{eq:more}), the mass of the \( \Omega_b^-(3/2^-) \) baryon can be expressed as  
\begin{eqnarray}
m &=& m_b + \overline{\Lambda} + \delta m 
\\ \nonumber &=& (4.66 \pm 0.03) + (1.53 \pm 0.10) + (0.16 \pm 0.02)~\mathrm{GeV}
\\ \nonumber &=& 6.35 \pm 0.13~\mathrm{GeV} \, .
\end{eqnarray}
The mass of its partner, the \( \Omega_b^-(5/2^-) \) baryon, is determined to be  
\begin{eqnarray}
m^\prime &=& 6.36\pm0.12~\mathrm{GeV} \, ,
\end{eqnarray}
and the corresponding mass splitting is  
\begin{eqnarray}
\Delta m = m - m^\prime &=& 10\pm4~\mathrm{MeV} \, .
\end{eqnarray}
It is important to emphasize that, although absolute mass predictions of singly heavy baryons from QCD sum rules typically carry sizable theoretical uncertainties (about 100~MeV for the above $\Omega_b^-(3/2^-)$ and $\Omega_b^-(5/2^-)$ states), the mass splittings within heavy-quark spin doublets are significantly more reliable (just 4~MeV in this case). This improvement stems from the fact that such splittings are largely independent of the heavy quark mass and are instead controlled by the dynamics of the light degrees of freedom. Our analysis indicates that the \( \rho \)-mode excitation lies below the \( \lambda \)-mode, contrary to conventional expectations based on constituent quark models. This inversion may be attributed to the relatively large uncertainties in the inter-multiplet mass differences predicted by QCD sum rules, which are comparable to those associated with the absolute mass values. In contrast, the intra-multiplet splittings exhibit much smaller theoretical uncertainties and can therefore be determined with greater precision.

\subsection{Light-cone sum rules}
\label{sec3.3}

The QCD sum rule technique has been extensively applied to a wide range of physical processes—including the determination of hadron spectra, decay constants, and electromagnetic form factors—and has achieved considerable success. This method has been further extended within the framework of QCD light-cone sum rules (LCSR). 
The central idea is to perform the operator product expansion (OPE) near the light-cone (\( x^2 = 0 \)), which reduces the conventional three-point correlation function to a more tractable two-point form, thereby significantly simplifying the calculation; see, {\it e.g.}, Refs.~\cite{Chen:2010ic,Huang:2010dc}, which study the decay properties of the $J^{PC} = 1^{-+}$ hybrid mesons using both three-point correlation functions and their corresponding two-point approximations.

In this subsection we consider again the \( P \)-wave bottom baryon \( \Omega_b^-({3/2}^-) \), belonging to the \( [\mathbf{6}_F, 2, 1, \lambda] \) doublet, as a representative example and evaluate its \( D \)-wave decay into \( \Xi_b^0(1/2^+) \) and \( K^-(0^-) \) using the light-cone sum rule approach. The effective interaction Lagrangian describing this transition is given by
\begin{eqnarray}
\mathcal{L}
= g \times {\bar \Omega_{b\mu}}(3/2^-) \gamma_\nu \gamma_5 \Xi_{b}(1/2^+) \, \partial^{\mu} \partial^{\nu} K(0^-) \, ,
\end{eqnarray}
where $g$ is the coupling constant responsible for the $D$-wave decay process $\Omega_b^-({3/2}^-) \to \Xi_b^0 K^-$.

To proceed, we examine the following two-point correlation function
\begin{eqnarray}
&& \Pi^\alpha(\omega, \omega^\prime)
\\ \nonumber &=& \int d^4 x~e^{-i k \cdot x}~\langle 0 | J^\alpha_{3/2,-,\Omega_b^-,2,1,\lambda}(0) \bar J_{\Xi_b^{0}}(x) | K^-(q) \rangle
\\ \nonumber &=& {1+v\!\!\!\slash\over2} G^\alpha (\omega, \omega^\prime) \, .
\end{eqnarray}
rather than the three-point function. The use of the two-point correlation function with the kaon field background becomes good for light Nambu-Goldstone boson. In this expression \( k^\prime = k + q \), \( \omega = v \cdot k \), and \( \omega^\prime = v \cdot k^\prime \). Here, \( k_\mu \), \( k^\prime_\mu \), and \( q_\mu \) denote the four-momenta of the final baryon \( \Xi_b^0(1/2^+) \), the initial baryon \( \Omega_b^-(3/2^-) \), and the emitted meson \( K^- \), respectively. 
The interpolating field $J^\alpha_{3/2,-,\Omega_b^-,2,1,\lambda}$ has already been defined in Eq.~(\ref{eq:example}), and the interpolating field $J_{\Xi_b^{0}}$ is defined as
\begin{eqnarray}
J_{\Xi_b^{0}} &=& \epsilon_{abc} \left[d^{aT} C\gamma_{5} s^{b}\right] h_{v}^{c} \, ,
\end{eqnarray}
which couples to the ground-state $\Xi_b^0(1/2^+)$ baryon via
\begin{eqnarray}
\langle 0| J_{\Xi_b^{0}} | \Xi_b^0(1/2^+) \rangle = f_{\Xi_b} \times u(x) \, ,
\end{eqnarray}
where the parameter \( f_{\Xi_b} \) denotes the decay constant, and \( u(x) \) is the Dirac spinor describing a spin-\( 1/2 \) fermion.

At the hadronic level, the correlation function \( G^{\alpha}(\omega, \omega^\prime) \) takes the form:
\begin{equation}
G^\alpha (\omega, \omega^\prime) = g \times { f_{\Omega_b} f_{\Xi_b} \over (\bar \Lambda_{\Omega_b} - \omega^\prime) (\bar \Lambda_{\Xi_b} - \omega)} \times q\!\!\!\slash \gamma_5 ~ q^\alpha + \cdots \, , 
\label{G0C1}
\end{equation}
where the ellipsis represents contributions from higher resonances and continuum states.
For clarity, we have used \( f \to f_{\Omega_b} \) to denote the decay constant of the \( \Omega_b^-({3/2}^-) \) baryon, as defined in Eq.~(\ref{eq:decayconstant}).

\begin{widetext}
At the quark-gluon level, we compute \( G^{\alpha}(\omega, \omega^\prime) \) using the method of operator product expansion (OPE):
\begin{eqnarray}
\label{eq:g1}
G^\alpha (\omega, \omega^\prime)
&=& \int_0^\infty dt \int_0^1 du e^{i (1-u) \omega^\prime t} e^{i u \omega t} \times 8 \times \Big (
\frac{f_K m_s u}{4\pi^2 t^2}\phi_{2;K}(u)+\frac{f_K m_s^2 u}{12(m_u+m_s)\pi^2 t^2}\phi_{3;K}^\sigma(u)
\\ \nonumber &&
+ \frac{f_K m_s^2 m_K^2 u}{48(m_u+m_s)\pi^2}\phi_{3;K}^\sigma(u)+\frac{f_K m_s u}{64\pi^2}\phi_{4;K}(u)+\frac{f_K u}{12}\langle \bar s s\rangle\phi_{2;K}(u)+\frac{f_K m_s m_K^2 u t^2}{288(m_u+m_s)}\langle s s\rangle\phi_{3;K}^\sigma(u)
\\ \nonumber &&
+ \frac{f_K u t^2}{192}\langle s s\rangle\phi_{4;K}(u)+\frac{f_K u t^2}{192}\langle g_s \bar s\sigma G s\rangle \phi_{2;K}(u)+\frac{f_K u t^4}{3072}\langle g_s \bar s \sigma G s\rangle\phi_{4;K}(u) \Big ) \times \gamma \cdot q~\gamma_5~q^\alpha
\\ \nonumber &-&
\int_0^\infty dt \int_0^1 du \int D \underline{\alpha} e^{i \omega^{\prime} t(\alpha_2 + u \alpha_3)} e^{i \omega t(1 - \alpha_2 - u \alpha_3)} \times \Big (\frac{f_{3K} u}{2\pi^2 t^2}\Phi_{3;K}(\underline{\alpha})-\frac{f_{3K}}{2\pi^2 t^2}\Phi_{3;K}(\underline{\alpha})
\\ \nonumber &&
+\frac{i f_{3K} u^2 \alpha_3}{2\pi^2 t v \cdot q}\Phi_{3;K}(\underline{\alpha})+\frac{i f_{3K} u \alpha_2}{2\pi^2 t v \cdot q}\Phi_{3;K}(\underline{\alpha})-\frac{i f_{3K} u}{2\pi^2 t v \cdot q}\Phi_{3;K}(\underline{\alpha})\Big ) \times \gamma \cdot q~\gamma_5~q^\alpha + \cdots\, .
\end{eqnarray}
The explicit forms of the light-cone distribution amplitudes (\( \phi_{2;K}(u) \), \( \phi_{3;K}^\sigma(u) \), \( \phi_{4;K}(u) \), and \( \Phi_{3;K}(\underline{\alpha}) \), etc.) and the values of the light-cone sum rule parameters (\( f_{3K} \), etc.) used in the above expressions can be found in Refs.~\cite{Ball:1998je,Ball:2006wn,Ball:2004rg,Ball:1998kk,Ball:1998sk,Ball:1998ff,Ball:2007rt,Ball:2007zt,Ball:2002ps}.

We then apply the Borel transformation to both Eq.~(\ref{G0C1}) at the hadronic level and Eq.~(\ref{eq:g1}) at the quark-gluon level:
\begin{eqnarray}
&& g \times f_{\Omega_b} f_{\Xi_b} \times e^{- {\bar \Lambda_{\Omega_b} / T_1}} e^{ - {\bar \Lambda_{\Xi_b} / T_2}}
\label{eq:621lambda}
\\ \nonumber &=& 8 \times \Big ( -\frac{i f_k m_s u_0}{4\pi^2}T^3 f_2({\omega_c \over T})\phi_{2;K}(u_0)-\frac{i f_K m_K^2 u_0}{12(m_u+m_s)\pi^2}T^3 f_2({\omega_c \over T})\phi_{3;K}^\sigma(u_0)+\frac{i f_K m_s u_0}{64\pi^2}T f_0({\omega_c \over T})\phi_{4;K}(u_0)
\\ \nonumber &&
+\frac{i f_K u_0}{12}\langle \bar s s\rangle T f_0({\omega_c \over T})\phi_{2;K}(u_0)-\frac{i f_K m_s u_0}{288(m_u+m_s)}\langle \bar s s\rangle {1\over T}\phi_{3;K}^\sigma(u_0)-\frac{i f_K u_0}{192}\langle \bar s s\rangle {1\over T}\phi_{4;K}(u_0)
\\ \nonumber &&
-\frac{i f_K u_0}{192}\langle g_s \bar s\sigma G s\rangle {1\over T}\phi_{2;K}(u_0)+\frac{i f_K u_0}{3072}\langle g_s \bar s \sigma G s\rangle{1\over T^3}\phi_{4;K}(u_0) \Big )
\\ \nonumber &-&
\Big(-\frac{i f_{3K}}{2\pi^2}T^3f_2({\omega_c \over T}) \int_0^{1 \over 2} d\alpha_2 \int_{{1 \over 2}-\alpha_2}^{1-\alpha_2} d\alpha_3 ({u_0 \over \alpha_3} \Phi_{3;K}(\underline{\alpha})-{1 \over \alpha_3} \Phi_{3;K}(\underline{\alpha}))
\\ \nonumber &&
+\frac{i f_{3K}}{2\pi^2}T^3f_2({\omega_c\over T}) \int_0^{1 \over 2} d\alpha_2 \int_{{1 \over 2}-\alpha_2}^{1-\alpha_2} d\alpha_3 {1\over \alpha_3}{\partial\over\partial\alpha_3}(\alpha_3 u_0\Phi_{3;K}(\underline{\alpha})+\alpha_2\Phi_{3;K}(\underline{\alpha})-\Phi_{3;K}(\underline{\alpha}))\Big ) \, .
\end{eqnarray}
\end{widetext}
In the above expressions, the variable \( \alpha = \{ \alpha_1, \alpha_2, \alpha_3 \} \), and the integration measure \( \int D \underline{\alpha} = \int_0^1 d\alpha_1 \int_0^1 d\alpha_2 \int_0^1 d\alpha_3 \) is subject to the constraint \( \delta(\alpha_1 + \alpha_2 + \alpha_3 - 1) \), which has already been integrated out in Eq.~(\ref{eq:g1}). The function \( f_n(x) \) is defined as
\begin{eqnarray}
f_n(x) \equiv 1 - e^{-x} \sum_{k=0}^n \frac{x^k}{k!} \, .
\end{eqnarray}
The parameters \( \omega \) and \( \omega^\prime \) are mapped to the Borel masses \( T_1 \) and \( T_2 \), respectively. We adopt the symmetric point \( T_1 = T_2 = 2T \), which implies \( u_0 = \frac{T_1}{T_1 + T_2} = \frac{1}{2} \). The continuum threshold is set to \( \omega_c = 1.665~\mathrm{GeV} \), corresponding to the average value derived from the mass sum rules for \( \Omega_b^-(3/2^-) \) and \( \Xi_b^0(1/2^+) \). The Borel parameter \( T \) is chosen within the range \( 0.26~\mathrm{GeV} < T < 0.37~\mathrm{GeV} \), as determined from the mass sum rule analysis of \( \Omega_b^-(3/2^-) \). 

\begin{figure}[hbt]
\centering
\scalebox{0.6}{\includegraphics{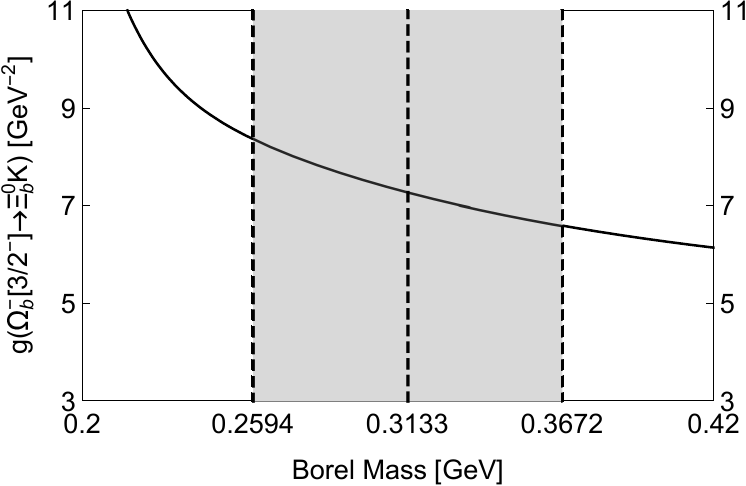}}
\caption{The coupling constant \( g \) as a function of the Borel mass \( T \), obtained for the baryon \( \Omega_b(3/2^-) \) in the \( [\mathbf{6}_F, 2, 1, \lambda] \) doublet.}
\label{fig:311R}
\end{figure}

The coupling constant $g$ is numerically calculated as
\begin{eqnarray}
g &=& 7.27~{^{+1.10}_{-0.68}}~{^{+2.22}_{-1.67}}~{^{+2.34}_{-1.66}}~{^{+1.96}_{-1.42}}~{\rm GeV}^{-2}
\\ \nonumber &=& 7.27~{^{+3.92}_{-2.83}}~{\rm GeV}^{-2} \, ,
\end{eqnarray}
where the uncertainties originate from four main sources: the variation of the Borel parameter \( T \), uncertainties in the input parameters associated with the \( \Xi_b(1/2^+) \) state, those related to the \( \Omega_b(3/2^-) \) state, and the QCD parameters specified in Eqs.~(\ref{eq:condensate}). The dependence of \( g \) on the Borel mass \( T \) is illustrated in Fig.~\ref{fig:311R}.

The partial decay width for the process \( \Omega_b(3/2^-) \to \Xi_b + K \) is evaluated using the following expression:
\begin{eqnarray}
&& \Gamma \left( \Omega_b^-({3/2}^-) \rightarrow \Xi_b^{0} + K^- \right)
\\ \nonumber &=& \frac{|\vec p_2|}{32\pi^2 m_0^2} \times g^2 \times p_{2,\mu}p_{2,\nu}p_{2,\rho}p_{2,\sigma}
\\ \nonumber &\times& {\rm Tr}\Big[ \gamma^\nu\gamma_5 \left( p\!\!\!\slash_1 + m_1 \right) \gamma^\sigma \gamma_5
\\ \nonumber && \Big(g^{\rho\mu}-{\gamma^\rho\gamma^\mu\over3}-{p_{0}^{\rho}\gamma^\mu-p_{0}^{\mu}\gamma^\rho \over 3m_0} -{2p_{0}^{\rho}p_{0}^{\mu} \over 3m_0^2}  \Big) ( p\!\!\!\slash_0 + m_0 ) \Big],
\end{eqnarray}
where the indices 0, 1, and 2 refer to the initial baryon \( \Omega_b^-(3/2^-) \), the final baryon \( \Xi_b^0(1/2^+) \), and the emitted meson \( K^- \), respectively. Substituting the numerical values, we obtain the partial decay width:
\begin{eqnarray}
\Gamma_{\Omega_b^-({3/2}^-) \rightarrow \Xi_b^{0}K^-}&=& 4.6~{^{+3.3}_{-1.9}}{\rm~MeV} \, .
\end{eqnarray}

It is worth noting that \( \mathcal{O}(1/m_c) \) corrections are not included in the light-cone sum rule calculation of this decay width, in contrast to the mass spectrum analysis where such corrections are considered using conventional QCD sum rules. This distinction arises from both theoretical and practical considerations. First, baryons typically decay via multiple channels, resulting in significantly more decay processes than mass states. Second, the computation of decay widths especially with higher-order corrections is considerably more complex. Third, and perhaps most importantly, the leading-order uncertainties in decay width calculations are already substantial. For charmed baryons, previous studies~\cite{Dai:1998ve,Zhu:2000py} suggest that \( 1/m_c \) corrections may contribute up to 30\%, which is comparable to or even smaller than the leading-order uncertainties. Thus, unless these uncertainties can be significantly reduced, incorporating subleading corrections may not lead to a meaningful improvement in predictive power.

\section{Summary for various theoretical results}
\label{sec4}

In this section we provide a concise review of QCD sum rule studies on $P$-wave singly heavy baryons. Since heavy quark effective theory (HQET) is more applicable to singly bottom baryons than to singly charmed baryons, we begin by discussing the singly bottom baryons belonging to the $SU(3)$ flavor $\mathbf{\bar{3}}_F$ representation in Sec.~\ref{sec4.1}, followed by their charmed counterparts in Sec.~\ref{sec4.2}, where mixing effects are also taken into account. Sections~\ref{sec4.3} and~\ref{sec4.4} cover the singly bottom baryons of the $SU(3)$ flavor $\mathbf{6}_F$ and their charmed counterparts, respectively.

\subsection{Singly bottom baryons of the $SU(3)$ flavor $\mathbf{\bar{3}}_F$}
\label{sec4.1}

\begin{table*}[ht]
\begin{center}
\renewcommand{\arraystretch}{1.5}
\caption{Summary of the mass spectra and decay properties of the $P$-wave bottom baryons of the $SU(3)$ flavor $\mathbf{\bar{3}}_F$, obtained using QCD sum rules and light-cone sum rules within the framework of heavy quark effective theory. Possible experimental candidates are listed in the last column for comparison. Data are adapted from Refs.~\cite{Wang:2024rai,Luo:2024jov}.}
\setlength{\tabcolsep}{0.1mm}{
\begin{tabular}{c|c| c | c | c | c | c |c c }
\hline\hline
 \multirow{2}{*}{B}&\multirow{2}{*}{Multiplet}&Baryon & ~~~~Mass~~~~ &~Splitting~ & ~~~~\multirow{2}{*}{Partial Decay Width} ~~~~ &~~\multirow{2}{*}{Total width}~~& \multirow{2}{*}{Candidate}
\\& & ($j^P$) & ({GeV}) & ({MeV}) & & &
\\ \hline\hline
\multirow{15}{*}{~~$\Lambda_b$~~}&\multirow{2}{*}{~~$[\mathbf{\bar 3}_F, 0, 1, \rho]$~~}
&~~\multirow{2}{*}{$\Lambda_b({1\over2}^-)$}~~     &\multirow{2}{*}{$5.92^{+0.17}_{-0.19}$} &\multirow{2}{*}{-}
&$\Gamma(\Lambda_b^0({1\over2}^-)\to \Sigma_b^0\gamma$)=$170^{+2700}_{-~130}$~keV       &\multirow{2}{*}{$0.5^{+9.0}_{-0.4}$~MeV }  &\multirow{2}{*}{-}
\\ 
&&&&&$\Gamma(\Lambda_b^0({1\over2}^-)\to \Sigma_b^{*0}\gamma$)=$280^{+6800}_{-~280}$~keV                        &
\\\cline{2-8}
&\multirow{4}{*}{~~$[\mathbf{\bar 3}_F, 1, 1, \rho]$~~} &
\multirow{2}{*}{~~$\Lambda_b({1\over2}^-)$~~}     &\multirow{2}{*}{$5.92^{+0.13}_{-0.10}$}      &\multirow{4}{*}{$8\pm 3$}
&~~~$\Gamma(\Lambda_b({1\over2}^-)\to \Sigma_b\pi\to\Lambda_b\pi\pi$)=$2^{+5}_{-2}$~keV ~~               &\multirow{2}{*}{~~$36^{+69}_{-32}$~keV~~}   &\multirow{2}{*}{~$\Lambda_b(5912)$}
\\
&&&&&~~~~~~$\Gamma(\Lambda_b^0({1\over2}^-)\to \Lambda_b^0\gamma)=34^{+64}_{-30}$~keV             &
\\ \cline{3-4} \cline{6-8}
&&\multirow{2}{*}{$\Lambda_b({3\over2}^-)$}&\multirow{2}{*}{$5.92^{+0.13}_{-0.10}$} &
&$\Gamma(\Lambda_b({3\over2}^-)\to \Sigma_b^*\pi\to\Lambda_b\pi\pi$)=$5^{+11}_{-~5}$~keV          &\multirow{2}{*}{$70^{+140}_{-~60}$~keV}                &\multirow{2}{*}{$\Lambda_b(5920)$}
\\
&&&&&$\Gamma(\Lambda_b^0({3\over2}^-)\to \Lambda_b^0 \gamma$)  =$65^{+130}_{-~55}$~keV          &
\\\cline{2-8}
&\multirow{3}{*}{~~$[\mathbf{\bar 3}_F, 2, 1, \rho]$~~} &
\multirow{2}{*}{$\Lambda_b({3\over2}^-)$}     &\multirow{2}{*}{$5.93^{+0.13}_{-0.13}$}      &\multirow{3}{*}{$17\pm8$}
&$\Gamma(\Lambda_b({3\over2}^-)\to\Sigma_b^*\pi$ )=$0.0^{+12.0}_{-~0.0}$~MeV
&\multirow{2}{*}{$0.0^{+12.0}_{-~0.0}$~MeV} &\multirow{2}{*}{-}
\\
&&&&&$\Gamma(\Lambda_b^0({3\over2}^-)\to\Sigma_b^0\gamma$ )=$73^{+520}_{-~73}$~keV                    &
\\ \cline{3-4} \cline{6-8}
&&\multirow{1}{*}{$\Lambda_b({5\over2}^-)$}     &\multirow{1}{*}{$5.93^{+0.13}_{-0.13}$}      &
&$\Gamma(\Lambda_b^0({5\over2}^-)\to\Sigma_b^{*0}\gamma$)             =$11^{+84}_{-11}$~keV                    &$11^{+84}_{-11}$~keV &\multirow{1}{*}{-}
\\\cline{2-8}
&\multirow{6}{*}{~~$[\mathbf{\bar 3}_F, 1, 0, \lambda]$~~} &
\multirow{3}{*}{$\Lambda_b({1\over2}^-)$}     &\multirow{3}{*}{$5.91^{+0.17}_{-0.13}$}      &\multirow{6}{*}{$4\pm2$}
&$\Gamma(\Lambda_b({1\over2}^-)\to\Sigma_b\pi\to\Lambda_b\pi\pi$ )=$0.0^{+32.0}_{-~0.0}$~GeV
&\multirow{3}{*}{$0.0^{+32.0}_{-~0.0}$~GeV} &\multirow{2}{*}{-}
\\
&&&&&$\Gamma(\Lambda_b^0({1\over2}^-)\to\Sigma_b^{*0}\gamma$ )=$0.1^{+3.4}_{-0.0}$~MeV                    &
\\
&&&&&$\Gamma(\Lambda_b^0({1\over2}^-)\to\Sigma_b^0\gamma$ )=$0.3^{+6.2}_{-0.3}$~MeV                    &
\\ \cline{3-4} \cline{6-8}
&&\multirow{3}{*}{$\Lambda_b({3\over2}^-)$}     &\multirow{3}{*}{$5.91^{+0.17}_{-0.13}$}      &
&$\Gamma(\Lambda_b^0({3\over2}^-)\to\Sigma_b^{*}\pi\to\Lambda_b\pi\pi$)    =$0.0^{+6.8}_{-0.0}$~GeV                    &\multirow{3}{*}{$0.0^{+6.8}_{-0.0}$~GeV} &\multirow{1}{*}{-}
\\
&&&&&$\Gamma(\Lambda_b^0({3\over2}^-)\to\Sigma_b^{*0}\gamma$ )=$9.2^{+250.0}_{-~~8.1}$~keV                    &
\\
&&&&&$\Gamma(\Lambda_b^0({3\over2}^-)\to\Sigma_b^0\gamma$ )=$0.2^{+3.5}_{-0.2}$~MeV                    &
\\ \hline \hline
\multirow{22}{*}{~~$\Xi_b$~~}&\multirow{3}{*}{~~$[\mathbf{\bar 3}_F, 0, 1, \rho]$~~}
&~~\multirow{3}{*}{$\Xi_b({1\over2}^-)$}~~     &\multirow{3}{*}{$6.10^{+0.08}_{-0.08}$} &\multirow{3}{*}{-}
&$\Gamma(\Xi_b({1\over2}^-)\to \Xi_b\pi$) =$12^{+10.0}_{-~9.3}$~GeV     &\multirow{3}{*}{$12^{+10.0}_{-~9.3}$~GeV }  &\multirow{3}{*}{-}
\\
&&&&&$\Gamma(\Xi_b^0({1\over2}^-)\to \Xi_b^0\gamma$)=$440^{+390}_{-440}$~keV                        &
\\ 
&&&&&$\Gamma(\Xi_b^0({1\over2}^-)\to \Xi_b^0\gamma$)=$720^{+640}_{-720}$~keV                        &
\\\cline{2-8}
&\multirow{5}{*}{~~$[\mathbf{\bar 3}_F, 1, 1, \rho]$~~}
&~~\multirow{2}{*}{$\Xi_b({1\over2}^-)$}~~     &\multirow{2}{*}{$6.09^{+0.13}_{-0.12}$} &\multirow{4}{*}{$7\pm3$}
&$\Gamma(\Xi_b({1\over2}^-)\to \Xi_b^{\prime}\pi$) =$3.7^{+9.5}_{-3.7}$~MeV     &\multirow{2}{*}{$3.7^{+9.5}_{-3.7}$~MeV }  &\multirow{2}{*}{$\Xi_b(6087)$}
\\
&&&&&$\Gamma(\Xi_b^0({1\over2}^-)\to \Xi_b^0\gamma$)=$41^{+78}_{-41}$~keV                        &
\\ \cline{3-4}\cline{6-8}
&&~~\multirow{3}{*}{$\Xi_b({3\over2}^-)$}~~     &\multirow{3}{*}{$6.10^{+0.13}_{-0.12}$} &
&$\Gamma(\Xi_b({3\over2}^-)\to \Xi_b^*\pi$) =$640^{+1700}_{-~640}$~keV            &\multirow{3}{*}{$730^{+1900}_{-~730}$~keV}  &\multirow{3}{*}{$\Xi_b(6095)$}
\\
&&&&&$\Gamma(\Xi_b({3\over2}^-)\to \Xi_b^{\prime}\pi$) =$2^{+4}_{-2}$~keV                         &
\\
&&&&&$\Gamma(\Xi_b^0({3\over2}^-)\to \Xi_b^0\gamma$)=$85^{+160}_{-~85}$~keV                        &
\\\cline{2-8}
&\multirow{8}{*}{$[\mathbf{\bar 3}_F, 2, 1, \rho]$}
&~~\multirow{4}{*}{$\Xi_b({3\over2}^-)$}~~    &\multirow{4}{*}{$6.10^{+0.15}_{-0.10}$}      &\multirow{8}{*}{$14\pm7$}
&$\Gamma(\Xi_b({3\over2}^-)\to\Xi_b\pi$)          = $1.0^{+8.3}_{-1.0}$~MeV                        &\multirow{4}{*}{~$1.5^{+12.0}_{-~1.5}$~MeV~}    &\multirow{4}{*}{-}
\\
&&&&&$\Gamma(\Xi_b({3\over2}^-)\to\Xi_b^{\prime}\pi$)          = $0^{+390}_{-~~0}$~keV                       &
\\
&&&&&$\Gamma(\Xi_b({3\over2}^-)\to\Xi_b^*\pi$)          = $370^{+3260}_{-~370}$~keV                        &    &
\\
&&&&&$\Gamma(\Xi_b^0({3\over2}^-)\to\Xi_b^{\prime0}\gamma$)          = $140^{+510}_{-140}$~keV                       &
\\  \cline{3-4}\cline{6-8}
&&~~\multirow{4}{*}{$\Xi_b({5\over2}^-)$}~~    &\multirow{4}{*}{$6.11^{+0.15}_{-0.10}$}     &
&$\Gamma(\Xi_b({5\over2}^-)\to\Xi_b\pi$)          = $1.0^{+7.2}_{-1.0}$~MeV                      &\multirow{4}{*}{$1.0^{+7.4}_{-1.0}$~MeV}       &\multirow{4}{*}{-}
\\
&&&&&$\Gamma(\Xi_b({5\over2}^-)\to\Xi_b^{\prime}\pi$)          = $0^{+180}_{-~~0}$~keV                      &       &
\\
&&&&&$\Gamma(\Xi_b({5\over2}^-)\to\Xi_b^{*}\pi$)          = $0^{+10}_{-~0}$~keV                     &       &
\\
&&&&&$\Gamma(\Xi_b^0({5\over2}^-)\to\Xi_b^{*0}\gamma$)          = $19^{+74}_{-19}$~keV                        &       &
\\\cline{2-8}
&\multirow{6}{*}{$[\mathbf{\bar 3}_F, 1, 0, \lambda]$}
&~~\multirow{3}{*}{$\Xi_b({1\over2}^-)$}~~    &\multirow{3}{*}{$6.10^{+0.20}_{-0.10}$}      &\multirow{6}{*}{$4\pm2$}
&$\Gamma(\Xi_b({1\over2}^-)\to\Xi_b^{\prime}\pi$)          = $2.8^{+8.2}_{-2.8}$~GeV                        &\multirow{3}{*}{~$2.8^{+8.2}_{-2.8}$~GeV~}    &\multirow{4}{*}{-}
\\
&&&&&$\Gamma(\Xi_b^0({1\over2}^-)\to\Xi_b^{\prime0}\gamma$)          = $1.2^{+11.0}_{-~1.2}$~MeV                       &
\\
&&&&&$\Gamma(\Xi_b^0({1\over2}^-)\to\Xi_b^{\star0}\gamma$)          = $0.5^{+6.4}_{-0.5}$~MeV                       &
\\  \cline{3-4}\cline{6-8}
&&~~\multirow{3}{*}{$\Xi_b({3\over2}^-)$}~~    &\multirow{3}{*}{$6.11^{+0.20}_{-0.10}$}     &
&$\Gamma(\Xi_b({3\over2}^-)\to\Xi_b^{\prime}\pi$)          = $0.0^{+2.0}_{-0.0}$~MeV                      &\multirow{3}{*}{$0.3^{+2.0}_{-0.3}$~GeV}       &\multirow{4}{*}{-}
\\
&&&&&$\Gamma(\Xi_b({3\over2}^-)\to\Xi_b^{*}\pi$)          = $0.3^{+2.0}_{-0.3}$~GeV                     &       &
\\
&&&&&$\Gamma(\Xi_b^0({3\over2}^-)\to\Xi_b^{\prime0}\gamma$)          = $0.7^{+5.8}_{-0.7}$~MeV                        &       &
\\ \hline\hline
\end{tabular}}
\label{tab:result1}
\end{center}
\end{table*}

$P$-wave bottom baryons belonging to the $SU(3)$ flavor $\mathbf{\bar{3}}_F$ representation have been systematically studied in Refs.~\cite{Wang:2024rai,Mao:2015gya,Yang:2019cvw,Luo:2024jov} using QCD sum rules and light-cone sum rules within the framework of heavy quark effective theory. As illustrated in Fig.~\ref{fig:categorization}, there are four baryon multiplets in total: $[\mathbf{\bar{3}}_F, 1, 0, \lambda]$, $[\mathbf{\bar{3}}_F, 0, 1, \rho]$, $[\mathbf{\bar{3}}_F, 1, 1, \rho]$, and $[\mathbf{\bar{3}}_F, 2, 1, \rho]$. As summarized in Table~\ref{tab:result1}, our results suggest that:
\begin{itemize}

\item The $[\mathbf{\bar{3}}_F, 1, 0, \lambda]$ doublet contains four excited bottom baryons: $\Lambda_b(1/2^-)$, $\Lambda_b(3/2^-)$, $\Xi_b(1/2^-)$, and $\Xi_b(3/2^-)$. The first two $\Lambda_b$ states can be used to explain the observed $\Lambda_b(5912)^0$ and $\Lambda_b(5920)^0$, while the latter two $\Xi_b$ states do not correspond well to $\Xi_b(6087)^0$ and $\Xi_b(6095)^0/\Xi_b(6100)^-$.

\item The \([\mathbf{\bar{3}}_F, 0, 1, \rho]\) singlet contains two excited bottom baryons: \( \Lambda_b(1/2^-) \) and \( \Xi_b(1/2^-) \). In particular, the decay width of the latter state is calculated to be rather large, suggesting that it would be difficult to observe experimentally.

\item The $[\mathbf{\bar{3}}_F, 1, 1, \rho]$ doublet contains four excited bottom baryons: $\Lambda_b(1/2^-)$, $\Lambda_b(3/2^-)$, $\Xi_b(1/2^-)$, and $\Xi_b(3/2^-)$. This doublet can collectively account for the observed states $\Lambda_b(5912)^0$, $\Lambda_b(5920)^0$, $\Xi_b(6087)^0$, and $\Xi_b(6095)^0/\Xi_b(6100)^-$.

\item The $[\mathbf{\bar{3}}_F, 2, 1, \rho]$ doublet contains four excited bottom baryons: $\Lambda_b(3/2^-)$, $\Lambda_b(5/2^-)$, $\Xi_b(3/2^-)$, and $\Xi_b(5/2^-)$. This doublet can be used to further predict two $\Lambda_b$ states and two $\Xi_b$ states. Taking into account the experimental measurements reported in Refs.~\cite{LHCb:2012kxf,CDF:2013pvu,CMS:2021rvl,LHCb:2023zpu}, these baryons are expected to be:
\begin{enumerate}[label=\alph*)]

\item $\Lambda_b^0(3/2^-)$: Its mass is approximately 5930~MeV, and its decay width is nearly zero. This state can be searched for in the $\Lambda_b \pi \pi$ decay channel.

\item $\Lambda_b^0(5/2^-)$: Its mass is approximately 5947~MeV, and its decay width is also nearly zero.

\item $\Xi_b^0(3/2^-)$: Its mass is approximately 6102~MeV, and its decay width is about 1.5~MeV. Its isospin partner, $\Xi_b^-(3/2^-)$, has a mass of approximately 6108~MeV. These two states can be searched for in the $\Xi_b \pi$ and $\Xi_b^* \pi$ decay channels.

\item $\Xi_b^0(5/2^-)$: Its mass is approximately 6116~MeV, and its decay width is about 1.0~MeV. Its isospin partner, $\Xi_b^-(5/2^-)$, has a mass of approximately 6122~MeV. These two states can be searched for in the $\Xi_b \pi$ decay channel.

\end{enumerate}

\end{itemize}

Another QCD sum rule analysis was conducted in Ref.~\cite{Xin:2023usd}, where the authors focused on $P$-wave bottom baryons of the $\lambda$-type. This study was performed not within the framework of HQET, but rather using full QCD sum rules. Similar to our findings, their results also support the interpretation of the $\Lambda_b(5912)^0$, $\Lambda_b(5920)^0$, $\Xi_b(6087)^0$, and $\Xi_b(6095)^0/\Xi_b(6100)^-$ as $P$-wave bottom baryons with spin-parity $1/2^-$ and $3/2^-$. However, their results favor the $\lambda$-mode excitation. Moreover, they propose that the $\Lambda_b(5920)^0$ and $\Xi_b(6095)^0/\Xi_b(6100)^-$ may each consist of two significant substructures.

\subsection{Singly charmed baryons of the $SU(3)$ flavor $\mathbf{\bar3}_F$}
\label{sec4.2}

\begin{table*}[ht]
\begin{center}
\renewcommand{\arraystretch}{1.5}
\caption{Summary of the mass spectra and decay properties of the $P$-wave charmed baryons of the $SU(3)$ flavor $\mathbf{\bar{3}}_F$. The first column lists the charmed baryons categorized according to heavy quark effective theory, while the third column presents the results after accounting for mixing effects. Possible experimental candidates are listed in the last column for comparison. The data are adapted from Ref.~\cite{Luo:2025jpn}.}
\scalebox{0.925}{\begin{tabular}{   c|c | c | c | c | c | c | c}
\hline\hline
  \multirow{2}{*}{HQET state}&\multirow{2}{*}{Mixing}&\multirow{2}{*}{Mixed state} & Mass & Difference & ~~~~~~~~~~~\multirow{2}{*}{Partial Decay Width }~~~~~~~~~~~ & Width  & \multirow{2}{*}{Candidate}
\\  &&&   ({GeV}) & ({MeV}) & & ({MeV}) &
\\ \hline\hline
$[\Lambda_c({1\over2}^-),0,1,\rho]$&--&$|\Lambda_c({1\over2}^-)\rangle$&$2.61^{+0.13}_{-0.13}$
&\multirow{1}{*}{--}&
$\begin{array}{l}
\Gamma\left(\Lambda_c^+({1/2}^-)\to\Sigma_c^+ \gamma\right)=690^{+490}_{-460}~\rm keV\\
\Gamma\left(\Lambda_c^+({1/2}^-)\to\Sigma_c^{*+} \gamma\right)=440^{+590}_{-380}~\rm keV\\
\end{array}$&$1.1^{+1.0}_{-0.8}$ &\multirow{1}{*}{--}
\\ \hline
$[\Lambda_c({1\over2}^-),1,1,\rho]$&--&$|\Lambda_c({1\over2}^-)\rangle$&$2.59^{+0.19}_{-0.15}$
&\multirow{4}{*}{$36^{+17}_{-18}$}&
$\begin{array}{l}
\Gamma^S\left(\Lambda_c({1/2}^-)\to\Sigma_c \pi\to\Lambda_c\pi\pi\right)=0.1^{+0.5}_{-0.1}~\rm MeV\\
\Gamma\left(\Lambda_c^+({1/2}^-)\to\Sigma_c^+ \gamma\right)=3.5^{+9.4}_{-3.5}~\rm keV\\
\Gamma\left(\Lambda_c^+({1/2}^-)\to\Lambda_c^+ \gamma\right)=8.3^{+9.5}_{-6.6}~\rm keV
\end{array}$&$0.1^{+0.5}_{-0.1}$ &\multirow{1}{*}{$\Lambda_c(2595)$}
\\ \cline{1-4} \cline{6-8}
$[\Lambda_c({3\over2}^-),1,1,\rho]$&\multirow{6}{*}{$\theta={37\pm5^\circ}$}
&$|\Lambda_c({3\over2}^-)\rangle_1$
&$2.63^{+0.48}_{-0.47}$&&
$\begin{array}{l}
\Gamma^S({\Lambda_c({3/2}^-)\to\Sigma^{*}_c\pi\to\Lambda_c\pi\pi})=0.20^{+0.35}_{-0.13}~{\rm MeV} \\
\Gamma^D({\Lambda_c({3/2}^-)\to\Sigma_c\pi})=9.3^{+6.7}_{-5.6}~{\rm keV}\\
\Gamma({\Lambda^{+}_c({3/2}^-)\to\Sigma^{+}_c\gamma})=160^{+430}_{-100}~{\rm keV} \\
\end{array}$&$0.36^{+0.78}_{-0.23}$& \multirow{1}{*}{$\Lambda_c(2625)$}
\\ \cline{1-1} \cline{3-8}
$[\Lambda_c({3\over2}^-),2,1,\rho]$&&$|\Lambda_c({3\over2}^-)\rangle_2$&$2.67^{+0.46}_{-0.51}$
&\multirow{4}{*}{$80^{+18}_{-18}$}&
$\begin{array}{l}
\Gamma^S({\Lambda_c({3/2}^-)\to\Sigma^{*}_c\pi\to\Lambda_c\pi\pi})=0.1^{+2.1}_{-0.1}~{\rm MeV} \\
\Gamma^D({\Lambda_c({3/2}^-)\to\Sigma^{*}_c\pi\to\Lambda_c\pi\pi})=0.0^{+6.7}_{-0.0}~{\rm MeV} \\ \Gamma^D({\Lambda_c({3/2}^-)\to\Sigma_c\pi})=0.0^{+2.5}_{-0.0}~{\rm MeV}\\
\Gamma({\Lambda^{+}_c({3/2}^-)\to\Sigma^{+}_c\gamma})=0.3^{+6.7}_{-0.3}~{\rm MeV} \\
\Gamma({\Lambda^+_c({3/2}^-)\to\Sigma^{*+}_c\gamma})=0.0^{+0.4}_{-0.0}~{\rm MeV} \\
\end{array}$&$0.4^{+18.4}_{-~0.4}$&--
\\ \cline{1-4} \cline{6-8}
$[\Lambda_c({5\over2}^-),2,1,\rho]$&--&$|\Lambda_c({5\over2}^-)\rangle$&$2.74^{+0.20}_{-0.16}$&&
$\begin{array}{l}
\Gamma^D\left(\Lambda_c({5/2}^-)\to\Sigma_c^{*} \pi\right)=7^{+19}_{-~7}~\rm MeV\\
\Gamma^D\left(\Lambda_c({5/2}^-)\to\Sigma_c \pi\right)=2^{+12}_{-~2}~\rm MeV\\
\Gamma\left(\Lambda_c^+({5/2}^-)\to\Sigma_c^{*+} \gamma\right)=0.7^{+2.0}_{-0.7}~\rm MeV\\
\end{array}$&$10^{+33}_{-10}$&--
\\ \hline
$[\Lambda_c({1\over2}^-),1,0,\lambda]$&--&$|\Lambda_c({1\over2}^-)\rangle$&$2.66^{+0.22}_{-0.19}$
&\multirow{4}{*}{$30^{+5}_{-5}$}&
$\begin{array}{l}
\Gamma^S\left(\Lambda_c({1/2}^-)\to\Sigma_c \pi\to\Lambda_c\pi\pi\right)=0.0^{+15.0}_{-~0.0}~\rm GeV\\
\Gamma\left(\Lambda_c^+({1/2}^-)\to\Sigma_c^+ \gamma\right)=2.8^{+2.4}_{-1.7}~\rm MeV\\
\Gamma\left(\Lambda_c^+({1/2}^-)\to\Sigma_c^{*+} \gamma\right)=0.6^{+0.9}_{-0.5}~\rm MeV\\
\end{array}$&$0\sim 10^4$ &\multirow{1}{*}{--}
\\ \cline{1-4} \cline{6-8}
$[\Lambda_c({3\over2}^-),1,0,\lambda]$&\multirow{1}{*}{--}
&$|\Lambda_c({3\over2}^-)\rangle$
&$2.69^{+0.23}_{-0.19}$&&
$\begin{array}{l}
\Gamma^S({\Lambda_c({3/2}^-)\to\Sigma^{*}_c\pi\to\Lambda_c\pi\pi})=0.0^{+4.8}_{-0.0}~{\rm GeV} \\
\Gamma^D({\Lambda_c({3/2}^-)\to\Sigma_c\pi})=10^{+40}_{-10}~{\rm MeV}\\
\Gamma({\Lambda^{+}_c({3/2}^-)\to\Sigma^{+}_c\gamma})=1.9^{+1.4}_{-1.0}~{\rm MeV} \\
\end{array}$&$0\sim10^3$& \multirow{1}{*}{--}
\\ \hline
$[\Xi_c({1\over2}^-),0,1,\rho]$&--&$|\Xi_c({1\over2}^-)\rangle$&$2.81^{+0.06}_{-0.06}$
&\multirow{1}{*}{--}&
$\begin{array}{l}
\Gamma\left(\Xi_c({1/2}^-)\to\Xi_c\pi\right)=5.7^{+3.6}_{-3.8}~\rm GeV\\
\Gamma\left(\Xi_c^+({1/2}^-)\to\Xi_c^+ \gamma\right)=2.1^{+2.8}_{-1.3}~\rm MeV\\
\Gamma\left(\Xi_c^+({1/2}^-)\to\Xi_c^{*+} \gamma\right)=2.1^{+3.2}_{-2.1}~\rm MeV\\
\end{array}$&$0\sim10^3$ &\multirow{1}{*}{--}
\\ \hline
$[\Xi_c({1\over2}^-),1,1,\rho]$&--&$|\Xi_c({1\over2}^-)\rangle$&$2.80^{+0.08}_{-0.08}$
&\multirow{5}{*}{$27^{+15}_{-15}$}&
$\begin{array}{l}
\Gamma^S({\Xi_c({1/2}^-)\to\Xi^{\prime}_c\pi})=4.5^{+34.5}_{-~4.5}~{\rm MeV} \\
\Gamma({\Xi^{+}_c({1/2}^-)\to\Xi^{+}_c\gamma})=18^{+25}_{-15}~{\rm keV} \\
\Gamma({\Xi^{+}_c({1/2}^-)\to\Xi^{*+}_c\gamma})=4.8^{+8.9}_{-4.8}~{\rm keV} \\
\end{array}$&$4.5^{+34.5}_{-~4.5}$&$\Xi_c(2790)$
\\ \cline{1-4}\cline{6-8}
$[\Xi_c({3\over2}^-),1,1,\rho]$&\multirow{6}{*}{$\theta={37\pm5^\circ}$}
&$|\Xi_c({3\over2}^-)\rangle_1$
&$2.83^{+0.22}_{-0.27}$&&
$\begin{array}{l}
\Gamma^S({\Xi_c({3/2}^-)\to\Xi^{*}_c\pi})=3.8^{+17.7}_{-~3.0}~{\rm MeV} \\
\Gamma({\Xi^{+}_c({3/2}^-)\to\Xi^{\prime+}_c\gamma})=0.4^{+1.8}_{-0.4}~{\rm MeV} \\
\Gamma({\Xi^{+}_c({3/2}^-)\to\Xi^{*+}_c\gamma})=23^{+180}_{-~23}~{\rm keV} \\
\end{array}$&$4.2^{+19.5}_{-~3.4}$
& $\Xi_c(2815) $
\\ \cline{1-1}\cline{3-8}
$[\Xi_c({3\over2}^-),2,1,\rho]$&&$|\Xi_c({3\over2}^-)\rangle_2$&$2.86^{+0.33}_{-0.27}$
&\multirow{4}{*}{$70^{+8}_{-8}$}&
$\begin{array}{l}
\Gamma^S({\Xi_c({3/2}^-)\to\Xi^{*}_c\pi})=0.0^{+0.5}_{-0.0}~{\rm MeV} \\
\Gamma^D({\Xi_c({3/2}^-)\to\Xi^{\prime}_c\pi})=0.6^{+3.3}_{-0.6}~{\rm MeV} \\
\Gamma({\Xi^{+}_c({3/2}^-)\to\Xi^{\prime+}_c\gamma})=0.7^{+4.4}_{-0.7}~{\rm MeV} \\
\Gamma({\Xi^{+}_c({3/2}^-)\to\Xi^{*+}_c\gamma})=19^{+220}_{-~19}~{\rm keV} \\
\end{array}$&$1.4^{+8.2}_{-1.4}$&--
\\ \cline{1-4} \cline{6-8}
$[\Xi_c({5\over2}^-),2,1,\rho]$&--&$|\Xi_c({5\over2}^-)\rangle$
&$2.93^{+0.10}_{-0.09}$&&
$\begin{array}{l}
\Gamma^D({\Xi_c({5/2}^-)\to\Xi_c\pi})=7^{+3}_{-5}~{\rm MeV} \\
\Gamma^D({\Xi_c({5/2}^-)\to\Xi^{*}_c\pi})=0.01^{+0.06}_{-0.01}~{\rm MeV} \\
\Gamma^D({\Xi_c({5/2}^-)\to\Xi^{\prime}_c\pi})=0.5^{+0.5}_{-0.3}~{\rm MeV} \\
\Gamma({\Xi_c^+({5/2}^-)\to\Xi^{*+}_c\pi})=250^{+120}_{-120}~{\rm keV} \\
\end{array}$&$8.0^{+3.5}_{-5.3}$
&--
\\ \hline
$[\Xi_c({1\over2}^-),1,0,\lambda]$&--&$|\Xi_c({1\over2}^-)\rangle$&$2.79^{+0.09}_{-0.10}$
&\multirow{4}{*}{$30^{+3}_{-3}$}&
$\begin{array}{l}
\Gamma^S({\Xi_c({1/2}^-)\to\Xi^{\prime}_c\pi})=0\sim16~{\rm GeV} \\
\Gamma({\Xi^{+}_c({1/2}^-)\to\Xi^{+}_c\gamma})=3.8^{+1.5}_{-1.9}~{\rm MeV} \\
\Gamma({\Xi^{+}_c({1/2}^-)\to\Xi^{*+}_c\gamma})=1.0^{+0.5}_{-0.5}~{\rm MeV} \\
\end{array}$&$0\sim10^4$&--
\\ \cline{1-4}\cline{6-8}
$[\Xi_c({3\over2}^-),1,0,\lambda]$&\multirow{1}{*}{--}
&$|\Xi_c({3\over2}^-)\rangle$
&$2.82^{+0.09}_{-0.10}$&&
$\begin{array}{l}
\Gamma^S({\Xi_c({3/2}^-)\to\Xi^{*}_c\pi})=1.1^{+1.1}_{-1.0}~{\rm GeV} \\
\Gamma({\Xi^{+}_c({3/2}^-)\to\Xi^{\prime+}_c\gamma})=2.8^{+10.0}_{-~1.5}~{\rm MeV} \\
\end{array}$&$0\sim10^3$
& --
\\ \hline \hline
\end{tabular}}
\label{tab:result2}
\end{center}
\end{table*}

$P$-wave charmed baryons belonging to the $SU(3)$ flavor $\mathbf{\bar{3}}_F$ representation have been systematically studied in Refs.~\cite{Luo:2025jpn,Chen:2015kpa,Chen:2017sci} using QCD sum rules and light-cone sum rules within the framework of heavy quark effective theory. The resulting predictions are comparable to those for the $P$-wave bottom baryons presented in Sec.~\ref{sec4.1}; therefore, in this subsection, we restrict our attention to the $[\mathbf{\bar{3}}_F, 1, 1, \rho]$ and $[\mathbf{\bar{3}}_F, 2, 1, \rho]$ doublets. However, it is necessary to further account for mixing effects, since heavy quark effective theory, being an effective theory, is more reliable for bottom baryons than for charmed baryons. As a result, the $J^P = 1/2^-$ charmed baryons can mix with one another, and similarly, the $J^P = 3/2^-$ charmed baryons can also undergo mixing. In particular, the mixing between the $[\mathbf{\bar{3}}_F, 1, 1, \rho]$ and $[\mathbf{\bar{3}}_F, 2, 1, \rho]$ doublets was explicitly investigated in Ref.~\cite{Luo:2025jpn}:
\begin{eqnarray}
&& \left(\begin{array}{c}
|\Lambda_c(3/2^-)\rangle_1\\
|\Lambda_c(3/2^-)\rangle_2
\end{array}\right) =
\\ \nonumber && ~~~~~~~~~~
\left(\begin{array}{cc}
\cos\theta & \sin\theta \\
-\sin\theta & \cos\theta
\end{array}\right)
~
\left(\begin{array}{c}
|\Lambda_c(3/2^-),1,1,\rho\rangle\\
|\Lambda_c(3/2^-),2,1,\rho\rangle
\end{array}\right) \, ,
\\ && 
\left(\begin{array}{c}
|\Xi_c(3/2^-)\rangle_1\\
|\Xi_c(3/2^-)\rangle_2
\end{array}\right) =
\\ \nonumber && ~~~~~~~~~~
\left(\begin{array}{cc}
\cos\theta & \sin\theta \\
-\sin\theta & \cos\theta
\end{array}\right)
~
\left(\begin{array}{c}
|\Xi_c(3/2^-),1,1,\rho\rangle\\
|\Xi_c(3/2^-),2,1,\rho\rangle
\end{array}\right) \, .
\end{eqnarray}
The mixing angle $\theta = 37 \pm 5^\circ$ is the same as that adopted in Sec.~\ref{sec4.4} for the mixing between the charmed baryon doublets $[\mathbf{6}_F, 1, 1, \lambda]$ and $[\mathbf{6}_F, 2, 1, \lambda]$, although these two mixing angles do not necessarily have to be the same in principle. This mixing angle has been fine-tuned to reduce the mass splitting within the $[\mathbf{\bar{3}}_F, 1, 1, \rho]$ doublet. The resulting predictions show improved agreement with the experimentally measured mass differences between $\Lambda_c(2595)^+$ and $\Lambda_c(2625)^+$, as well as between $\Xi_c(2790)^{0/+}$ and $\Xi_c(2815)^{0/+}$~\cite{PDG2024}.

The principal results of Refs.~\cite{Luo:2025jpn,Chen:2015kpa,Chen:2017sci} are summarized in Table~\ref{tab:result2}, and they suggest the following conclusions:
\begin{itemize}

\item The modified $[\mathbf{\bar{3}}_F, 1, 1, \rho]$ doublet comprises four excited charmed baryons: $\Lambda_c(1/2^-)$, $|\Lambda_c(3/2^-)\rangle_1$, $\Xi_c(1/2^-)$, and $|\Xi_c(3/2^-)\rangle_1$. This configuration provides a coherent theoretical framework that consistently accounts for the four experimentally observed states: $\Lambda_c(2595)^+$, $\Lambda_c(2625)^+$, $\Xi_c(2790)^{0/+}$, and $\Xi_c(2815)^{0/+}$.

\item The modified $[\mathbf{\bar{3}}_F, 2, 1, \rho]$ doublet also includes four excited charmed baryons: $|\Lambda_c(3/2^-)\rangle_2$, $\Lambda_c(5/2^-)$, $|\Xi_c(3/2^-)\rangle_2$, and $\Xi_c(5/2^-)$. This structure enables further predictions of two $\Lambda_c$ and two $\Xi_c$ states. As $\rho$-mode excitations, the experimental identification of these states would not only provide substantial support for the present theoretical model but also serve as compelling evidence for the existence of the $\rho$-mode in the charmed baryon spectrum.

\end{itemize}

One of the early studies based on full QCD sum rules was carried out in 2010, as reported in Ref.~\cite{Wang:2010fq}, where the authors derived the masses of the $\Lambda_c$ and $\Xi_c$ baryons with $J^P = 1/2^-$ to be $2.61 \pm 0.21$~GeV and $2.76 \pm 0.18$~GeV, respectively. These results provide an explanation for the $\Lambda_c(2595)^+$ and $\Xi_c(2790)^{0/+}$, respectively. This study does not discuss in detail the $\rho$-mode and $\lambda$-mode excitations.

Another full QCD sum rule calculation was conducted in Ref.~\cite{Zhang:2008pm}, where the authors performed a systematic study of charmed and bottom baryons. In particular, the masses of the $\Lambda_c$ baryons with $J^P = 1/2^-$ and $3/2^-$ were derived to be $2.53 \pm 0.22$~GeV and $2.58 \pm 0.24$~GeV, respectively, while the masses of the $\Xi_c$ baryons with $J^P = 1/2^-$ and $3/2^-$ were found to be $2.65 \pm 0.27$~GeV and $2.69 \pm 0.29$~GeV, respectively. These results can be used to interpret the $\Lambda_c(2595)^+$, $\Lambda_c(2625)^+$, $\Xi_c(2790)^{0/+}$, and $\Xi_c(2815)^{0/+}$. The interpolating currents employed in this study include $\rho$-mode orbital excitations, making their results largely consistent with ours.

\subsection{Singly bottom baryons of the $SU(3)$ flavor $\mathbf{6}_F$}
\label{sec4.3}

\begin{table*}[ht]
\begin{center}
\renewcommand{\arraystretch}{1.3}
\caption{Summary of the mass spectra and decay properties of the $P$-wave bottom baryons in the $[\mathbf{6}_F, 1, 1, \lambda]$ and $[\mathbf{6}_F, 2, 1, \lambda]$ doublets, obtained using QCD sum rules and light-cone sum rules within the framework of heavy quark effective theory. Possible experimental candidates are listed in the last column for comparison. Data are adapted from Ref.~\cite{Luo:2024jov}.}
\setlength{\tabcolsep}{0.1mm}{
\begin{tabular}{c|c|c|c|c|c| c |c c }
\hline\hline
 \multirow{2}{*}{B}&\multirow{2}{*}{Multiplet}&Baryon & Mass & ~~Splitting~~ &\multirow{2}{*}{ Partial Decay Width} &\multirow{2}{*}{Total width} & \multirow{2}{*}{Candidate}
\\& & ($j^P$) & ({GeV}) & ({MeV}) & & &
\\ \hline\hline
\multirow{17}{*}{$\Sigma_b$}&\multirow{8}{*}{$[\mathbf{ 6}_F, 1, 1, \lambda]$} &
~\multirow{4}{*}{$\Sigma_b({1\over2}^-)$}~    &\multirow{4}{*}{$6.06\pm 0.13$}      &\multirow{8}{*}{$6\pm 3$}
&~~~~$\Gamma(\Sigma_b({1\over2}^-)\to \Sigma_b\pi$)   = $14^{+21}_{-11}$~MeV~~~~
&\multirow{4}{*}{$14^{+21}_{-11}$~MeV}&\multirow{4}{*}{-}
\\
&&&&&$\Gamma(\Sigma_b({1\over2}^-)\to \Sigma_b^{*}\pi$)   = $76^{+144}_{-~76}$~keV           &
\\
&&&&&$\Gamma(\Sigma_b({1\over2}^-)\to \Lambda_b\rho\to\Lambda_b\pi\pi$)   = $87$~keV               &
\\
&&&&&$\Gamma(\Sigma_b^+({1\over2}^-)\to\Sigma_b^{*+}\gamma$)   =$20^{+57}_{-20}$~keV            &
\\ \cline{3-4} \cline{6-8}
&&\multirow{4}{*}{$\Sigma_b({3\over2}^-)$}&\multirow{4}{*}{$6.07\pm 0.07$} &
&$\Gamma(\Sigma_b({3\over2}^-)\to \Sigma_b^{*}\pi$)   = $4.0^{+5.8}_{-2.9}$~MeV        &\multirow{4}{*}{$4.8^{+5.9}_{-2.9}$~MeV}        &\multirow{4}{*}{-}
\\
&&&&&$\Gamma(\Sigma_b({3\over2}^-)\to\Sigma_b\pi$ ) = $550^{+740}_{-360}$~keV             &
\\
&&&&&$\Gamma(\Sigma_b({3\over2}^-)\to \Lambda_b\rho\to\Lambda_b\pi\pi$)   = $230$~keV                    &
\\
&&&&&$\Gamma(\Sigma_b^+({3\over2}^-)\to \Sigma_b^{*+}\gamma$)   = $21^{+53}_{-21}$~keV                &
\\ \cline{2-8}
&\multirow{8}{*}{$[\mathbf{ 6}_F, 2, 1, \lambda]$} &
\multirow{5}{*}{$\Sigma_b({3\over2}^-)$}     &\multirow{5}{*}{$6.11\pm 0.16$}      &\multirow{9}{*}{$12\pm 5$}
&~~~$\Gamma(\Sigma_b({3\over2}^-)\to \Lambda_b\pi$ )  = $49^{+76}_{-33}$~MeV
&\multirow{5}{*}{$51^{+76}_{-33}$~MeV} &
~\multirow{5}{*}{$\Sigma_b(6097)$}
\\
&&&&&~$\Gamma(\Sigma_b({3\over2}^-)\to \Sigma_b\pi$)   = $1.6^{+3.2}_{-1.1}$~MeV              &
\\
&&&&&~$\Gamma(\Sigma_b({3\over2}^-)\to \Sigma_b^*\pi$)   = $250^{+370}_{-160}$~keV              &
\\
&&&&&~$\Gamma(\Sigma_b({3\over2}^-)\to \Sigma_b\rho\to\Sigma_b\pi\pi$)   = $0.14$~keV~              &
\\
&&&&&$\Gamma(\Sigma_b^+({3\over2}^-)\to\Sigma_b^{+}\gamma$ )         =$120^{+1100}_{-~120}$~keV            &
\\ \cline{3-4} \cline{6-8}
&&\multirow{4}{*}{$\Sigma_b({5\over2}^-)$}&\multirow{4}{*}{$6.12\pm 0.15$} &
&$\Gamma(\Sigma_b({5\over2}^-)\to\Lambda_b\pi$) = $21^{+24}_{-14}$~MeV   &\multirow{4}{*}{$22^{+24}_{-14}$~MeV}        &\multirow{4}{*}{-}
\\
&&&&&$\Gamma(\Sigma_b({5\over2}^-)\to \Sigma_b^{*}\pi$)   = $1.1^{+1.8}_{-0.8}$~MeV                         &
\\
&&&&&$\Gamma(\Sigma_b({5\over2}^-)\to \Sigma_b^\pi$)   =  $360^{+710}_{-240}$~keV            &
\\
&&&&&$\Gamma(\Sigma_b^+({5\over2}^-)\to \Sigma_b^{*+}\gamma$)   = $58^{+180}_{-~58}$~keV                &
\\ \hline \hline
\multirow{21}{*}{~~$\Xi_b^\prime$~~}&\multirow{9}{*}{$[\mathbf{ 6}_F, 1, 1, \lambda]$}
&\multirow{4}{*}{$\Xi_b^{\prime}({1\over2}^-)$} &\multirow{4}{*}{$6.21\pm 0.11$}      &\multirow{9}{*}{$7\pm 2$}
&$\Gamma(\Xi_b^{\prime}({1\over2}^-)\to \Xi_b^{\prime}\pi$)  =$4.5^{+5.8}_{-3.3}$~MeV                    &\multirow{4}{*}{$4.7^{+6.0}_{-3.4}$~MeV } &\multirow{4}{*}{-}
\\
&&&&&$\Gamma(\Xi_b^{\prime}({1\over2}^-)\to\Xi_b^{{\*}}\pi$) =$160^{+180}_{-100}$~keV            &
\\
&&&&&$\Gamma(\Xi_b^{\prime}({1\over2}^-)\to\Xi_b\rho\to\Xi_b\pi\pi$) =$43$~keV            &
\\
&&&&&$\Gamma(\Xi_b^{\prime-}({1\over2}^-)\to\Xi_b^{-}\gamma$) =$130^{+260}_{-110}$~keV         &
\\ \cline{3-4}\cline{6-8}
&&\multirow{5}{*}{$\Xi_b^{\prime}({3\over2}^-)$}&\multirow{5}{*}{$6.22\pm 0.11$}
&&$\Gamma(\Xi_b({3\over2}^-)\to\Xi_b^{*}\pi$)  =$1.4^{+1.0}_{-0.9}$~MeV     &\multirow{5}{*}{$1.8^{+1.4}_{-1.1}$~MeV} &\multirow{4}{*}{-}
\\
&&&&&$\Gamma(\Xi_b^{\prime}({3\over2}^-)\to \Xi_b^{\prime}\pi$) =$340^{+350}_{-200}$~keV              &
\\
&&&&&$\Gamma(\Xi_b^{\prime}({3\over2}^-)\to\Xi_b\rho\to\Xi_b\pi\pi$) =$78$~keV            &
\\
&&&&&$\Gamma(\Xi_b^{\prime}({3\over2}^-)\to\Xi_b^\prime\rho\to\Xi_b^\prime\pi\pi$) =$0.006$~keV            &
\\
&&&&&$\Gamma(\Xi_b^{\prime-}({3\over2}^-)\to \Xi_b^{-}\gamma$)  =$51^{+93}_{-51}$~keV                           &
\\ \cline{2-8}
&\multirow{12}{*}{$[\mathbf{ 6}_F, 2, 1, \lambda]$}
&\multirow{6}{*}{$\Xi_b^{\prime}({3\over2}^-)$}  &\multirow{6}{*}{$6.23\pm 0.15$}      &\multirow{12}{*}{$11\pm 5$}
&$\Gamma(\Xi_b^{\prime}({3\over2}^-)\to \Xi_b\pi$)=$19^{+26}_{-13}$~MeV                    &\multirow{6}{*}{$27^{+29}_{-14}$~MeV} &\multirow{6}{*}{~$\Xi_b(6227)$}
\\
&&&&&$\Gamma(\Xi_b^{\prime}({3\over2}^-)\to \Lambda_bK$)  =$7.4^{+11.0}_{-~4.8}$~MeV             &
\\
&&&&&$\Gamma(\Xi_b^{\prime}({3\over2}^-)\to \Xi_b^\prime\pi$ ) =$790^{+1100}_{-~790}$~keV    &
\\
&&&&&$\Gamma(\Xi_b^{\prime}({3\over2}^-)\to \Xi_b^*\pi$)  =$130^{+170}_{-~80}$~keV    &
\\
&&&&&$\Gamma(\Xi_b^{\prime}({3\over2}^-)\to\Xi_b^\prime\rho\to\Xi_b^\prime\pi\pi$) =$0.56$~keV            &
\\
&&&&&$\Gamma(\Xi_b^{\prime-}({3\over2}^-)\to\Xi_b^{{\prime}-}\gamma$) =$46^{+250}_{-~46}$~keV         &
\\ \cline{3-4}\cline{6-8}
&&\multirow{6}{*}{$\Xi_b^{\prime}({5\over2}^-)$}  &\multirow{6}{*}{$6.24\pm 0.14$}
&&$\Gamma(\Xi_b({5\over2}^-)\to\Xi_b\pi$)  =$8.1^{+11.2}_{-~5.7}$~MeV     &\multirow{6}{*}{~$12.3^{+12.3}_{-~6.1}$~MeV~}   &\multirow{4}{*}{-}
\\
&&&&&$\Gamma(\Xi_b({5\over2}^-)\to\Lambda_b K$)  =$3.4^{+5.1}_{-2.2}$~MeV              &
\\
&&&&&$\Gamma(\Xi_b({5\over2}^-)\to\Xi_b^\prime\pi$)  =$170^{+240}_{-110}$~keV              &
\\
&&&&&$\Gamma(\Xi_b({5\over2}^-)\to\Xi_b^*\pi$)  =$580^{+800}_{-380}$~keV              &
\\
&&&&&$\Gamma(\Xi_b^{\prime}({5\over2}^-)\to\Xi_b^*\rho\to\Xi_b^*\pi\pi$) =$0.06$~keV            &
\\
&&&&&$\Gamma(\Xi_b^{\prime-}({5\over2}^-)\to \Xi_b^{*-}\gamma$)  =$26^{+92}_{-26}$~keV                           &
\\ \hline\hline
\multirow{8}{*}{$\Omega_b$~}&\multirow{2}{*}{~$[\mathbf{6}_F, 1, 0, \rho]$~}
&\multirow{1}{*}{$\Omega_b({1\over2}^-)$}&\multirow{1}{*}{~$6.32\pm 0.11$~} &\multirow{2}{*}{~~$2\pm 1$~~}&$\Gamma(\Omega_b^-({1\over2}^-)\to\Omega_b^-\gamma$)=$ 60^{+130}_{-~55}$~keV
&\multirow{1}{*}{~$ 60^{+130}_{-~55}$~keV}&~\multirow{2}{*}{$\Omega_b(6316)$}
\\ \cline{3-4}\cline{6-7}
&&\multirow{1}{*}{$\Omega_b({3\over2}^-)$}     &\multirow{1}{*}{$6.32\pm 0.11$}      &
&$\Gamma(\Omega_b^-({3\over2}^-)\to\Omega_b^-\gamma$)        = $9.1^{+20.6}_{-~5.0}$~keV                      &\multirow{1}{*}{$ 9.1^{+20.6}_{-~5.0}$~keV } &
\\\cline{2-8}
&\multirow{2}{*}{$[\mathbf{ 6}_F, 1, 1, \lambda]$}
&\multirow{1}{*}{$\Omega_b({1\over2}^-)$}     &\multirow{1}{*}{$6.34\pm 0.11$}      &\multirow{2}{*}{$6\pm 2$}
& $\Gamma(\Omega_b^-({1\over2}^-)\to\Omega_b^-\gamma$)=$12^{+21}_{-12}$~keV                              &\multirow{1}{*}{$ 12^{+21}_{-12}$~keV  }&~\multirow{1}{*}{$\Omega_b(6330)$}
\\  \cline{3-4} \cline{6-8}
&&\multirow{1}{*}{$\Omega_b({3\over2}^-)$}&\multirow{1}{*}{$6.34\pm 0.09$} &
&$\Gamma(\Omega_b^-({3\over2}^-)\to\Omega_b^-\gamma$)  =$6.9^{+10.0}_{-~6.4}$~keV                         &\multirow{1}{*}{$6.9^{+10.0}_{-~6.4}$~keV} &~\multirow{1}{*}{$\Omega_b(6340)$}
\\\cline{2-8}
&\multirow{4}{*}{$[\mathbf{ 6}_F, 2, 1, \lambda]$}
&\multirow{2}{*}{$\Omega_b({3\over2}^-)$}     &\multirow{2}{*}{$6.35\pm 0.13$}      &\multirow{4}{*}{$10\pm 4$}
&$\Gamma(\Omega_b({3\over2}^-)\to\Xi_b K$)  =$ 4.6^{+3.3}_{-1.9}$~MeV       &\multirow{2}{*}{$ 4.6^{+3.3}_{-1.9}$~MeV}&~\multirow{2}{*}{$\Omega_b(6350)$}
\\
&&&&&$\Gamma(\Omega_b^-({3\over2}^-)\to\Omega_b^-\gamma$)=$17^{+52}_{-17}$~keV                          &  &
\\  \cline{3-4} \cline{6-8}
&&\multirow{2}{*}{$\Omega_b({5\over2}^-)$}&\multirow{2}{*}{$6.36\pm 0.12$} &
&$\Gamma(\Omega_b({5\over2}^-)\to\Xi_b K$ )      =$ 2.5^{+3.5}_{-1.6}$~MeV                      &\multirow{2}{*}{$ 2.5^{+3.5}_{-1.6}$~MeV} &\multirow{2}{*}{-}
\\
&&&&&$\Gamma(\Omega_b^-({5\over2}^-)\to\Omega_b^{*-}\gamma$)  =$0.83^{+3.45}_{-0.83}$~keV         &   &
\\ \hline \hline
\end{tabular}}
\label{tab:result3}
\end{center}
\end{table*}

$P$-wave bottom baryons belonging to the $SU(3)$ flavor $\mathbf{6}_F$ representation have been systematically investigated in Refs.~\cite{Yang:2020zrh,Mao:2015gya,Yang:2019cvw,Luo:2024jov} using QCD sum rules and light-cone sum rules within the framework of heavy quark effective theory. As illustrated in Fig.~\ref{fig:categorization}, a total of four baryon multiplets are identified: $[\mathbf{6}_F, 1, 0, \rho]$, $[\mathbf{6}_F, 0, 1, \lambda]$, $[\mathbf{6}_F, 1, 1, \lambda]$, and $[\mathbf{6}_F, 2, 1, \lambda]$. The obtained results suggest that:
\begin{itemize}

\item The $[\mathbf{6}_F, 1, 0, \rho]$ doublet contains six bottom baryons: $\Sigma_b(1/2^-)$, $\Sigma_b(3/2^-)$, $\Xi'_b(1/2^-)$, $\Xi'_b(3/2^-)$, $\Omega_b(1/2^-)$, and $\Omega_b(3/2^-)$. The total decay widths of $\Sigma_b(1/2^-)$, $\Sigma_b(3/2^-)$, $\Xi'_b(1/2^-)$, and $\Xi'_b(3/2^-)$ are all predicted to be very large, whereas those of $\Omega_b(1/2^-)$ and $\Omega_b(3/2^-)$ are predicted to be vanishingly small.

\item The $[\mathbf{6}_F, 0, 1, \lambda]$ singlet contains three bottom baryons: $\Sigma_b(1/2^-)$, $\Xi'_b(1/2^-)$, and $\Omega_b(1/2^-)$. Their total decay widths are predicted to be large, rendering them difficult to observe experimentally.

\item The $[\mathbf{6}_F, 1, 1, \lambda]$ doublet contains six bottom baryons: $\Sigma_b(1/2^-)$, $\Sigma_b(3/2^-)$, $\Xi'_b(1/2^-)$, $\Xi'_b(3/2^-)$, $\Omega_b(1/2^-)$, and $\Omega_b(3/2^-)$. The total decay widths of all six states are predicted to be less than 100~MeV.

\item The $[\mathbf{6}_F, 2, 1, \lambda]$ doublet also contains six bottom baryons: $\Sigma_b(3/2^-)$, $\Sigma_b(5/2^-)$, $\Xi'_b(3/2^-)$, $\Xi'_b(5/2^-)$, $\Omega_b(3/2^-)$, and $\Omega_b(5/2^-)$. Similarly, the total decay widths of all members in this doublet are estimated to be below 100~MeV.

\end{itemize}
Summarizing the above results, we identify a total of four $\Sigma_b$, four $\Xi'_b$, and six $\Omega_b$ baryons with relatively narrow widths (less than 100~MeV), making them promising candidates for experimental observation. Their masses, mass splittings within the same multiplets, and decay properties are summarized in Table~\ref{tab:result2}. Possible experimental counterparts are also provided in the last column for comparison. 

Note that within our QCD sum rule approach, the $\rho$-mode doublet $[\mathbf{6}_F, 1, 0, \rho]$ is predicted to lie below the two $\lambda$-mode doublets, $[\mathbf{6}_F, 1, 1, \lambda]$ and $[\mathbf{6}_F, 2, 1, \lambda]$, in mass. This ordering differs from the expectations of conventional quark models~\cite{Copley:1979wj,Yoshida:2015tia}. Such difference would be tested by future experimental data. To test the existence of the $\rho$-mode doublet $[\mathbf{6}_F, 1, 0, \rho]$, we propose the following experimental investigations: first, determine the spin-parity quantum number of the $\Omega_b(6316)^-$; second, examine whether it can be resolved into two nearly degenerate states; and third, search for its corresponding $\Sigma_b$ and $\Xi'_b$ partner states.

A full QCD sum rule analysis was presented in Ref.~\cite{XuYongJiang:2020cht}, where all $P$-wave $\Omega_b$ states represented by derivative interpolating currents were studied. The corresponding masses and pole residues were calculated, and the results are consistent with those in Ref.~\cite{Chen:2020mpy}. Specifically, the $\Omega_b(6350)^-$ is suggested to be a $P$-wave $\Omega_b$ baryon with $J^P = 3/2^-$ and a $\lambda$-mode excitation; the $\Omega_b(6330)^-$ and $\Omega_b(6340)^-$ are identified as partner states with $\lambda$-mode excitations, having $J^P = 1/2^-$ and $3/2^-$, respectively; while the $\Omega_b(6316)^-$ is likely a $P$-wave $\Omega_b$ baryon with either $J^P = 1/2^-$ or $3/2^-$, associated with a $\rho$-mode excitation.

Another QCD sum rule analysis was conducted in Ref.~\cite{Agaev:2017ywp}, where both orbitally and radially excited $\Omega_b$ baryons were investigated. The masses of the $J^P = 1/2^-$ and $3/2^-$ states were calculated to be $6336 \pm 183$~MeV and $6301 \pm 193$~MeV, respectively. These results may account for two of the four excited $\Omega_b$ baryons observed by LHCb~\cite{LHCb:2020tqd}. In a subsequent study~\cite{Aliev:2018lcs}, the same group examined the excited $\Xi_b^\prime$ baryons and suggested that the $\Xi_b(6227)^-$ can be interpreted as the $P$-wave excitation with $J^P = 3/2^-$. Neither of these two studies discusses the $\rho$-mode or $\lambda$-mode excitations in detail.

\subsection{Singly charmed baryons of the $SU(3)$ flavor $\mathbf{6}_F$}
\label{sec4.4}

\begin{table*}[hbpt]
\begin{center}
\renewcommand{\arraystretch}{1.25}
\caption{Summary of the mass spectra and decay properties of the $P$-wave charmed baryons in the $[\mathbf{6}_F, 1, 1, \lambda]$ and $[\mathbf{6}_F, 2, 1, \lambda]$ doublets. Possible experimental candidates are listed in the last column for comparison. Data are adapted from Refs.~\cite{Yang:2021lce,Luo:2025pzb}.}
\scalebox{0.925}{\begin{tabular}{ c |c | c | c | c | c | c | c}
\hline\hline
\multirow{2}{*}{HQET state}&\multirow{2}{*}{Mixing}&\multirow{2}{*}{Mixed state} & Mass & Difference & ~~~~~~~~~~\multirow{2}{*}{Decay channel}~~~~~~~~~~ & Width  & \multirow{2}{*}{Candidate}
\\  &&&   ({GeV}) & ({MeV}) & & ({MeV}) &
\\ \hline\hline
$[\Sigma_c({1\over2}^-),1,1,\lambda]$& -- &$[\Sigma_c({1\over2}^-),1,1,\lambda]$&$2.73^{+0.17}_{-0.18}$&\multirow{4}{*}{$23^{+19}_{-43}$}&
$\begin{array}{l}
\Gamma_S\left(\Sigma_c({1/2}^-)\to \Lambda_c\rho\to\Lambda_c\pi\pi\right)=9.2^{+37.0}_{-~9.2}~{\rm MeV}\\
\Gamma_S\left(\Sigma_c({1/2}^-)\to \Sigma_c\rho\to\Sigma_c\pi\pi\right)=1.2^{+2.1}_{-1.0}~{\rm MeV}\\
\Gamma_S\left(\Sigma_c({1/2}^-)\to\Sigma_c\pi\right)=37^{+60}_{-28}~{\rm MeV}\\
\Gamma\left(\Sigma_c^0({1/2}^-)\to\Sigma_c^0 \gamma\right)=11^{+30}_{-11}~{\rm keV}\\
\Gamma\left(\Sigma_c^0({1/2}^-)\to\Sigma_c^{*0} \gamma\right)=3.5^{+15.0}_{-~3.5}~{\rm keV}
\end{array}$&$48^{+70}_{-29}$ &-
\\ \cline{1-4} \cline{6-8}
$[\Sigma_c({3\over2}^-),1,1,\lambda]$&\multirow{7}{*}{$\theta={37\pm5^\circ}$}&$|\Sigma_c({3\over2}^-)\rangle_1$&$2.75^{+0.17}_{-0.17}$&&
$\begin{array}{l}
\Gamma_D\left(\Sigma_c({3/2}^-)\to\Lambda_c\pi\right)=13^{+20}_{-~9}~\rm MeV\\
\Gamma_D\left(\Sigma_c({3/2}^-)\to\Sigma_c\pi\right)=3.3^{+4.2}_{-2.2}~\rm MeV\\
\Gamma_S\left(\Sigma_c({3/2}^-)\to\Sigma_c^{*}\pi\right)
=6.4^{+10.3}_{-~4.7}~\rm MeV\\
\Gamma({\Sigma^0_c({3/2}^-)\to\Sigma^0_c\gamma})=53^{+130}_{-49}~{\rm keV} \\
\Gamma({\Sigma^0_c({3/2}^-)\to\Sigma^{*0}_c\gamma})=13^{+46}_{-13}~{\rm keV} \\
\end{array}$&$24^{+23}_{-10}$& \multirow{7}{*}{$\Sigma_c(2800)$}
\\ \cline{1-1} \cline{3-7}
$[\Sigma_c({3\over2}^-),2,1,\lambda]$&&$|\Sigma_c({3\over2}^-)\rangle_2$&$2.80^{+0.14}_{-0.12}$&\multirow{4}{*}{$68^{+51}_{-51}$}&
$\begin{array}{l}
\Gamma_D\left(\Sigma_c({3/2}^-)\to\Lambda_c\pi\right)=23^{+35}_{-16}~\rm MeV\\
\Gamma_S\left(\Sigma_c({3/2}^-)\to\Sigma_c^{*}\pi\right)
=3.5^{+6.1}_{-2.7}~\rm MeV\\
\Gamma({\Sigma^0_c({3/2}^-)\to\Sigma^0_c\gamma})=54^{+76}_{-41}~{\rm keV} \\ \Gamma({\Sigma^0_c({3/2}^-)\to\Sigma^{*0}_c\gamma})=7.9^{+15.3}_{-~6.7}~{\rm keV} \\
\end{array}$&$28^{+36}_{-16}$&
\\ \cline{1-4} \cline{6-8}
$[\Sigma_c({5\over2}^-),2,1,\lambda]$&--&$[\Sigma_c({5\over2}^-),2,1,\lambda]$&$2.87^{+0.12}_{-0.11}$&&
$\begin{array}{l}
\Gamma_D\left(\Sigma_c({5/2}^-)\to\Lambda_c\pi\right)=12^{+18}_{-~8}~\rm MeV\\
\Gamma_D\left(\Sigma_c({5/2}^-)\to \Sigma_c\pi\right)=0.39^{+0.72}_{-0.32}~\rm MeV\\
\Gamma_D\left(\Sigma_c({5/2}^-)\to\Sigma_c^{*}\pi\right)
=0.61^{+1.14}_{-0.50}~\rm MeV\\
\Gamma\left(\Sigma_c^0({1/2}^-)\to\Sigma_c^{*0} \gamma\right)=8.5^{+9.2}_{-6.8}~\rm keV\\
\end{array}$&$13^{+18}_{-~8}$&-
\\ \hline
$[\Xi_c^\prime({1\over2}^-),1,1,\lambda]$& -- &$[\Xi_c^\prime({1\over2}^-),1,1,\lambda]$&$2.91^{+0.13}_{-0.12}$&\multirow{5}{*}{$27^{+16}_{-27}$}&
$\begin{array}{l}
\Gamma_S\left(\Xi_c^{\prime}({1/2}^-)\to\Xi_c\rho\to\Xi_c\pi\pi\right)
=1.7^{+7.6}_{-1.7}~\rm MeV\\
\Gamma({\Xi^{0}_c({3/2}^-)\to\Xi^{\prime0}_c\gamma})=35^{+61}_{-30}~{\rm keV} \\
\Gamma_S\left(\Xi_c^{\prime}({1/2}^-)\to \Xi_c^{\prime}\pi\right)=12^{+15}_{-~8}~\rm MeV\\
\Gamma({\Xi^{0}_c({3/2}^-)\to\Xi^{*0}_c\gamma})=13^{+27}_{-13}~{\rm keV} \\
\end{array}$&$14^{+17}_{-~8}$&$\Xi_c(2923)$
\\ \cline{1-4}\cline{6-8}
$[\Xi_c^{\prime}({3\over2}^-),1,1,\lambda]$&\multirow{8}{*}{$\theta={37\pm5^\circ}$}&$|\Xi_c^\prime({3\over2}^-)\rangle_1$&$2.94^{+0.12}_{-0.11}$&&
$\begin{array}{l}
\Gamma_D\left(\Xi_c^{\prime}({3/2}^-)\to \Lambda_c \bar K\right)=2.3^{+4.3}_{-1.7}~\rm MeV\\
\Gamma_D\left(\Xi_c^{\prime}({3/2}^-)\to \Xi_c\pi\right)=4.6^{+8.1}_{-3.3}~\rm MeV\\
\Gamma_D\left(\Xi_c^{\prime}({3/2}^-)\to\Xi_ c^{\prime}\pi\right)=2.0^{+2.2}_{-1.2}~\rm MeV\\
\Gamma_S\left(\Xi_c^{\prime}({3/2}^-)\to \Xi_c^{*}\pi\right)=2.1^{+2.6}_{-1.5}~\rm MeV\\
\Gamma({\Xi^+_c({3/2}^-)\to\Xi^{\prime+}_c\gamma})=52^{+58}_{-33}~{\rm keV} \\
\Gamma({\Xi^+_c({3/2}^-)\to\Xi^{*+}_c\gamma})=19^{+30}_{-15}~{\rm keV} \\
\Gamma({\Xi^{0}_c({3/2}^-)\to\Xi^{\prime0}_c\gamma})=130^{+150}_{-86}~{\rm keV} \\
\Gamma({\Xi^{0}_c({3/2}^-)\to\Xi^{*0}_c\gamma})=41^{+69}_{-30}~{\rm keV} \\
\end{array}$&$12^{+10}_{-~4}$
& $\Xi_c(2939)$
\\ \cline{1-1}\cline{3-8}
$[\Xi_c^\prime({3\over2}^-),2,1,\lambda]$&&$|\Xi_c^\prime({3\over2}^-)\rangle_2$&$2.97^{+0.24}_{-0.15}$&\multirow{4}{*}{$56^{+30}_{-35}$}&
$\begin{array}{l}
\Gamma_D\left(\Xi_c^{\prime}({3/2}^-)\to \Lambda_c \bar K\right)=6.3^{+11.6}_{-~4.7}~\rm MeV\\
\Gamma_D\left(\Xi_c^{\prime}({3/2}^-)\to \Xi_c\pi\right)=11^{+19}_{-~8}~\rm MeV\\
\Gamma_S\left(\Xi_c^{\prime}({3/2}^-)\to \Xi_c^{*}\pi\right)= 1.3^{+1.80}_{-0.94}~\rm MeV\\
\Gamma({\Xi^+_c({3/2}^-)\to\Xi^{\prime+}_c\gamma})=22^{+54}_{-17}~{\rm keV} \\
\Gamma({\Xi^+_c({3/2}^-)\to\Xi^{*+}_c\gamma})=4.6^{+15.5}_{-~4.2}~{\rm keV} \\
\Gamma({\Xi^{0}_c({3/2}^-)\to\Xi^{\prime0}_c\gamma})=63^{+160}_{-~50}~{\rm keV} \\
\Gamma({\Xi^{0}_c({3/2}^-)\to\Xi^{*0}_c\gamma})=12^{+44}_{-12}~{\rm keV} \\
\end{array}$&$19^{+22}_{-~9}$&$\Xi_c(2965)$
\\ \cline{1-4} \cline{6-8}
$[\Xi^\prime_c({5\over2}^-),2,1,\lambda]$&--&$[\Xi^\prime_c({5\over2}^-),2,1,\lambda]$&$3.02^{+0.23}_{-0.14}$&&
$\begin{array}{l}
\Gamma_D\left(\Xi_c^{\prime}({5/2}^-)\to \Lambda_c \bar K\right)=6.3^{+11.4}_{-~4.6}~\rm MeV\\
\Gamma_D\left(\Xi_c^{\prime}({5/2}^-)\to \Xi_c \pi\right)=9.6^{+15.8}_{-~6.8}~\rm MeV\\
\Gamma_D\left(\Xi_c^{\prime}({5/2}^-)\to \Xi_c^{*} \pi\right)=1.5^{+2.6}_{-1.1}~\rm MeV\\
\Gamma({\Xi^{0}_c({3/2}^-)\to\Xi^{*0}_c\gamma})=44^{+100}_{-~44}~{\rm keV}
\end{array}$&$18^{+20}_{-~8}$
&--
\\ \hline
$[\Omega_c({1\over2}^-),1,0,\rho]$& -- &$[\Omega_c({1\over2}^-),1,0,\rho]$&$2.99^{+0.15}_{-0.15}$& \multirow{2}{*}{$12^{+5}_{-5}$} & $\Gamma\left(\Omega_c^0({1/2}^-)\to\Omega_c^0 \gamma\right) =76^{+170}_{-~76}~\rm keV$ &$\neq 0$&\multirow{2}{*}{$\Omega_c(3000)$}
\\ \cline{1-4} \cline{6-7}
$[\Omega_c({3\over2}^-),1,0,\rho]$& -- &$[\Omega_c({3\over2}^-),1,0,\rho]$&$3.00^{+0.15}_{-0.15}$&& $\Gamma\left(\Omega_c^0({3/2}^-)\to\Omega_c^0 \gamma\right) 20^{+43}_{-20}~\rm keV$ &$\neq 0$&
\\ \hline
$[\Omega_c({1\over2}^-),1,1,\lambda]$& -- &$[\Omega_c({1\over2}^-),1,1,\lambda]$&$3.04^{+0.11}_{-0.09}$& \multirow{2}{*}{$27^{+15}_{-23}$} & $\begin{array}{l}
\Gamma\left(\Omega_c^0({1/2}^-)\to\Omega_c^0 \gamma\right)=48^{+69}_{-41}~\rm MeV\\
\Gamma\left(\Omega_c^0({1/2}^-)\to\Omega_c^{*0} \gamma\right)=17^{+29}_{-15}~\rm keV
\end{array}$ &$\neq 0$&$\Omega_c(3050)$
\\ \cline{1-4} \cline{6-8}
$[\Omega_c({3\over2}^-),1,1,\lambda]$&\multirow{4}{*}{$\theta =37\pm5^\circ$}&$|\Omega_c({3\over2}^-)\rangle_1$&$3.06^{+0.10}_{-0.09}$&  &
$\begin{array}{l}
\Gamma_D\left(\Omega_c({3/2}^-)\to \Xi_c \bar K\right)=2.0^{+3.5}_{-1.5}~\rm MeV\\
\Gamma({\Omega^{0}_c({3/2}^-)\to\Omega^{0}_c\gamma})=160^{+140}_{-~86}~{\rm keV} \\
\Gamma({\Omega^{0}_c({3/2}^-)\to\Omega^{*0}_c\gamma})=49^{+62}_{-31}~{\rm keV} \\
\end{array}$&$2.0^{+3.5}_{-1.5}$&$\Omega_c(3066)$
\\ \cline{1-1} \cline{3-8}
$[\Omega_c({3\over2}^-),2,1,\lambda]$&&$|\Omega_c({3\over2}^-)\rangle_2$&$3.09^{+0.22}_{-0.17}$&\multirow{3}{*}{$51^{+26}_{-29}$}&
$\begin{array}{l}
\Gamma_D\left(\Omega_c({3/2}^-)\to \Xi_c \bar K\right)=6.3^{+11.2}_{-~4.8}~\rm MeV\\
\Gamma({\Omega^{0}_c({3/2}^-)\to\Omega^{0}_c\gamma})=18^{+36}_{-16}~{\rm keV} \\
\Gamma({\Omega^{0}_c({3/2}^-)\to\Omega^{*0}_c\gamma})=3.7^{+9.8}_{-3.5}~{\rm keV} \\
\end{array}$&$6.4^{+11.2}_{-~4.8}$&$\Omega_c(3090)$
\\ \cline{1-4} \cline{6-8}
$[\Omega_c({5\over2}^-),2,1,\lambda]$&--&$[\Omega_c({5\over2}^-),2,1,\lambda]$&$3.14^{+0.21}_{-0.15}$&&
$\begin{array}{l}
\Gamma_D\left(\Omega_c({5/2}^-)\to \Xi_c \bar K\right)=5.5^{+9.6}_{-4.1}~\rm MeV\\
\Gamma\left(\Omega_c^0({5/2}^-)\to \Omega_c^{*0} \gamma\right)=5.2^{+14.6}_{-~5.2}~\rm keV
\end{array}$&$5.5^{+9.6}_{-4.1}$&$\Omega_c(3119)$
\\ \hline\hline
\end{tabular}}
\label{tab:result4}
\end{center}
\end{table*}

$P$-wave charmed baryons belonging to the $SU(3)$ flavor $\mathbf{6}_F$ representation have been systematically investigated in Refs.~\cite{Yang:2021lce,Chen:2015kpa,Chen:2017sci,Yang:2020zjl} using QCD sum rules and light-cone sum rules within the framework of heavy quark effective theory. The resulting predictions are comparable to those obtained for the $P$-wave bottom baryons discussed in Sec.~\ref{sec4.3}. Nevertheless, it is essential to explicitly take into account the mixing effects between the $[\mathbf{6}_F, 1, 1, \lambda]$ and $[\mathbf{6}_F, 2, 1, \lambda]$ doublets, as was similarly done in Sec.~\ref{sec4.2}:
\begin{eqnarray}
&& \left(\begin{array}{c}
|\Sigma_c(3/2^-)\rangle_1\\
|\Sigma_c(3/2^-)\rangle_2
\end{array}\right) =
\\ \nonumber && ~~~~~~~~~~
\left(\begin{array}{cc}
\cos\theta & \sin\theta \\
-\sin\theta & \cos\theta
\end{array}\right)
\left(\begin{array}{c}
|\Sigma_c(3/2^-),1,1,\lambda\rangle\\
|\Sigma_c(3/2^-),2,1,\lambda\rangle
\end{array}\right) \, ,
\\ &&
\left(\begin{array}{c}
|\Xi_c^\prime(3/2^-)\rangle_1\\
|\Xi_c^\prime(3/2^-)\rangle_2
\end{array}\right) =
\\ \nonumber && ~~~~~~~~~~
\left(\begin{array}{cc}
\cos\theta & \sin\theta \\
-\sin\theta & \cos\theta
\end{array}\right)
\left(\begin{array}{c}
|\Xi_c^\prime(3/2^-),1,1,\lambda\rangle\\
|\Xi_c^\prime(3/2^-),2,1,\lambda\rangle
\end{array}\right) \, ,
\\ &&
\left(\begin{array}{c}
|\Omega_c(3/2^-)\rangle_1\\
|\Omega_c(3/2^-)\rangle_2
\end{array}\right) =
\\ \nonumber && ~~~~~~~~~~
\left(\begin{array}{cc}
\cos\theta & \sin\theta \\
-\sin\theta & \cos\theta
\end{array}\right)
\left(\begin{array}{c}
|\Omega_c(3/2^-),1,1,\lambda\rangle\\
|\Omega_c(3/2^-),2,1,\lambda\rangle
\end{array}\right) \, .
\end{eqnarray}
The mixing angle $\theta = 37 \pm 5^\circ$ is the same as that used in Sec.~\ref{sec4.2} for the mixing between the charmed baryon doublets $[\mathbf{\bar{3}}_F, 1, 1, \rho]$ and $[\mathbf{\bar{3}}_F, 2, 1, \rho]$, even though, in principle, the two mixing angles need not be identical. Notably, this mixing angle has been fine-tuned to significantly affect the decay widths of the two $\Xi_c^\prime(3/2^-)$ states and to reduce the mass splitting within the $[\mathbf{6}_F, 1, 1, \lambda]$ doublet, thereby improving agreement with the experimental results reported by LHCb~\cite{LHCb:2020iby}.

As illustrated in Fig.~\ref{fig:categorization}, four baryon multiplets are identified: $[\mathbf{6}_F, 1, 0, \rho]$, $[\mathbf{6}_F, 0, 1, \lambda]$, $[\mathbf{6}_F, 1, 1, \lambda]$, and $[\mathbf{6}_F, 2, 1, \lambda]$. The results from Refs.~\cite{Yang:2021lce,Chen:2015kpa,Chen:2017sci,Yang:2020zjl} suggest that
\begin{itemize}

\item The $[\mathbf{6}_F, 1, 0, \rho]$ doublet contains six charmed baryons: $\Sigma_c(1/2^-)$, $\Sigma_c(3/2^-)$, $\Xi'_c(1/2^-)$, $\Xi'_c(3/2^-)$, $\Omega_c(1/2^-)$, and $\Omega_c(3/2^-)$. The $\Omega_c(3000)^0$ can be interpreted as a $P$-wave $\Omega_c$ baryon with either $J^P = 1/2^-$ or $3/2^-$, belonging to this doublet. However, the total decay widths of the $\Sigma_c(1/2^-)$, $\Sigma_c(3/2^-)$, $\Xi'_c(1/2^-)$, and $\Xi'_c(3/2^-)$ states are predicted to be quite large.

\item The $[\mathbf{6}_F, 0, 1, \lambda]$ singlet contains three charmed baryons: $\Sigma_c(1/2^-)$, $\Xi'_c(1/2^-)$, and $\Omega_c(1/2^-)$. Their total decay widths are predicted to be very large, making them difficult to observe experimentally.

\item The $[\mathbf{6}_F, 1, 1, \lambda]$ doublet contains six charmed baryons: $\Sigma_c(1/2^-)$, $\Sigma_c(3/2^-)$, $\Xi'_c(1/2^-)$, $\Xi'_c(3/2^-)$, $\Omega_c(1/2^-)$, and $\Omega_c(3/2^-)$. 
The $[\mathbf{6}_F, 2, 1, \lambda]$ doublet also includes six charmed baryons: $\Sigma_c(3/2^-)$, $\Sigma_c(5/2^-)$, $\Xi'_c(3/2^-)$, $\Xi'_c(5/2^-)$, $\Omega_c(3/2^-)$, and $\Omega_c(5/2^-)$. 
Our results suggest the following interpretations:
\begin{enumerate}[label=\alph*)]

\item The $\Xi_c(2923)^0$ and $\Omega_c(3050)^0$ can be interpreted as $P$-wave $\Xi'_c$ and $\Omega_c$ baryons with $J^P = 1/2^-$, belonging to the $[\mathbf{6}_F, 1, 1, \lambda]$ doublet.
    
\item The $\Omega_c(3119)^0$ can be interpreted as a $P$-wave $\Omega_c$ baryon with $J^P = 5/2^-$, belonging to the $[\mathbf{6}_F, 2, 1, \lambda]$ doublet.
    
\item The $\Xi_c(2939)^0$, $\Xi_c(2965)^0$, $\Omega_c(3066)^0$, and $\Omega_c(3090)^0$ can be interpreted as $P$-wave $\Xi'_c$ and $\Omega_c$ baryons with $J^P = 3/2^-$, arising from the mixing of the $[\mathbf{6}_F, 1, 1, \lambda]$ and $[\mathbf{6}_F, 2, 1, \lambda]$ doublets.

\item The $\Sigma_c(2800)^0$ can be interpreted as a combination of the $P$-wave $\Sigma_c$ baryons from the $[\mathbf{6}_F, 1, 1, \lambda]$ and $[\mathbf{6}_F, 2, 1, \lambda]$ doublets. We propose further investigation of this state to assess whether it can be resolved into multiple excited charmed baryons.

\end{enumerate}

\end{itemize}
Summarizing the above results, we identify four $\Sigma_c$, four $\Xi'_c$, and six $\Omega_c$ baryons with relatively narrow decay widths (less than 100~MeV), making them viable candidates for experimental observation. Their masses, mass splittings within the same multiplets, and decay properties are summarized in Table~\ref{tab:result4}, with possible experimental counterparts also listed for comparison. Note that there is still room for the $\Xi_c(2882)^0$, observed by LHCb in 2023~\cite{LHCb:2022vns}, while the interpretations of $\Xi_c(2923)^0$, $\Xi_c(2939)^0$, and $\Xi_c(2965)^0$ require slight adjustments.

We note once again that, within our QCD sum rule approach, the $\rho$-mode multiplet $[\mathbf{6}_F, 1, 0, \rho]$ is predicted to lie below the $\lambda$-mode multiplets $[\mathbf{6}_F, 1, 1, \lambda]$ and $[\mathbf{6}_F, 2, 1, \lambda]$, in agreement with the results presented in Sec.~\ref{sec4.3}. In this context, we propose a detailed experimental investigation of the $\Omega_c(3000)^0$ as a potential $\rho$-mode candidate to confirm the existence of $\rho$-mode heavy baryons.

A full QCD sum rule analysis was conducted in Ref.~\cite{Aliev:2018ube}, where the authors considered the $\Xi_c(2930)^0$ as both an orbitally excited state and a radially excited state. Notably, this state was later resolved into two distinct resonances, $\Xi_c(2923)^0$ and $\Xi_c(2939)^0$, by the LHCb Collaboration~\cite{LHCb:2020iby}. In their work, the authors calculated the mass and decay width of the $\Xi_c(2930)^0$ to the $\Lambda_c K$ channel. Their results favor interpreting it as an orbitally excited $\Xi_c$ baryon with $J^P = 1/2^-$. A similar study was conducted by the same group to investigate the excited $\Omega_c$ baryons~\cite{Agaev:2017lip}. Their results suggest assigning the $\Omega_c(3000)^0$, $\Omega_c(3050)^0$, and $\Omega_c(3119)^0$ to the excited $\Omega_c$ baryons with quantum numbers $(1P,\,1/2^-)$, $(1P,\,3/2^-)$, and $(2S,\,3/2^+)$, respectively. The $(2S,\,1/2^+)$ baryon can be assigned either to the $\Omega_c(3066)^0$ or the $\Omega_c(3090)^0$. Concerning the spin and parity quantum numbers, the naive quark model assigns, from lower to higher states, $1/2^-$, $1/2^-$, $3/2^-$, $3/2^-$, $5/2^-$~\cite{Luo:2023sra,Yoshida:2015tia}, while recent coupled channel model predict $1/2^-$, $3/2^-$, $3/2^-$, $5/2^-$, $3/2^-$~\cite{Zhang:2025gar}. The different predictions by different approaches should be tested by experiments or by first-principle calculations of lattice simulations.

Another full QCD sum rule analysis was conducted in Ref.~\cite{Wang:2017xam}, where the authors studied the $1S$, $1P$, $2S$, and $2P$ $\Omega_c$ states with $J = 1/2$ and $3/2$. Their results support assigning the $\Omega_c(3000)^0$ to the $1P$ state with $J^P = 1/2^-$, the $\Omega_c(3090)^0$ to either the $1P$ state with $J^P = 3/2^-$ or the $2S$ state with $J^P = 1/2^+$, and the $\Omega_c(3119)^0$ to the $2S$ $\Omega_c$ state with $J^P = 3/2^+$. The above three studies~\cite{Aliev:2018ube,Agaev:2017lip,Wang:2017xam} do not provide detailed discussions on the $\rho$-mode and $\lambda$-mode excitations, as the authors assume that a baryonic current simultaneously couples to both positive- and negative-parity states. This technique is also employed in our recent study of excited $\Omega_c$ baryons~\cite{Su:2024lzy}.

\section{Summary and outlook}
\label{sec5}

Recent years have seen significant advances in the study of singly heavy baryons. In particular, several experiments have unveiled the fine structure of the heavy baryon system, revealing their rich internal dynamics. For example, in 2017 the LHCb collaboration reported the simultaneous observation of five excited $\Omega_c$ baryons in the $\Xi_c K$ invariant mass spectrum~\cite{LHCb:2017uwr}. Later in 2020, LHCb observed four excited $\Omega_b$ baryons in the $\Xi_b K$ invariant mass spectrum~\cite{LHCb:2020tqd}, and in the same year, LHCb also observed three excited $\Xi_c$ baryons in the $\Lambda_c K$ invariant mass spectrum~\cite{LHCb:2020iby}. Additionally, many other collaborations, such as ATLAS, Belle-II, BESIII, and CMS, have made significant contributions to this field.

Some of the excited heavy baryons discussed above are good candidates for $P$-wave heavy baryons. Various theoretical approaches and phenomenological models have been developed to investigate their properties, with the QCD sum rule method being one of the most widely applied methods. Building on our previous reviews~\cite{Chen:2016spr,Chen:2022asf}, this paper provides a concise overview of these QCD sum rule studies. In particular, over the past decade, we have systematically studied $P$-wave singly heavy baryons using QCD sum rules and light-cone sum rules within the framework of heavy quark effective theory. Our results indicate the following:
\begin{itemize}

\item The $\Lambda_b(5912)^0$, $\Lambda_b(5920)^0$, $\Xi_b(6087)^0$, and $\Xi_b(6095)^0/\Xi_b(6100)^-$ can be interpreted as $P$-wave bottom baryons of the $SU(3)$ flavor $\mathbf{\bar{3}}_F$.

\item The $\Lambda_c(2595)^+$, $\Lambda_c(2625)^+$, $\Xi_c(2790)^{0/+}$, and $\Xi_c(2815)^{0/+}$ are their charmed counterparts.

\item The $\Sigma_b(6097)^\pm$, $\Xi_b(6227)^-$, $\Omega_b(6316)^-$, $\Omega_b(6330)^-$, $\Omega_b(6340)^-$, and $\Omega_b(6350)^-$ can be interpreted as $P$-wave bottom baryons of the $SU(3)$ flavor $\mathbf{6}_F$.

\item The $\Sigma_c(2800)^0$, $\Xi_c(2923)^0$, $\Xi_c(2939)^0$, $\Xi_c(2965)^0$, $\Omega_c(3000)^0$, $\Omega_c(3066)^0$, $\Omega_c(3090)^0$, $\Omega_c(3050)^0$, and $\Omega_c(3119)^0$ can be interpreted as $P$-wave charmed baryons of the $SU(3)$ flavor $\mathbf{6}_F$, with room for the $\Xi_c(2882)^0$ still remaining.

\end{itemize}
Furthermore, our results suggest the possible existence of several $P$-wave singly heavy baryons with relatively narrow decay widths, making them viable candidates for experimental observation. These include two $\Lambda_b$ states, two $\Xi_b$ states, two $\Lambda_c$ states, two $\Xi_c$ states, three $\Sigma_b$ states, three $\Xi_b^\prime$ states, two $\Omega_b$ states, three $\Sigma_c$ states, and one $\Omega_c$ state, as summarized in Tables~\ref{tab:result1}-\ref{tab:result4}.

Besides the QCD sum rule method, the constituent quark model has also been systematically employed to study singly heavy baryons. Predictions from these two approaches show reasonable agreement, although minor discrepancies remain~\cite{Chen:2022asf}. In the quark model the $\lambda$-mode orbital excitation leads to five possible $P$-wave $\Omega_c$ baryons: two with $J^P = 1/2^-$, two with $J^P = 3/2^-$, and one with $J^P = 5/2^-$, as illustrated in Fig.~\ref{fig:categorization}. While these five states could potentially account for the five excited $\Omega_c$ baryons observed by LHCb~\cite{LHCb:2017uwr}, a subsequent measurement~\cite{LHCb:2021ptx} disfavors this scenario. Consistently, both constituent quark model~\cite{Wang:2017hej,Liang:2020hbo} and QCD sum rule calculations~\cite{Yang:2020zrh,Yang:2021lce} suggest that one of the $\lambda$-mode states---specifically the one with $J^P = 1/2^-$---may have a broad decay width, rendering it difficult to observe experimentally. This discrepancy raises a key question: \emph{how should the remaining $\Omega_c$ state observed by LHCb be interpreted, given that only four can be accommodated within the $\lambda$-mode framework?} Plausible interpretations have been proposed: a radial ($2S$) excitation, a $P$-wave excitation of the $\rho$-mode, or a coupled channel scenario of the three-quark states and meson-baryon molecular states~\cite{Zhang:2025gar}. A precise experimental determination of the quantum numbers of these states would be essential to distinguish among these scenarios.

Note that within our QCD sum rule approach, the $\rho$-mode excitations are predicted to lie below the $\lambda$-mode excitations in terms of mass. This behavior differs from the expectations of conventional quark models~\cite{Copley:1979wj,Yoshida:2015tia}. To verify the existence of $\rho$-mode heavy baryons, we propose several investigations, as outlined in Sec.~\ref{sec4}. We can also pose a broader question: \emph{Can all possible $P$-wave singly heavy baryons be observed experimentally}, considering that some are predicted to have large decay widths based on QCD sum rule calculations? This leads to an even more fundamental question: \emph{What is the shortest possible lifetime of an observable particle}~\cite{Chen:2023zlj}? The various approximate flavor symmetries within the heavy baryon system—arising from the differences among the up, down, strange, charm, and bottom quarks—are key to addressing this question. Moreover, they provide an opportunity to explore the following fundamental problem: \emph{How can one generally describe approximate (flavor) symmetries}~\cite{Chen:2025way}?

The synergy of theoretical and experimental approaches has greatly advanced our understanding of the fine structure of singly heavy baryons, revealing their complex internal dynamics. Experimental observations, phenomenological analyses, and lattice QCD simulations have all made substantial contributions to this field, which remains both challenging and highly significant. We anticipate that the continued collaboration of these complementary approaches will provide deeper insights into singly heavy baryons, as well as other hadrons, and help clarify their nature in the near future.

\section*{Acknowledgments}

We are grateful for the inspiring and beneficial collaborative works with
Wei Chen,
Er-Liang Cui,
Veljko Dmitra{\v{s}}inovi{\'c},
Li-Sheng Geng,
Philipp Gubler,
Ding-Kun Lian,
Xiang Liu,
Yan-Rui Liu,
Qiang Mao,
Keitaro Nagata,
Wei-Han Tan,
Yi-Jie Wang,
Dan Zhou,
Zhi-Yong Zhou, and
Shi-Lin Zhu,
on the relevant topics.
H.X.C. is supported by the National Natural Science Foundation of China under Grant No.~12075019 and
the Jiangsu Provincial Double-Innovation Program under Grant No.~JSSCRC2021488.
A.H. is supported in part by the Grants-in-Aid for Scientific Research [Grant No.~24K07050(C)].
N.S. is supported by the National Natural Science Foundation of China under Grant No.~12505156.
H.M.Y. is supported by the China Postdoctoral Science Foundation under Grants No.~2024M750049.

\bibliographystyle{elsarticle-num}
\bibliography{ref}

\end{document}